\documentclass[twocolumn]{pnas-new}

\templatetype{pnasresearcharticle} 
\newcommand{\beginSI}{%
        \setcounter{table}{0}
        \renewcommand{\thetable}{S\arabic{table}}%
        \setcounter{equation}{0}
        \renewcommand{\theequation}{S\arabic{equation}}%
        \setcounter{figure}{0}
        \renewcommand{\thefigure}{S\arabic{figure}}%
         }

\usepackage{multirow}

\title{Planets Across Space and Time (PAST). \uppercase\expandafter{\romannumeral5}. The evolution of hot Jupiters revealed by the age distribution of their host stars}

\author[a,b,c]{Di-Chang Chen}
\author[a,b,1]{Ji-Wei Xie} 
\author[a,b,1]{Ji-Lin Zhou}
\author[d,e,1]{Subo Dong}
\author[a,b]{Jia-Yi Yang}
\author[f]{Wei Zhu}
\author[g]{Chao Liu} 
\author[h,g]{Yang Huang}
\author[g]{Mao-Sheng Xiang}
\author[i]{Hai-Feng Wang}
\author[j]{Zheng Zheng}
\author[g]{Ali Luo}
\author[g]{Jing-Hua Zhang}
\author[a,b]{Zi Zhu}

\affil[a]{School of Astronomy and Space Science, Nanjing University, Nanjing 210023, China}
\affil[b]{Key Laboratory of Modern Astronomy and Astrophysics, Ministry of Education, Nanjing 210023, China}
\affil[c]{LAMOST Fellow}
\affil[d]{Department of Astronomy, School of Physics, Peking University, Yiheyuan Rd. 5, Haidian District, Beijing, China, 100871}
\affil[e]{Kavli Institute for Astronomy and Astrophysics, Peking University, Beijing 100871, China}
\affil[f]{Department of Astronomy, Tsinghua University, Beijing, China, Beijing 100084}
\affil[g]{National Astronomical Observatories, Chinese Academy of Sciences, Beijing 100012, China}
\affil[h]{University of Chinese Academy of Sciences, Beijing 100049, China}
\affil[i]{Centro Ricerche Enrico Fermi, Via Pansiperna 89a, I-00184 Rome, Italy}
\affil[j]{Department of Physics and Astronomy, University of Utah, Salt Lake City, UT 84112}

\leadauthor{Chen} 

\significancestatement{{Hot Jupiters are the first exoplanet population discovered around main-sequence stars, which challenges the classical planetary formation theory based on our solar system. 
Even after a quarter century since the discovery of the first hot Jupiter, 51 Peg b, the origin and evolution of hot Jupiters remain puzzled.
Using a sample with stellar kinematic properties derived from large surveys of astrometry (e.g., Gaia) and spectroscopy (e.g., LAMOST), we characterize the kinematic ages of stars hosting hot and warm/cold Jupiters with the age-velocity dispersion relation, confirming the result of the previous study that hot Jupiter hosts are on average younger than field stars using a relative age proxy (Hamer \& Schlaufman 2019). 
Furthermore, we find that the frequency of hot Jupiters declines with age but that of warm/cold Jupiters does not vary with age. Such trends are naturally expected from tidal decays of hot Jupiters' orbits.} Our derived age and metallicity distributions of hot Jupiter hosts can contribute to helping explain puzzles involving hot Jupiter frequency. These include the long-standing discrepancy between the hot Jupiter frequencies inferred from RV and transit survey and the null detection of hot Jupiters in globular clusters.
}

\authorcontributions{J.-W.X. proposed this idea and initiated the collaboration. 
J.-W.X., J.-L.Z., S.D., and D.-C.C. designed the research.
D.-C.C., J.-W.X., and S.D. led the data analyses.
D.-C.C., J.-W.X., and S.D. analyzed the results and drafted the manuscript.
All authors contributed to discussing the results, editing, and revising the manuscript.}
\authordeclaration{The authors declare no conflict of interest.}
\correspondingauthor{\textsuperscript{1}To whom correspondence should be addressed. E-mail: jwxie@nju.edu.cn; zhoujl@nju.edu.cn; dongsubo@pku.edu.cn}


\begin{abstract}
The unexpected discovery of hot Jupiters challenged the classical theory of planet formation inspired by our solar system. Until now, the origin and evolution of hot Jupiters are still uncertain.
Determining their age distribution and temporal evolution can provide more clues into the mechanism of their formation and subsequent evolution. 
Using a sample of 383 giant planets around Sun-like stars collected from the kinematic catalogs of the Planets Across Space and Time (PAST) project, we find that hot Jupiters are preferentially hosted by relatively younger stars in the  Galactic thin disk.
We subsequently find that the frequency of hot Jupiters declines with age as {$F_{\rm HJ} \propto {\rm exp}(-0.20 \pm 0.06 \times \frac{t}{\rm Gyr})$}. In contrast, the frequency of warm/cold Jupiters shows no significant dependence on age. Such a trend is expected from the tidal evolution of hot Jupiters' orbits, and our result offers supporting evidence using a large sample. We also perform a joint analysis on the planet frequencies in the stellar age-metallicity plane. The result suggests that the frequencies of hot Jupiters and warm/cold Jupiters, after removing the age dependence are both correlated with stellar metallicities as $F_{\rm HJ} \propto 10^{\rm 1.6^{+0.3}_{-0.3} \times [Fe/H]}$ and $F_{\rm WJ/CJ} \propto 10^{\rm 1.1^{+0.2}_{-0.3} \times [Fe/H]}$, respectively. 
Moreover, we show that the above correlations can explain the bulk of the discrepancy in hot Jupiter frequencies inferred from the transit and radial velocity (RV) surveys, given that RV targets tend to be more metal-rich and younger than transits.
\end{abstract}

\dates{This manuscript was compiled on \today}
\doi{\url{www.pnas.org/cgi/doi/10.1073/pnas.XXXXXXXXXX}}

\begin{document}

\maketitle
\thispagestyle{firststyle}
\ifthenelse{\boolean{shortarticle}}{\ifthenelse{\boolean{singlecolumn}}{\abscontentformatted}{\abscontent}}{}

\dropcap{H}ot Jupiters generally refer to Jupiter-size planets with orbital periods $\lesssim$ 10 days around host stars. The existence of Jovian planets on such short periods poses significant challenges to the classical planet formation theories based on our solar system. Since the discovery of the prototype 51 Pegasi b in 1995 \citep{1995Natur.378..355M}, hot Jupiters have been one of the most studied exoplanet populations, with a number of fascinating properties uncovered. For example, in contrast to the co-planar planetary orbits in the solar system, the orbits of some hot Jupiters are found to be inclined with respect to their stars' equators  \citep[see a recent review by][]{2022PASP..134h2001A}. Statistical works have explored how hot Jupiters distribute and their dependence on the properties of host stars. Remarkably, hot Jupiters are preferentially found around metal-rich stars  \citep{2004A&A...415.1153S, 2005ApJ...622.1102F}. Despite efforts over nearly three decades, the origin of hot Jupiters remains puzzling \citep[see the review by][]{2018ARA&A..56..175D}.

An under-studied probe into the formation and evolution history of hot Jupiters is their dependence on the hosts' ages: Are there any differences in the hosts' age distributions between hot Jupiters and their counterparts at longer periods (i.e., warm/cold Jupiters)?
Does the frequency of hot Jupiters evolve with age? And if so, how?
The answers to these questions can constrain their origin, especially revealing whether/how the tidal interactions with host stars shape their orbits.

The main bottleneck of investigating the temporal evolution of hot Jupiters has been the difficulty of making decent stellar age estimates for the hosts. The commonly used isochrone fitting method has a typical age uncertainty exceeding $\sim 50\%$  \citep[e.g.,][]{2020AJ....159..280B} for an individual main-sequence host. Alternatively, the velocity dispersion for an assemble of stars is known to correlate with age. 
{The velocity dispersions have been previously used as a relative age proxy to study the hosts of hot Jupiters \citep{2013ApJ...772..143S, 2019AJ....158..190H, 2022AJ....164...26H}.
Specifically, Hamer \& Schlaufman (2019) found that the hosts of hot Jupiters are on average younger than the field stars, which can be interpreted by the tidal inspiral of hot Jupiters around hosts with modified stellar tidal quality factor $Q^{'}_{*} \lesssim 10^{7}$ \citep{2019AJ....158..190H}.
In a subsequent work, they showed that hot Jupiter host stars with larger obliquities are older compared to the aligned systems, suggesting those misaligned hot Jupiters arrived at their presently short-period orbits at late times \citep{2022AJ....164...26H}.}

\begin{figure*}[!t]
\centering
\includegraphics[width=0.95\linewidth]{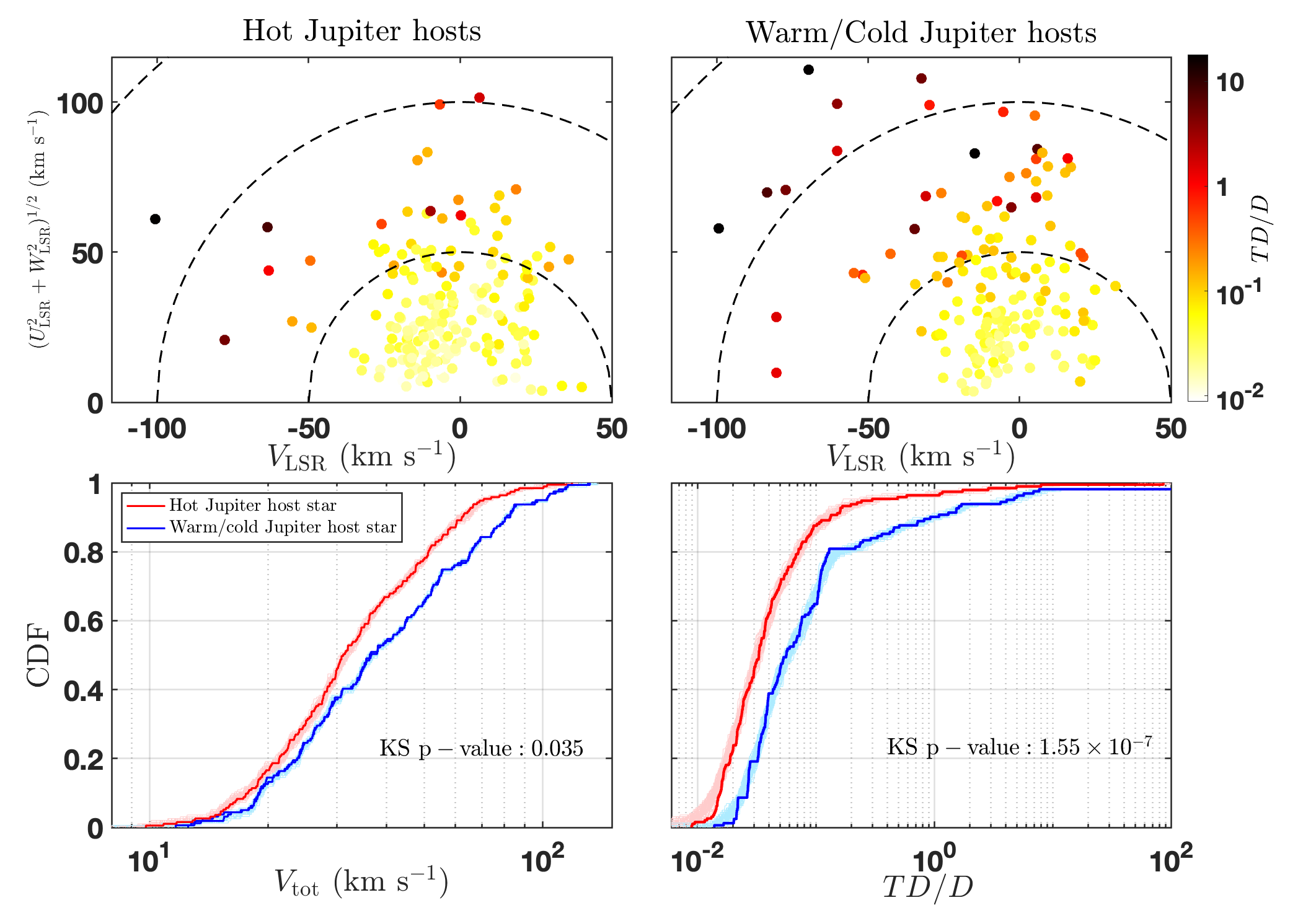}
\caption{
Top panels: {The Toomre diagrams color-coded by the relative probabilities of thick disk (TD) over thin disk (D), $TD/D$} for the hot Jupiter host stars (Top-Left panel) and warm/cold Jupiter host stars(Top-Right panel).
Dotted lines show constant values of the total  Galactic velocity $V_{\rm tot} = (U_{\rm LSR}^2+V_{\rm LSR}^2+W_{\rm LSR}^2)^{1/2}$ in steps of 50 $\rm km \ s^{-1}$.
Bottom panels: The cumulative distributions of the total velocities $V_{\rm tot}$ (Bottom-Left panel) and the relative probabilities between thick disk (TD) to thin disk (D), $TD/D$ (Bottom-Right panel) for the hot Jupiter host stars (red) and warm/cold Jupiter host stars (blue).
The two-sample K-S test $p-$ values are plotted at the lower right corner of each panel.
{To evaluate the significance of the differences in $V_{\rm tot}$ and $TD/D$ between hot Jupiter hosts and warm/cold Jupiter hosts, we also plotted the cumulative distributions of the 1,000 sets of resampled data considering their uncertainties in light red/blue colors.}
\label{figToomrediagramVtotTDDHJCJ}}
\end{figure*}

The average age of an assemble of stars can be estimated statistically from their kinematics using the Age-Velocity dispersion Relation (AVR) \citep[e.g.,][]{1977A&A....60..263W,2009A&A...501..941H}. In the first two papers of the Planet Across Space and Time series \citep[hereafter referred to as PAST \uppercase\expandafter{\romannumeral1} and \uppercase\expandafter{\romannumeral2};][]{2021ApJ...909..115C,2022AJ....163..249C},
we have refined the AVR to derive the kinematic ages with inner uncertainties of $\sim 10\%-20\%$ and constructed catalogs of stellar kinematic properties (e.g., Galactic position, velocity, and component membership) by combining data from the LAMOST, Gaia, APOGEE, RAVE, {\it Kepler} and NASA Exoplanet Archive \footnote{\url{https://exoplanetarchive.ipac.caltech.edu}}.
In this work, benefiting from the kinematic methods and catalogs of the PAST series, we perform a statistical investigation into the age distribution and temporal evolution of hot Jupiters. {Specifically, we aim to characterizing the kinematic ages of stars hosting hot Jupiters and warm/cold Jupiters and deriving the frequency of hot Jupiters as functions of stellar age and metallicity. }

\section*{Analyses and Results}

We collect our stellar sample by selecting Sun-like stars in the Galactic disk from the planet host stellar catalog of PAST \uppercase\expandafter{\romannumeral1} \citep{2021ApJ...909..115C}.
We then crossmatch with the catalogs of confirmed planets and the {\it Kepler} DR 25 candidates from NASA exoplanet archive \citep[https://exoplanetarchive.ipac.caltech.edu;][]{2013PASP..125..989A} and select giant planets as our planetary sample.
Our sample contains 355 stars hosting 383 giant planets (See SI, Table \ref{tab:sampleprocedure}, Figure \ref{figPMplanetsample}).
Among these planets, 193 are hot Jupiters.
In \S \ref{sec.meth.sample} of SI Appendix, we provide more details on the sample selection.
In our sample, giant planets are discovered by various facilities (ground/space),  with radial velocity (RV) or transit methods. 
In specific, our planetary sample consists of 29 hot Jupiters and 40 warm/cold Jupiters detected by space-based facilities with transit method (ST), 147 hot Jupiters and 3 warm/cold Jupiters detected by ground-based facilities with transit method (GT), 17 hot Jupiters and 147 warm/cold Jupiters detected with RV method (RV) (see Table S2 of SI). \\

\noindent \textbf{Hot Jupiters are preferentially hosted by younger stars in the thin disk.} 
We first compare the kinematic properties of the hot Jupiter host stars to warm/cold Jupiter host stars.
Figure \ref{figToomrediagramVtotTDDHJCJ} displays their Toomre diagrams colored by the relative probabilities between thick disk to thin disk $TD/D$.
As can be seen, compared to warm/cold Jupiter hosts, hot Jupiter hosts have smaller total velocities $V_{\rm tot}$ and $TD/D$.
We perform the two-sample Kolmogorov-Smirnov (K-S) tests to evaluate the significance.
As shown in the bottom panels of Figure \ref{figToomrediagramVtotTDDHJCJ}, the resulting $p-$values are $0.035$ and $1.55 \times 10^{-7}$ for the $V_{\rm tot}$ and $TD/D$.
{We also resample the observed velocities from the normal distribution for 1,000 times and perform the two sample K-S tests for the resampled data.
Out of the 1,000 times of resampled data, the resulted $p-$values of $V_{\rm tot}$ and $TD/D$ are less than 0.05 and 0.003 for 971 and 999 times respectively, }
suggesting that hot Jupiter host stars statistically tend to belong to the thin disk with smaller Galactic velocity compared to the warm/cold Jupiter hosts. 

\begin{figure}[!t]
\centering
\includegraphics[width=\linewidth]{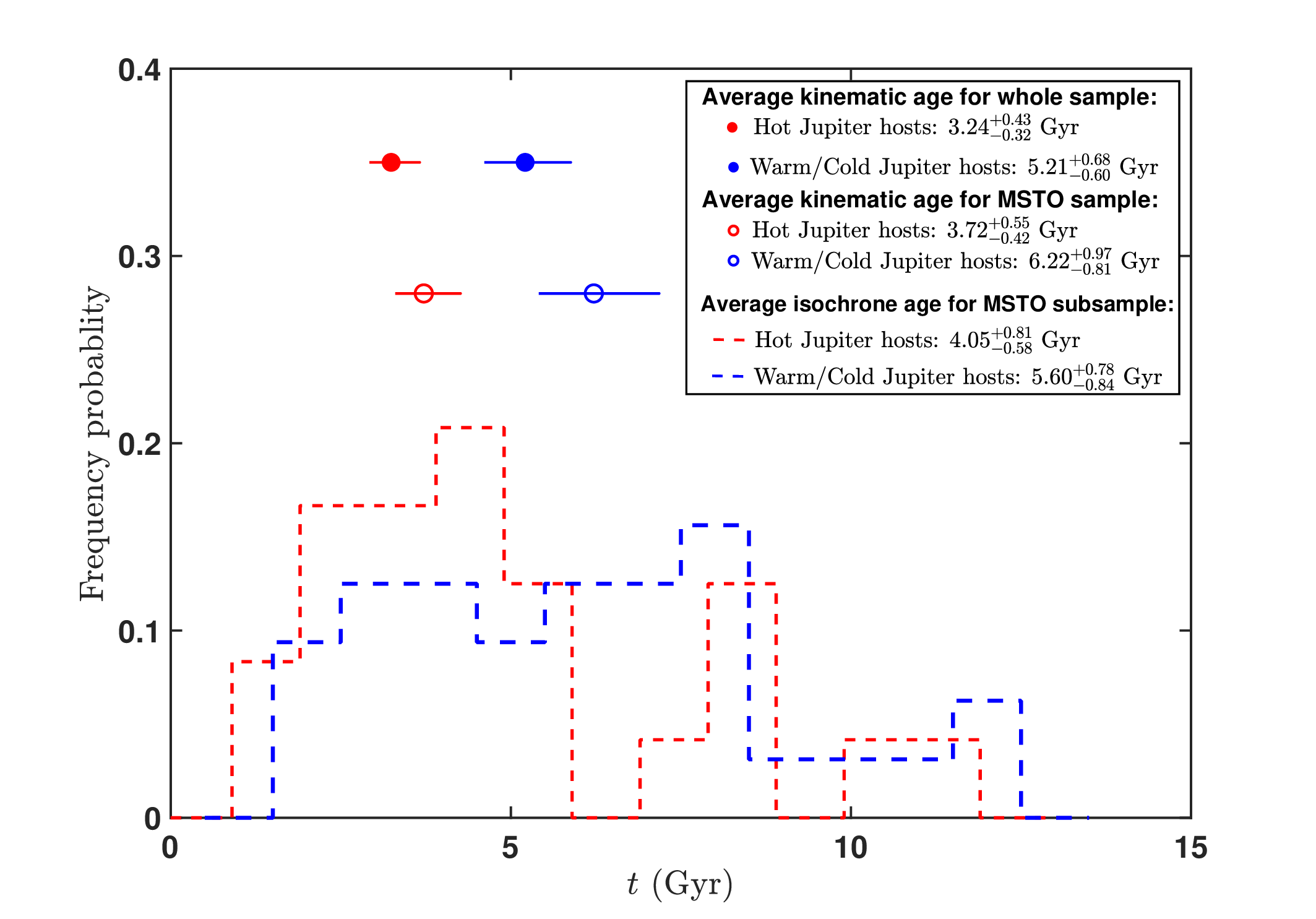}
\caption{The distributions of isochrone ages for the main-sequence turn-off (MSTO) subsamples of hot Jupiter hosts (dotted red histogram) and warm Jupiter hosts (dotted blue line histogram).
For comparisons, the average kinematic ages of the hot Jupiter and warm/cold Jupiter hosts in the MSTO subsample and the whole sample are plotted as hollow points and solid points, respectively.
As can be seen, the MSTO subsample are systematically older than the whole sample as expected, and hot Jupiter hosts are older than warm/cold Jupiter hosts in all the cases.
\label{figAgeHJCJ}}
\end{figure}

Previous work \citep{2019AJ....158..190H} using Gaia DR2 data found that the velocity dispersion of stars hosting hot Jupiter had smaller velocity dispersion than the field population, implying the former had a younger age on average. 
We calculate the average kinematic ages from the vertical velocity dispersions of hot Jupiter and warm Jupiter hosts with the refined AVR (see SI, \S~2.1). The resulting average ages are $3.24^{+0.43}_{-0.32}$ Gyr and $5.21^{+0.68}_{-0.60}$ Gyr for the hot Jupiter hosts and warm/cold Jupiter hosts, respectively (see the solid points in Figure \ref{figAgeHJCJ}).
Hot Jupiter host stars are $1.97^{+0.75}_{-0.72}$ Gyr younger than the warm/cold Jupiter host stars.
In order to evaluate the significance of the age difference, we resample the data (i.e., Galactic velocities and coefficients of AVR) from their uncertainties and recalculate the kinematic ages.
We find that the average kinematic ages of hot Jupiter host stars are younger 9,972 times out of 10,000 sets of resampled data, corresponding to a confidence level of 99.72\%.
It is worth noting that the kinematic age represents the average age of a group of stars.
To further verify the above result, we also consider a subsample of main-sequence turn-off (MSTO) stars having relatively well-determined individual ages with a typical uncertainty for an isochrone age of $\sim 20\%-30\%$ \citep[e.g.,][]{2017ApJS..232....2X,2004ApJS..155..667D} (see \S~2.2 of SI).
As shown in Figure \ref{figAgeHJCJ}, for the MSTO subsample, hot Jupiter hosts generally have younger isochrone ages compared to warm/cold Jupiter hosts.
The average isochrone age of hot Jupiter hosts ($4.05^{+0.81}_{-0.56}$ Gyr) is smaller than that of warm/cold Jupiter hosts ($5.60^{+0.78}_{-0.84}$ Gyr) by $\sim 1.55$ Gyr, which is consistent with the kinematic results from the MSTO subsample, i.e., $3.72^{+0.55}_{-0.42}$ Gyr for hot Jupiters and $6.22^{+0.97}_{-0.81}$ Gyr for warm/cold Jupiters. 
In addition, we also compare the age distributions for hot Jupiter hosts and warm/cold Jupiter hosts from other sources (see \S~2.2 and Figure \ref{figAgeHJCJ_othersource} in SI), which show the similar results.

In \S~\ref{sec.dis} of SI, we discuss the influences of other stellar properties (e.g., $\rm [Fe/H]$ and phase space density, Figure S27-S29) and show that they cannot (mainly) account for the above-mentioned observed age differences. 
Based on the above analyses, we conclude that hot Jupiters around Sun-like stars are preferentially hosted by younger stars in the Galactic thin disk. \\

\begin{figure}[!t]
\centering
\includegraphics[width=\linewidth]{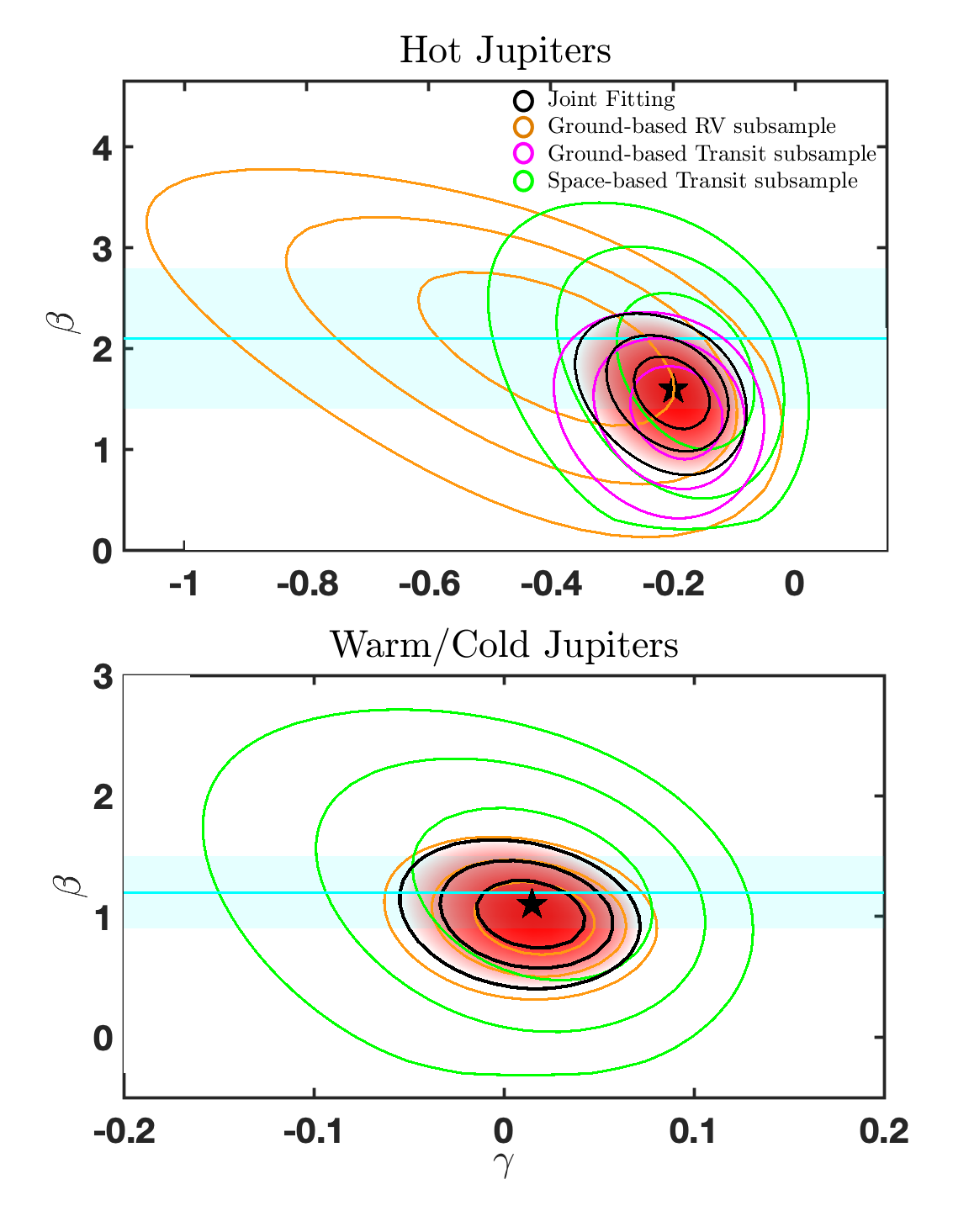}
\caption{Marginalized posterior probability density distributions for the metallicity and age exponents $(\beta, \gamma)$ of the planet frequency function for hot Jupiters (Top panel) and warm/cold Jupiters (bottom panel). The $1-\sigma$, $2-\sigma$, and $3-\sigma$ contours are displayed for RV, GT, and ST subsamples, and the joint fit in brown, purple, green, and black, respectively.
The solid pentagrams denote the best-fit values of the joint fit.
The cyan lines and regions represent the best-fit parameters and uncertainties of $\beta$ from previous studies \citep{2010PASP..122..905J,2017ApJ...838...25G}.
\label{figFHJCJ_JointGBSBRVTS_ABFittting}}
\end{figure}

\noindent \textbf{The frequency of hot Jupiters as a function of age.}
The above age difference implies that  the frequency of hot Jupiters evolves with age.
However, this result derived from planet hosts does not consider the detection efficiencies of field stars with null detection (which are the majority of survey targets) and may lead to biased conclusions.
Therefore, in this section, we further correct the detection biases and explore the temporal evolution of the frequencies (i.e., the intrinsic fractions of stars hosting planets) of hot Jupiters and warm/cold Jupiters. 
As mentioned before, our sample consists of three subsamples, i.e., RV, GT, and ST.
We construct the parent stellar samples for the three subsamples by using data from the Lick Planet Search, the LCES HIRES/Keck Precision Radial Velocity Exoplanet Survey, the public HAPRS RV database, Tycho-2/SuperWASP, and {\it Kepler}, respectively \citep[see \S~3.1.1, 3.2.1, 3.3.1 of SI;][]{2014ApJS..210....5F,2017AJ....153..208B,2020A&A...636A..74T,2021AJ....162..100C,2000A&A...355L..27H,2007ASPC..366..187W,2007MNRAS.381..851C}.
For each subsample, we divide the giant planets and corresponding parent stars into several bins according to their kinematic ages and calculate the frequencies of hot Jupiters ($F_{\rm HJ}$) and warm/cold Jupiters ($F_{\rm WJ/CJ}$) in each bin by correcting the geometric effect and detection efficiency (see \S~3.1.2, 3.2.2 and 3.3.2 of SI).
Figures \ref{figGBRVOccurrencerateDE}, \ref{figFHJGBTSGCDE} and \ref{figKeplerOccurrencerateDE} show $F_{\rm HJ}$ and $F_{\rm WJ/CJ}$ as a function of average kinematic age derived from RV, GT, and ST subsamples, respectively.

\begin{figure*}[!t]
\centering
\includegraphics[width=\linewidth]{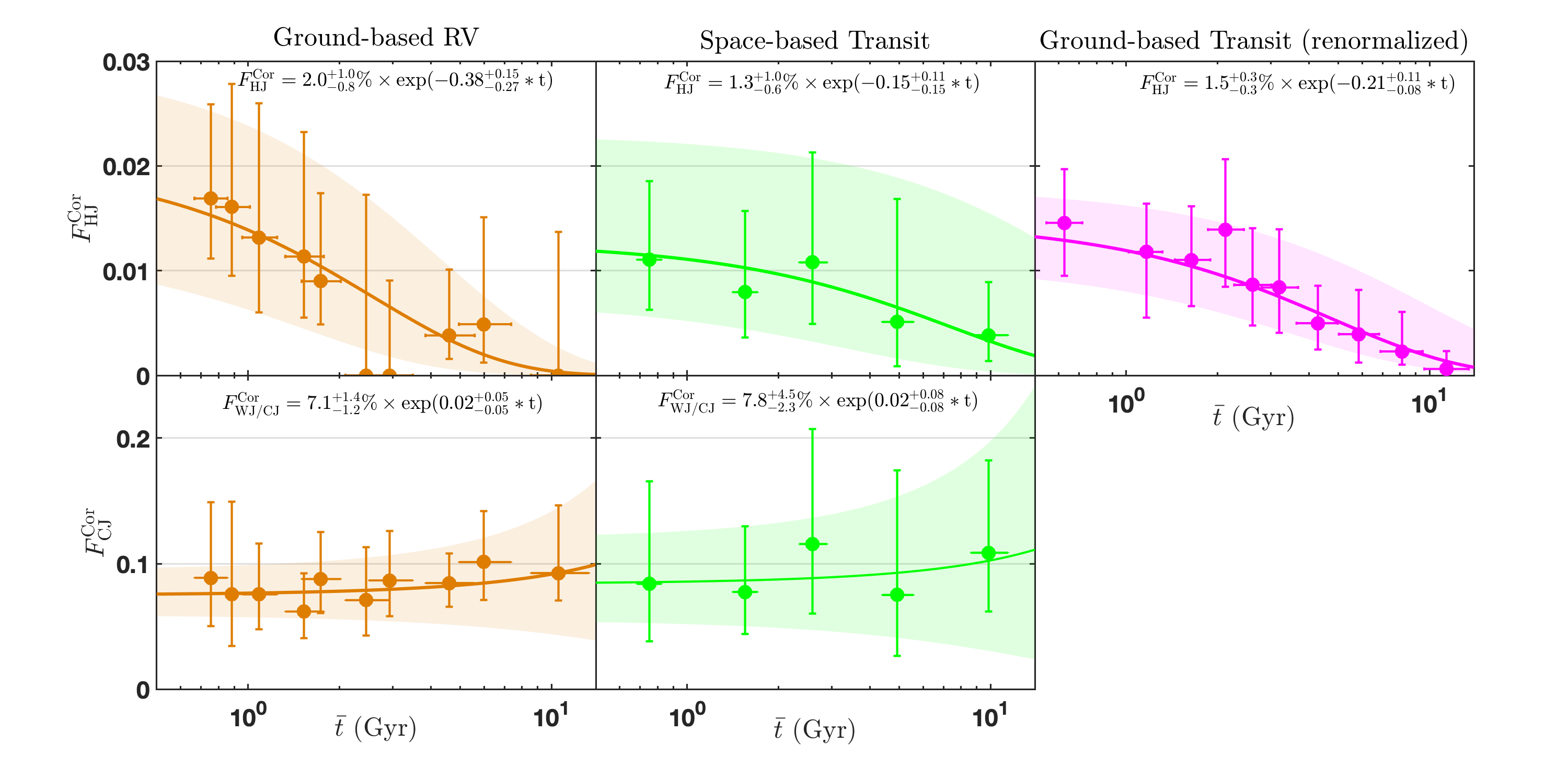}
\caption{The frequencies of hot Jupiters (Top panels) and warm/cold Jupiters (Bottom panels) as functions of average kinematic age for the ground-based RV (Left), space-based transit (Middle) and ground-based transit subsamples (Right), after normalizing the frequencies in each age bin to solar metallicity by applying the best-fit $\beta$.
The solid lines and regions denote the best-fit models and $1-\sigma$ intervals.
For the ground-based transit, we do not attempt to derive the absolute frequency (see \S \ref{sec.obs.GBTS}), and we re-normalized the amplitude to the intersection of the best fits of ground-based RV and
space-based transit for display-purpose only.
\label{figFHJCJRVTSSBGB}}
\end{figure*}

The kinematic age is known to be correlated with other stellar properties such as mass and metallicities \citep[e.g.,][]{2014A&A...562A..71B,2021AJ....162..100C}, which can also affect the frequency of giant planets \citep[e.g.][]{2010PASP..122..905J,2017ApJ...838...25G,2022arXiv221008313G}.
Since we only study Sun-like hosts, the stellar masses in different age bins do not differ significantly, while the stellar metallicities $\rm 
 [Fe/H]$ decrease with increasing age (see SI, Figure \ref{figGBRVTeffFeH}). 
 To qualify the dependence of $F_{\rm HJ}$ and $F_{\rm WJ/CJ}$ on stellar age/$\rm [Fe/H]$, we perform the Bayesian analysis for each of the three subsamples by modeling the planet frequency using three models: hybrid model $\rm [Fe/H]$ ($F({\rm [Fe/H]}, t) = C \times 10^{\beta \rm [Fe/H]} \times {\rm exp}(\gamma \times t)$), single-age model ($F({\rm [Fe/H]}, t) = C \times {\rm exp}(\gamma \times t)$, i.e., $\beta = 0$) and single-metallicity model ($F({\rm [Fe/H]}, t) = C \times 10^{\beta \rm [Fe/H]}$, i.e., $\gamma = 0$) (see \S~3.1-3.3 of SI). 
 We also make a joint fit for the coefficients $\beta$ and $\gamma$ by combining the data from the three subsamples (note that we focus on deriving the relative frequency as functions of age and metallicity, and the normalization factors $C$ for all subsamples are set as free parameters, see \S~3.4 of SI).
 The hybrid model is preferred comparing to the single-age model and single-metallicity model with smaller Akaike information criterion  \citep[AIC;][]{CAVANAUGH1997201} scores for all the cases, demonstrating that the trends of $F_{\rm HJ}$ and $F_{\rm WJ/CJ}$ are the combined effects of varying age and $\rm [Fe/H]$.
The best-fit parameters and their $1-\sigma$ intervals for two-parameter samples are displayed in Figure \ref{figFHJfittingFeHAgeGBRV}, \ref{figFCJfittingFeHAgeGBRV}, \ref{figFHJfittingFeHAgeGBTS}, \ref{figFHJfittingFeHAgeKepler}, \ref{figFCJfittingFeHAgeKepler} and summarized in Table~\ref{tab:fittingparamodel} of SI.
The best fits of the hybrid model can be mathematically expressed as:
{
\begin{eqnarray}
{F_{\rm HJ}} \propto 10^{(1.6^{+0.3}_{-0.3} \times \rm [Fe/H])} \times {\rm exp}(-0.20^{+0.06}_{-0.06} \times \frac{t}{\rm Gyr}),
\end{eqnarray}
\begin{eqnarray}
{F_{\rm WJ/CJ}} \propto 10^{(1.1^{+0.2}_{-0.3} \times \rm [Fe/H])} \times {\rm exp}(0.02^{+0.02}_{-0.03} \times \frac{t}{\rm Gyr}),
\end{eqnarray}}

Figure \ref{figFHJCJ_JointGBSBRVTS_ABFittting} shows the marginalized posterior probability density distributions of $F_{\rm HJ}$ and $F_{\rm WJ/CJ}$ from the joint fit and each of the three subsamples. 
For each subsample, the frequencies of hot Jupiters and warm/cold Jupiters are positively correlated ($\beta>0$) with stellar metallicity $\rm [Fe/H]$ at the $\gtrsim 3 \sigma$ levels. Moreover, the dependence of $F_{\rm HJ}$ on stellar $\rm [Fe/H]$ ($\beta \sim 1.5-1.8$) is stronger than that of $F_{\rm WJ/CJ}$ ($\beta \sim 1.0-1.2$). 
That is to say, giant planets, especially hot Jupiters, tend to be hosted by metal-richer stars, which is in general agreement with previous studies \citep{2010PASP..122..905J,2017ApJ...838...25G}.

To isolate the effect of stellar age, in each age bin, we calculate the planet frequency normalized to the solar metallicity (i.e., $\rm [Fe/H]=0$) according to the best-fits of $\beta$ (see Equation~\ref{normfeh} and \ref{corrfeh} in \S~3.1.3 of SI). 
Figure \ref{figFHJCJRVTSSBGB} shows the metallicity-``corrected'' frequencies of hot Jupiters $F^{\rm Cor}_{\rm HJ}$ and warm/cold Jupiters $F^{\rm Cor}_{\rm WJ/CJ}$ for RV (brown), ST (green), 
and GT (purple), respectively. 
As can be seen, as kinematic age increases, the frequencies of hot Jupiters decline at confidence levels of 98.45\%, 94.23\%, 99.60\%, and 99.99\% for RV, GT, ST, and the joint fit, respectively.
In comparison, warm/cold Jupiter frequencies are consistent with having no dependence on age as the fitted $\gamma$ for $F_{\rm WJ/CJ}$ are consistent with zero within $1\,\sigma$.


\section*{Discussions and Conclusions}
In this work, based on the giant planet sample selected from PAST \uppercase\expandafter{\romannumeral1},
we find that hot Jupiter host stars tend to be in the Galactic thin disk and are younger than warm/cold Jupiter hosts by $\sim 2$\,Gyr on average (Figure \ref{figToomrediagramVtotTDDHJCJ} and \ref{figAgeHJCJ}).
Then we derive the frequencies of hot Jupiters and warm/cold Jupiters as functions of stellar age and metallicities.
We find that the frequency of warm/cold Jupiters shows no significant dependence on age while the frequency of hot Jupiters decreases with age significantly (Equation 1 and 2).
Our sample shows the frequencies of hot Jupiters and warm/cold Jupiters exhibit positive correlations with stellar metallities, which are in qualitative agreement with previous studies. 
Nevertheless, the best-fit slope $\beta$ $(\sim1.6)$ of $F_{\rm HJ}$ is somewhat shallower than that of previous work \citep[$\sim2.1$;][]{2017ApJ...838...25G}, which is expected because the age dependence was not considered previously and thus $\beta$ was over-estimated.

{Age distribution of hot Jupiters could have important implications on their formation and evolution. 
There have been various proposed mechanisms (e.g., in-situ formation, disk migration, and high-eccentricity migration) to form hot Jupiters and they operate on different timescales (see the review by \citep{2018ARA&A..56..175D} and the references therein).
Specifically, in the cases of in-situ formation and disk migration, hot Jupiters should be formed early before the gas disks dissipate (within $\lesssim 10$ Myr).
In contrast, in the case of high-eccentricity migration, hot Jupiters could be delivered into $\lesssim 0.1$ AU through the whole lifetime of the host stars.
The arrival timescales for hot Jupiters to reach within $\sim 0.1$ AU may impact how the observed hot Jupiter frequency depends on age: the observed hot Jupiter frequency declining with age may weaken the case for high-eccentricity migration to be the predominant channel if it could not deliver the majority of hot Jupiters in place at relatively early stage.
In other word, our results imply that the bulk of hot Jupiters may arrive relatively early since the birth of their hosts because otherwise, the late arrived hot Jupiter can lead to an increase in the hot Jupiter frequency hosted by relatively old stars (at least in certain age intervals).}

{Tidal effects (e.g., stellar equilibrium/dynamic tide, planetary equilibrium/obliquity tide) have been widely considered essential for the evolution of hot Jupiters' short-period orbits \citep[see, e.g.,][]{2008ApJ...678.1396J, 2009ApJ...692L...9L,2012MNRAS.423..486L,2018ApJ...869L..15M,2018ARA&A..56..175D,2022Univ....8..211E}.
The tidal inspiral timescale depends on the planetary mass, orbital period, stellar mass, radius, and tidal quality factor \citep[see, e.g.,][]{2008ApJ...678.1396J,2012MNRAS.423..486L}}.
{Many previous studies have looked for observational evidence for the tidal decays from individual \citep[e.g., the orbital decay of hot Jupiters;][]{2009ApJ...698.1357J,2020ApJ...888L...5Y,2021AJ....161...72T,2021AJ....162..210D,2022AJ....163..281T} or ensemble properties of hot Jupiter systems \citep[e.g., the observed distributions of orbital distance/eccentricity;][]{2012ApJ...750..106S, 2015ApJ...798...66D,2017ApJ...836L..24W,2019AJ....157..217B} as well as the properties (e.g., spin and Galactic velocity distributions) of hot Jupiter host stars \citep{2019AJ....158..190H,2021ApJ...919..138T}.
Nevertheless, the tidal evolution of hot Jupiters remains uncertain. The stellar tidal quality factors derived from different works vary on several orders of magnitude \citep[e.g., $\sim 10^5$ from the tidal decay of WASP-12b and $\sim 10^{7-8}$ from the distribution of assemble properties;][]{2020ApJ...888L...5Y,2021AJ....161...72T,2017ApJ...836L..24W,2019AJ....157..217B}. Due to tidal decay, some hot Jupiters may inject into the Roche limit within the stellar lifetimes and get tidally disrupted. Consequently, such a process could naturally result in a declining frequency of hot Jupiters as a function of age, like the observed trend found in our work. For warm/cold Jupiters at longer orbital periods, the tidal inspiral timescales are expected to be too long to play any (significant) role in their evolution, which agrees with our result that there is no observed change of their frequency with age. 
Moreover, one can even derive the stellar tidal factor by fitting the hot Jupiter frequency-age relation (Figure \ref{figFHJCJRVTSSBGB} and Equation 1) with tidal evolution models. 
Such analyses also rely on hot Jupiter formation mechanisms because different mechanisms operate on different timescales and they set the initial conditions for the tidal evolution afterwards. 
In summary, our observational results can potentially be used to simultaneously constrain stellar tidal quality factors and test various formation models (to be explored in a follow-up work Chen et al. in preparation).}

\begin{figure}[!t]
\centering
\includegraphics[width=\linewidth]{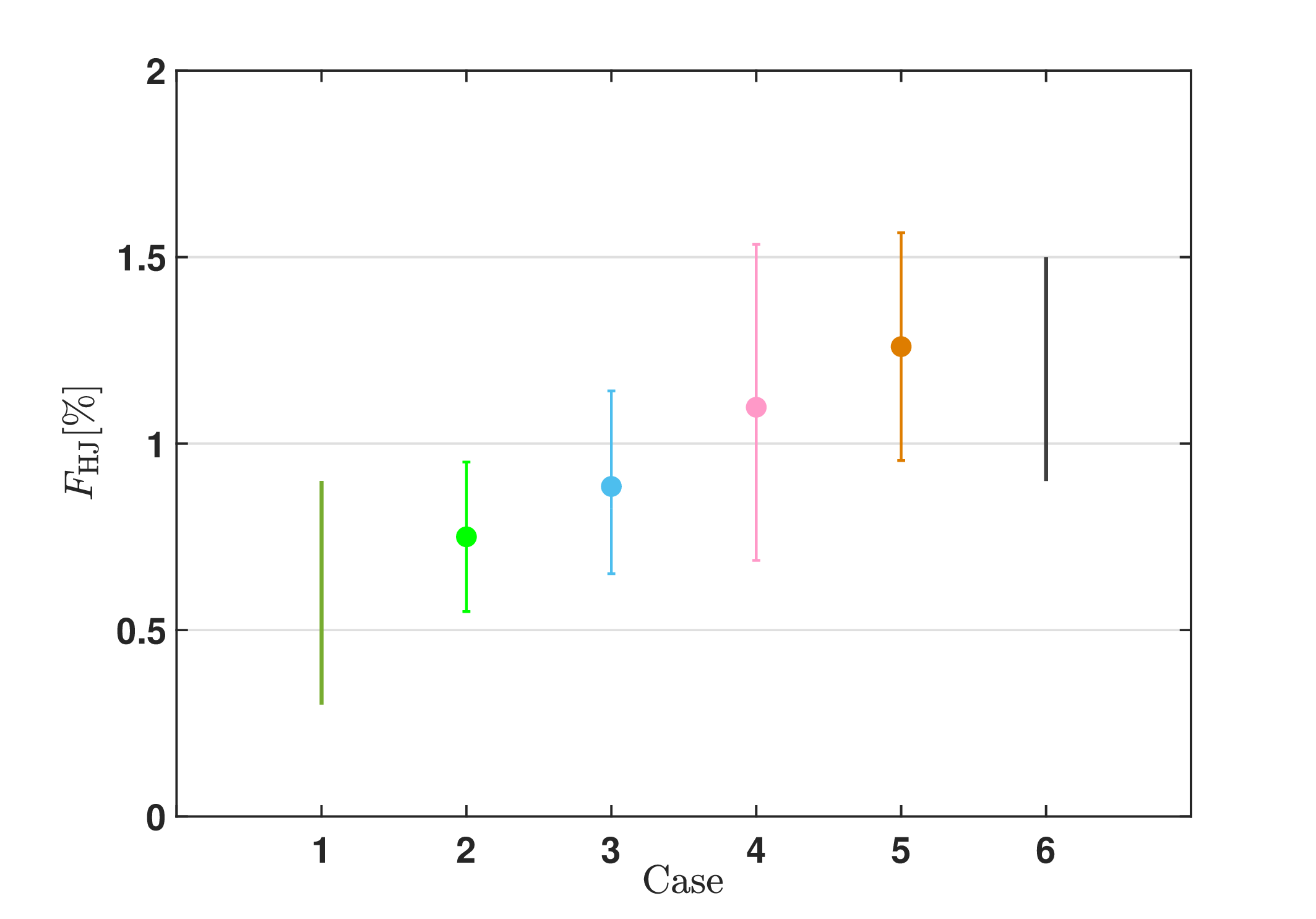}
\caption{The frequencies of hot Jupiters $F_{\rm HJ}$.
Case 1: Result derived in previous studies from transit surveys \citep[e.g.,][]{2006AcA....56....1G,2012ApJS..201...15H,2016A&A...587A..64S};
Case 2: Observed $F_{\rm HJ}$ derived in this work based on the {\it Kepler} sample;
Case 3: Modified $F_{\rm HJ}$ of the {\it Kepler} result by taking account of the $\rm [Fe/H]$ difference from the RV sample (see \S~4 of SI);
Case 4: Modified $F_{\rm HJ}$ of {\it Kepler} results by taking account of the $\rm [Fe/H]$ and kinematic age differences from the RV sample;
Case 5: Observed $F_{\rm HJ}$ derived in this work based on the RV sample;
Case 6: Results derived in previous studies from RV surveys \citep[e.g.,][]{2008PASP..120..531C,2011arXiv1109.2497M,2012ApJ...753..160W}.
\label{figFHJRVTSdiscrepency}}
\end{figure}

In the following, we discuss the implication of our results on addressing the discrepant frequencies of hot Jupiters derived from RV and transit surveys \citep[see, e.g.,][]{2021ARA&A..59..291Z}.
The frequencies of hot Jupiters inferred from RV surveys \citep[$\sim 0.9\%-1.5\%$; e.g.,][]{2008PASP..120..531C,2011arXiv1109.2497M,2012ApJ...753..160W} are significantly higher than those derived from transit surveys \citep[$\sim 0.3\%-0.8\%$; e.g.,][]{2006AcA....56....1G,2012ApJS..201...15H,2016A&A...587A..64S}.
Recently, some works have proposed that such a discrepancy could be potentially explained by the difference in the close binary fractions between RV and transit targets \citep{2021MNRAS.507.3593M,2022MNRAS.516...75B}. 
The close binaries may also influence warm Jupiters (giant planets within $\sim 0.1-1$\,AU) if this is the case.
Interestingly, there seems to be no significant difference in the frequencies of warm Jupiters between RV \citep[$\sim 4\%-5\%$;][]{2019ApJ...874...81F,2022AJ....164....5Z} and transit (e.g., $4.1 \pm 1.2\%$ from LAMOST-Gaia-Kepler catalog, see \S~4 of SI) surveys.
Here we investigate whether a combination of the metallicity and age effects could help address the frequency discrepancy. We compare the stellar and planetary properties of our RV and transit subsamples and find that (1) the metallicity of RV stars ($\rm [Fe/H] \sim 0.02$) are on average metal-richer than transit stars ($\rm [Fe/H] \sim -0.04$) (see SI, Figure \ref{figParentstarspropertiesSPGBTS}); (2) hot Jupiters discovered by transit method are on average $\sim 1$ Gyr older than RV subsamples (see SI, \ref{figGBTSstellarproperties}, \ref{figHoststarspropertiesSPGBTS} and \ref{figAgeHJRVTSEU}).
By using our derived $F_{\rm HJ}$ function (see Equation 1) and taking the age and metallicity differences from the RV surveys into account, we derive a modified $F_{\rm HJ}$ ($\sim 1.1\%$) from {\it Kepler} survey, which is consistent with the frequencies derived from RV surveys within $\sim1\,\sigma$ (as shown in Figure \ref{figFHJRVTSdiscrepency}; See detailed discussions in \S~4 of SI).

With the advantage of accurate stellar age and metallicity measurements, stellar clusters are attractive environments to search and study hot Jupiters. So far, searches in globular clusters composed of old stellar populations have yielded null detection, suggesting a lower frequency of hot Jupiters in globular clusters compared to those in field stars \citep{2000ApJ...545L..47G,2008ApJ...674.1117W,2012A&A...541A.144N,2017AJ....153..187M}.
In comparison, tens of hot Jupiters (candidates) have been discovered in open clusters containing young stars, and statistical studies mostly yield $F_{\rm HJ}$ higher than field stars after accounting for the effect of $\rm [Fe/H]$ \citep[see \S~5 of SI, Table \ref{tab:FHJClusters}; e.g.,][]{2012ApJ...756L..33Q,2014ApJ...787...27Q,2014A&A...561L...9B,2016A&A...592L...1B,2017A&A...603A..85B}. The above trends of hot Jupiter frequencies in globular and open clusters are in broad agreement with expected lower/higher frequencies of hot Jupiters with old/young stars from our results. In the following, we consider the null detection by Gilliland et al. (2003) from searching for hot Jupiters in the globular cluster 47 Tucanae. It was expected to detect $\sim 17$ hot Jupiters if assuming its hot Jupiter frequency to be the same as those in field stars, but zero was found \citep{2003ApJ...585.1056G}.
Masuda \& Winn (2017) revisited the problem by using the hot Jupiter frequency measured in the {\it Kepler} field and limiting their stellar masses to be in the same range as those in 47 Tucanae, and they found that the expected detection number is significantly reduced to $2.2^{+1.6}_{-1.1}$ \citep{2017AJ....153..187M}.
We find that by further considering the effect of age, the frequency of hot Jupiters in 47 Tucanae (age of $\sim 11.6$ Gyr) will decrease by another factor of $\sim 2.7$, and 
the expected number of detected hot Jupiters ($0.8^{+0.7}_{-0.5}$) is thus in little tension with the observed null result (see \S~5 of SI, Figure \ref{figNdet47Tuc}).

Future discoveries and analysis of hot Jupiters \citep[e.g.,][]{2022AJ....164...70Y,2023ApJS..265....1Y} may probe a broader range of stellar ages and metallicities in various stellar environments \citep[e.g., star formation regions, stellar clusters, associations, halo][]{2017MNRAS.467.1342Y,2018ARA&A..56..175D}, allowing for testing our results and offering better understanding on the age distribution and evolution of hot Jupiters.



\acknow{We thank Bo Ma, and Jing-Hua Zhang for helpful discussions and suggestions.
This work is supported by the National Key R\&D Program of China (No. 2019YFA0405100) and the National Natural Science Foundation of China (NSFC; grant No. 11933001, 11973028,  11903005, 11973028, 11933001, 11973028, 11933001, 12003027, 12150009, 12133005, 12173021) and the National Key R\&D Program of China (2019YFA0706601).
We also acknowledge the science research grants from the China Manned Space Project with NO.CMS-CSST-2021-B12 and CMS-CSST-2021-B09. 
J.-W.X. also acknowledges the support from the National Youth Talent Support Program.
D.-C.C. also acknowledges the Cultivation project for LAMOST Scientific Payoff, Research Achievement of CAMS-CAS and the fellowship of Chinese postdoctoral science foundation (2022M711566). 
S.D. acknowledges support by the New Cornerstone Science Foundation through the XPLORER PRIZE.
Funding for LAMOST (www.lamost.org) has been provided by the Chinese NDRC. LAMOST is operated and managed by the National Astronomical Observatories, CAS. This publication makes use of data products from the Two Micron All Sky Survey, which is a joint project of the University of Massachusetts and the Infrared Processing and Analysis Center/California Institute of Technology, funded by the National Aeronautics and Space Administration and the National Science Foundation. This research has made use of the NASA Exoplanet Archive, which is operated by the California Institute of Technology, under contract with the National Aeronautics and Space Administration under the Exoplanet Exploration Program. This paper makes use of data from the first public release of the WASP data (Butters et al. 2010 \citep{2010A&A...520L..10B}) as provided by the WASP consortium and services at the NASA Exoplanet Archive, which is operated by the California Institute of Technology, under contract with the National Aeronautics and Space Administration under the Exoplanet Exploration Program.
}

\showacknow


\clearpage

\beginSI 
\begin{center}
{ \LARGE  Supporting Information (SI)}\\[0.5cm]
\end{center}

\section{The sample}
\label{sec.meth.sample}
We initialize our stellar sample based on the planet host stellar catalog of PAST \uppercase\expandafter{\romannumeral1} \citep{2021ApJ...909..115C}, which provides stellar physical parameters, Galactic positions, Galactic velocities, membership probabilities of Galactic components for 2,174 stars hosting 2,872 planets.
The membership probabilities consist of the relative probabilities for the thick-disk-to-thin-disk $(TD/D)$, thick-disk to halo $(TD/H)$, Hercules-to-thin-disk $(Herc/D)$ and Hercules-to-thick-disk $(Herc/TD)$, where D, TD, Herc, and H denote the thin disk, the thick disk, the Hercules stream and the halo of the Milky Way galaxy, respectively.
The larger the relative probability is, the more likely a star belongs to the former component.
In this work, to avoid the influence of the stellar evolutionary stage, we only include Sun-like stars (i.e., single main-sequence FGK stars).
We select stars with effective temperatures $T_{\rm eff}$ in the range of $4700-6500$ K and surface gravities $\log g>4.0$. 
Besides, we only restrict to stars in the Galactic disk by excluding stars belonging to the halo and stellar streams since
the method to obtain kinematic age is only suitable for stars in the Galactic disk, as discussed in PAST I.
In specific, stars in the halo and Hercules stream are excluded by adopting $TD/D>1 \& Herc/D<0.5 \& Herc/TD<0.5$ according to Bensby et al. (2014) \citep{2014A&A...562A..71B}. 
We also removed other star streams (e.g., Arcturus, Sirius, and Pleiades/Hyades) by adopting the $U-V$ plane characteristics according to Kushniruk \& Bensby (2019) \citep{2019A&A...631A..47K}.

For the planetary sample, we select giant planets by adopting the following criterion: planetary mass ($M$ or $M\sin i$) in the range of $0.3-13$ Jupiter masses or planetary radius in the range of $6-20$ Earth radii.
Giant planets with orbital periods $P<10$ days are selected as hot Jupiters. 
After the above selections, there are 385 giant planets left.
Among them, most of the warm/cold Jupiters are discovered by RV, while the majority of hot Jupiters are discovered by transit. 
Only two warm/cold Jupiters are discovered by imaging.
Since the imaging subsamples are too small to carry out detailed exploration and the correction for their detection bias is difficult (e.g., collecting a homogeneous stellar sample), we also exclude these two warm/cold Jupiters. 
In finally, we are left with 355 stars hosting 383 giant planets (including 193 hot Jupiters and 190 warm/cold Jupiters with $P<30000$ days)

In Table \ref{tab:sampleprocedure}, we summarize the sample's composition after each step mentioned above.
Table \ref{tab:planetnumberdisc} shows the numbers of giant planets discovered by different methods (i.e., radial velocity (RV) and transit) and facilities (i.e., space-based and ground-based).

\begin{figure}[!t]
\centering
\includegraphics[width=\linewidth]{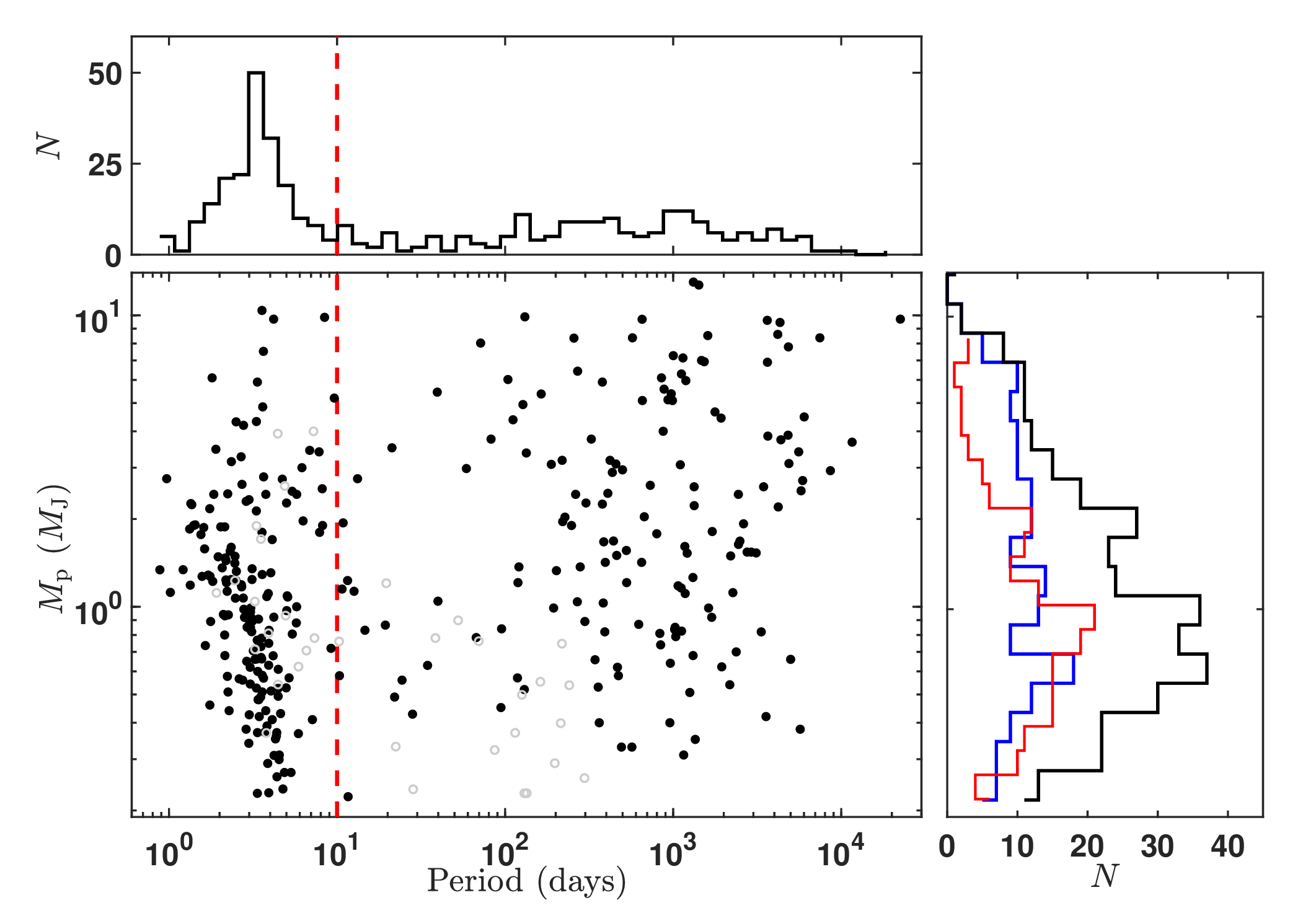}
\caption{The period-mass diagram for giant planetary sample in \S~1. 
Planets with mass measurements ($M$ or $M \sin i$) are plotted as solid black points.
{Open circles represent planets with no direct mass measurement, and their masses are estimated from their radii using a mass-radius relation.}
Histograms of masses for the hot Jupiters, warm/cold Jupiters and whole sample are plotted in red, blue and black respectively in the right panels.
Histogram of period is displayed in the toppest.
The vertical dashed lines represent where period $= 10$ days.
\label{figPMplanetsample}}
\end{figure}

In Figure \ref{figPMplanetsample}, we display the period-mass distribution of our planetary sample. 
For planets with radius measurements but no mass measurement (open circles), we estimate their masses from their radii by adopting the following formula: $\frac{M_{\rm p}}{M_{\rm J}} = \left( \frac{R_{\rm p}}{1.2 R_{\rm J}} \right)^{1/0.27}$ \citep{2016A&A...589A..75M}. 
As can be seen, the period distribution of hot Jupiters peaks around
$\sim 3-5$ days, which is consistent with previous studies \citep[e.g.,][]{2016A&A...587A..64S}.
It is also notable that hot Jupiters have a smaller median mass ($0.93 M_{\rm J}$) compared to that of the warm/cold Jupiters ($1.45 M_{\rm J}$). 

\section{Age distributions of host stars: hot Jupiters vs. warm/cold Jupiters}
\label{sec.obs.agedis}

\subsection*{2.1. Average kinematic ages}
\label{sec.obs.kineage}

In this section, we infer the age distributions of hot Jupiters and warm/cold Jupiter host stars from their kinematics by applying the Age-Velocity dispersion Relation (AVR) of PAST \uppercase\expandafter{\romannumeral1}, which gives
\begin{equation}
t = \left(\frac{\sigma}{k\rm \, km \ s^{-1}}\right)^{\frac{1}{b}}\, \rm Gyr,
\label{kineage}
\end{equation}
where t is stellar age, $\sigma$ is the velocity dispersion, $k$ and $b$ are two coefficients for AVR.
Throughout this work, we adopt the vertical velocity dispersion $\sigma_W$ to estimate the kinematic age. In the following, we explain our reasoning for this adoption. There may be some nearby star streams/associations in the solar neighborhood \citep[e.g.,][]{2016AstL...42...90B,2017A&A...608A..73K}.
Previous studies have shown that the distributions in $U_{\rm LSR}$, $V_{\rm LSR}$ and $L_Z$ of stars in the streams are clumpy \citep{2005A&A...430..165F,2012MNRAS.426L...1A,2017A&A...608A..73K,2018A&A...619A..72R}. Stars in a single stream (substream) have a very narrow range in $U_{\rm LSR}$ and $V_{\rm LSR}$ \citep[e.g.,][]{2019A&A...631A..47K}. As a result, there are relatively large deviations in AVR using $\sigma_U$ and $\sigma_V$ (e.g., Figure 9 of PAST \uppercase\expandafter{\romannumeral1}).
Therefore, we uniformly adopt the vertical velocity dispersion to estimate the kinematic age to avoid the effect of the deviations in $\sigma_U$ and $\sigma_V$ on the kinematic age caused by stellar streams.

To assess the kinematic ages' uncertainties, we use a Monte Carlo method by resampling the AVR coefficients ($k$ and $b$) and vertical velocity dispersions ($\sigma_W$) based on their uncertainties.
For $k$ and $b$, their values and uncertainties are adopted from Table 4 of PAST \uppercase\expandafter{\romannumeral1} \citep{2021ApJ...909..115C}, which are listed in Table \ref{tab:AVRkb}.
For $\sigma_W$, the value and uncertainty are calculated by resampling each star's Galactic vertical velocity from a normal distribution given its value and uncertainty.
The age uncertainty is set as the 50$\pm$34.1 percentiles in the resampled age distribution.

\begin{figure}[!t]
\centering
\includegraphics[width=\linewidth]{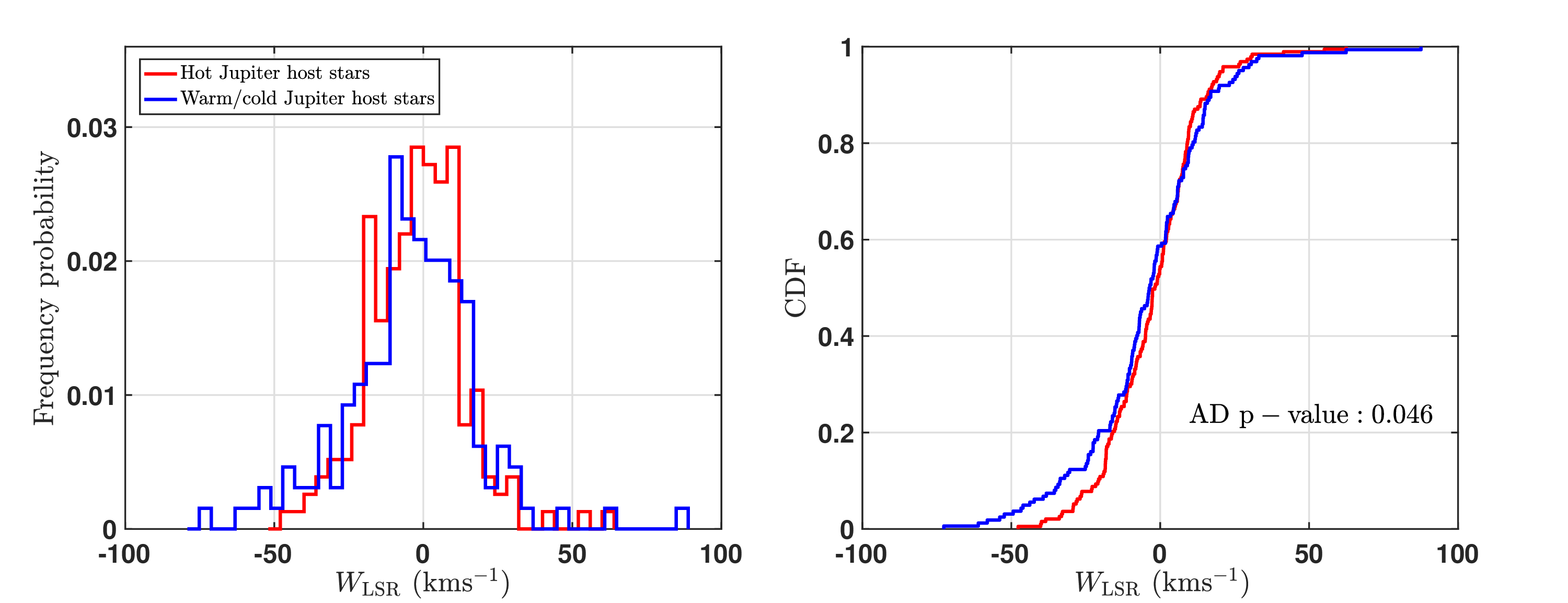}
\caption{The histograms (Left panel) and cumulative distributions (Right panel) of the vertical velocities $W_{\rm LSR}$ for hot Jupiter host stars (red) and warm/cold Jupiters (blue).
The two-sample Anderson-Darling (AD)  test $p-$ value is plotted at the lower right corner of the right panel.
Here we adopt the AD test since we focus on the dispersion of $W_{\rm LSR}$ (tailed distribution) and AD test is more sensitive at the tail than Kolmogorov-Smirnov test.
{The typical uncertainties of $W_{\rm LSR}$ are $\sim 1.8 \rm km \ s^{-1}$.}
\label{figWLSRHJCJ}}
\end{figure}

We first compare the distributions of vertical velocities $W_{\rm LSR}$ (as shown in Figure \ref{figWLSRHJCJ}) and
find that the $W_{\rm LSR}$ of hot Jupiters are distributed more concentrated around zero with a vertical velocity dispersion smaller than that of warm/cold Jupiter hosts, consistent with previous studies \citep{2019AJ....158..190H}.
Then we calculate their kinematic ages from vertical dispersions using Equation S1.
As shown in Figure \ref{figAgeHJCJ}, the kinematic ages are $3.24^{+0.43}_{-0.32}$ Gyr and $5.21^{+0.68}_{-0.60}$ Gyr for the hot Jupiter and warm/cold Jupiter hosts, respectively.
The difference between the kinematic ages of hot Jupiter and warm/cold Jupiter host stars, $\Delta t_{C-H} = 1.97^{+0.75}_{-0.72}$ Gyr. 
Out of the 10,000 sets of resampled data, the kinematic ages of hot Jupiter host stars are smaller than warm/cold Jupiters in 9,972 trials, corresponding to a confidence level of 99.72\%.

\begin{figure}[!t]
\centering
\includegraphics[width=\linewidth]{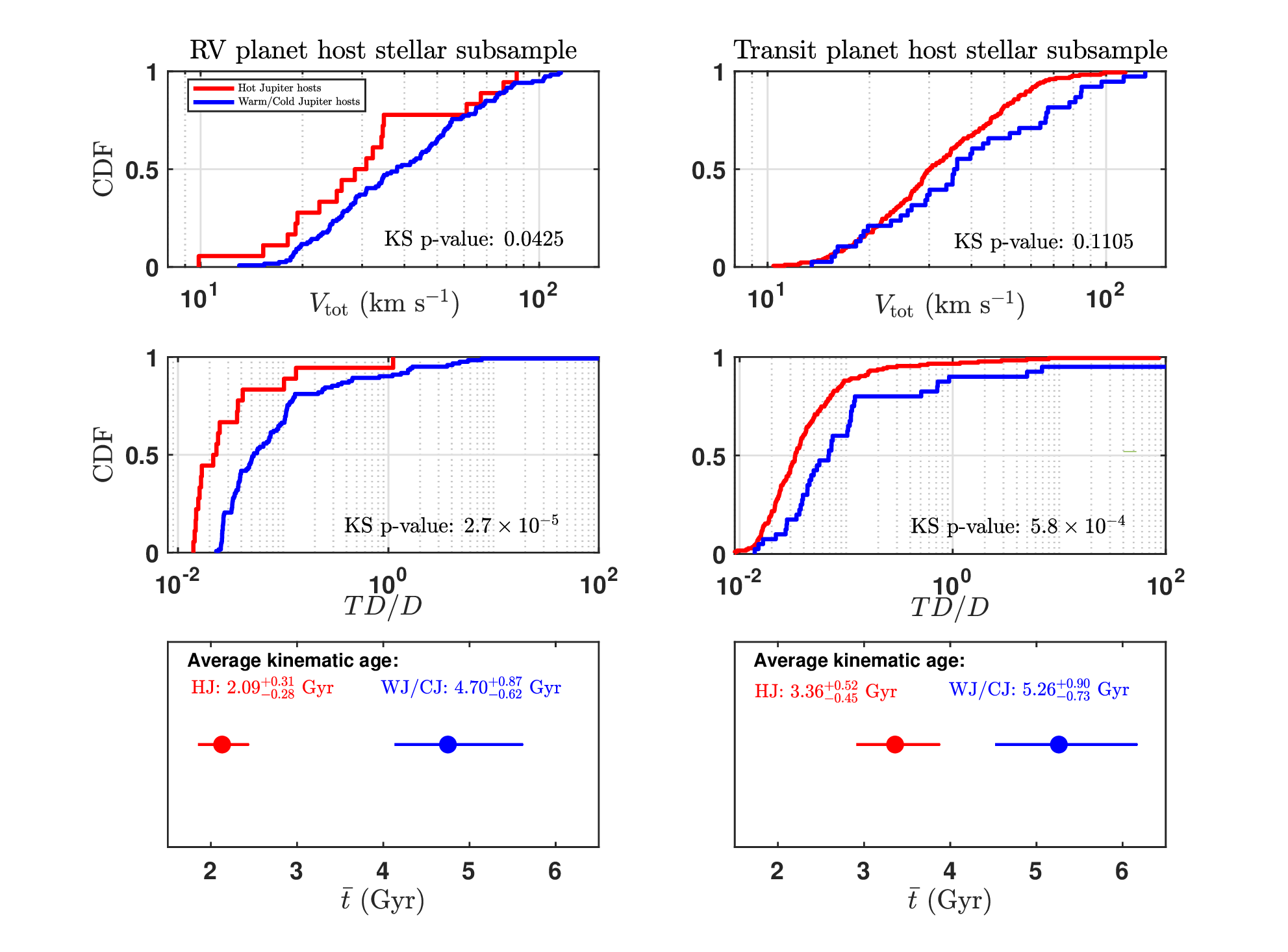}
\caption{The cumulative distributions of total velocities (Top), the probabilities of thick disk to thin disk TD/D (Middle) and kinematic (average) ages (Bottom) for the hot Jupiter host stars (red) and warm/cold Jupiter host stars (blue) derived from RV subsample (Left panel) and transit subsample (Right panel).
The typical (median) uncertainties of $V_{\rm tot}$ and $TD/D$ are $\sim 4 \rm km \ s^{-1}$ and $\sim 12\%$.
\label{figVtotTDDAgeHJCJ_RVTS}}
\end{figure}

{We also divide the whole planetary host sample into the RV and transit subsamples, and then compare the total Galactic velocities, the relative probabilities between thick disk to thin disk $TD/D$, and the average kinematic ages of hot Jupiters and warm/cold Jupiters.
As shown in Figure \ref{figVtotTDDAgeHJCJ_RVTS}, for both subsamples, hot Jupiter host stars have smaller total velocities $V_{\rm tot}$ and TD/D.
Furthermore, hot Jupiter hosts are on average kinematically younger than warm/cold Jupiter hosts 9,996 and 9,822 times out of 10,000 sets of resampled data for the RV and transit subsample respectively, corresponding to confidence levels of 99.96\% and 98.22\%.
These results are consistent with the results of combining the RV and transit subsamples in the whole sample (Figure 1 and 2), suggesting that the potential biases between RV and transit methods are unlikely to be responsible for the differences we see between the hot and warm Jupiters. }

In \S~\ref{sec.dis}, we also discuss the influences of other stellar properties (e.g., $\rm [Fe/H]$ and phase space density) and prove these are not the (main) reasons for the age differences. 
Based on the above analyses, we conclude that hot Jupiters around Sun-like stars are preferred around (kinematically) younger stars.

\subsection*{2.2 Age estimates with other methods}
\label{sec.obs.isoage}

\begin{figure}[!t]
\centering
\includegraphics[width=\linewidth]{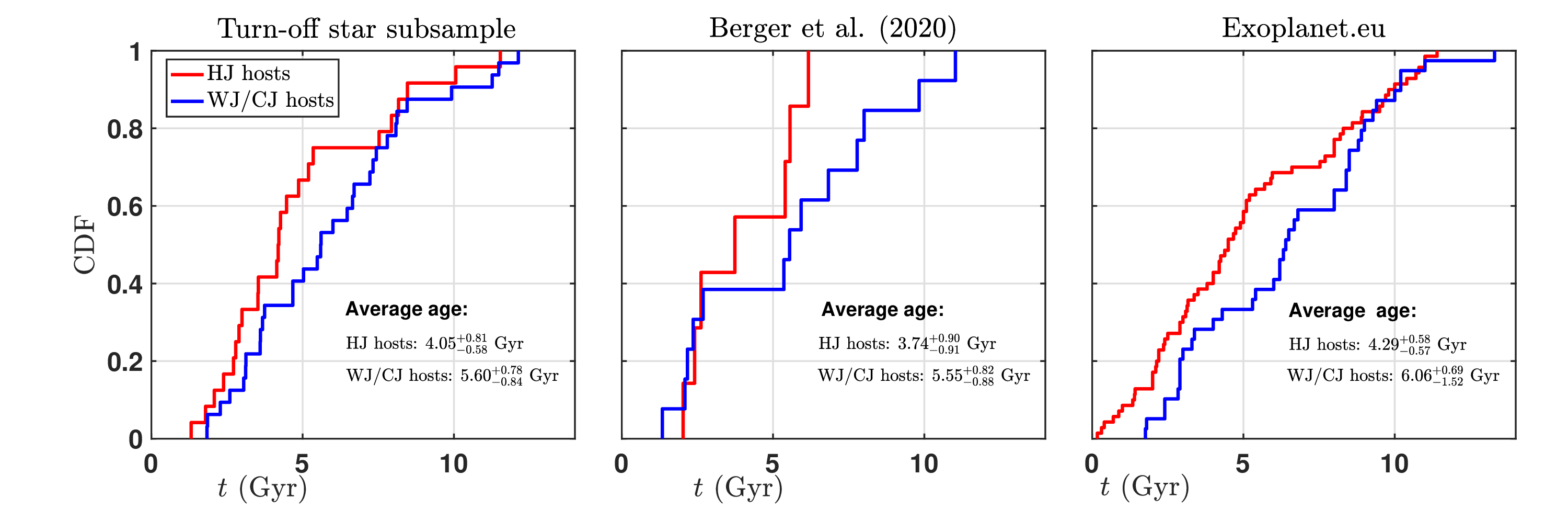}
\caption{The cumulative distributions of ages from different sources for hot Jupiter host stars (red) and warm/cold Jupiters (blue).
{The typical (median) uncertainties in ages are $\sim 25\%$, $\sim 30\%$, and $\sim 28\%$ for Turn-off subsample, selected giant planet hosts in Berger et al. (2020) and Exoplanet.eu, respectively.}
\label{figAgeHJCJ_othersource}}
\end{figure}

It is worth noting that the inferred kinematic ages only can represent the average ages of two populations.
To verify the inferred age distributions, we calculate and collect age estimates of giant planet host stars from other methods and sources: 

\begin{enumerate}

\item Isochrone ages for main-sequence turn-off stars.
By matching the observational parameters (e.g., effective temperature $T_{\rm eff}$, absolute magnitude in the V band $M_{\rm V}$ (or in other bands) and
metallicity $\rm [Fe/H]$) with the predictions of stellar evolutionary models in the Hertzsprung–Russell (HR) diagram, one can estimate the isochrone ages for large samples of stars.
This method is difficult for cool main-sequence or giant stars but works well for main-sequence turn-off stars or subgiants \citep[e.g.,][]{2010ARA&A..48..581S}.
Therefore, we select a subsample of turn-off stars from our stellar sample and calculate reliable stellar ages via isochrone fitting.

To do so, we first convert the Gaia DR2 $G$-band magnitudes to $V$-band with the formula: $\rm V = G+0.0176 +0.00686 \times (BP-RP)+0.1732 \times (BP-RP)^2$ according to the Gaia DR2 data release documentation \citep{2018A&A...616A...1G} and obtain the absolute magnitudes $M_{\rm V}$ using the Gaia distance.
Then we trace the locus of the turn-off stars in the $T_{\rm eff}-M_{V}$ plane with the same method as Xiang et al. (2017) \citep{2017ApJS..232....2X}.
The sample stars are then defined by requiring
\begin{equation}
  T_{\rm eff}>T^{\rm bRGB}_{\rm eff}+\Delta T_{\rm eff},
\end{equation}
\begin{equation}
  M_{\rm V}<M^{\rm TO}_{\rm V}+\Delta M_{\rm V},
\end{equation}
where $M_{\rm V}^{\rm TO}$ denotes the evolutionary trajectory of the main-sequence turn-off stars,
and $T^{\rm bRGB}_{\rm eff}$ denotes the evolutionary trajectory of the base of the red giant branch, which can be derived from $T_{\rm eff}$ and $M_{\rm V}$.
$\Delta M_{\rm V}$ and $\Delta T_{\rm eff}$ are set to reduce contamination from main-sequence turn-off and red giant branch stars (see \S~3 of Xiang et al. 2017).
In our planet host stellar sample, 55 stars are classified as turn-off stars (23 of them of hot Jupiter hosts).
By using the Yonsei–Yale ($Y^2$) isochrones \citep{2004ApJS..155..667D}, we make the age estimates with a typical uncertainty of $\sim 20-30\%$ for these turn-off stars from their $M_{\rm V}$, $T_{\rm eff}$, and $\rm [Fe/H]$.

\item Berger et al. (2020) \cite{2020AJ....159..280B} presents age estimates from isochrone fitting for 186,301 Gaia-Kepler stars with a median uncertainty of 56\%. 
We discard stars with Gaia DR2 re-normalized unit-weight error (RUWE) $>1.2$ to avoid likely binary systems. 
We only keep giant planet stars with relatively reliable isochrone ages
by applying the selection criteria, i.e., $\rm iso\_gof>0.99$, terminal age of the main sequence (TAMS) $<20$ Gyr.
In addition, we only leave Sun-like stars with relative age uncertainties smaller than 50\%. 
The final sample contains 20 giant planets, and 7 of them are hot Jupiters.

\item The Extrasolar Planets Encyclopaedia (hereafter, EU; http://exoplanet.eu/).
We select Sun-like stars hosting giant planets with relative uncertainties in ages less than $50\%$. 
The EU sample contains 177 giant planets, and 84 of them are hot Jupiters. 
It is worth noting that their age estimates are collected from various works and thus are in-homogeneous.

\item The {\it Kepler} asteroseismic LEGACY sample consists of 66 main sequence {\it Kepler} targets with asteroseismic age estimates \citep{2017ApJ...835..173S} . 
The average age uncertainty is 10\%. However, there is no star hosting giant planets. 
\end{enumerate}

Figure \ref{figAgeHJCJ_othersource} compares the cumulative distributions of ages from different sources for hot Jupiter hosts and warm/cold Jupiter hosts.
As can be seen, the results of different samples all show that hot Jupiter hosts are younger.
We also calculate the average ages for different sources.
To calculate the uncertainties of average ages, we resampled ages based on the reported ages and errors by assuming the Gaussian distribution $N (t, \sigma_t)$ for 1,000 sets and calculated the averages of resampled ages.
The uncertainties in average ages are set as the $50 \pm 34.1$ percentiles in the resample average ages.
As shown in Figure \ref{figAgeHJCJ_othersource}, hot Jupiter hosts have younger average ages by $\sim 1.5$ Gyr compared to those of warm/cold Jupiter hosts, which are consistent with the kinematic age differences within $1-\sigma$ errorbar.

\section{The frequencies of hot Jupiters and warm/cold Jupiters as a function of age} 
\label{sec.obs.evolution}
To explore the evolution of hot Jupiter and warm/cold Jupiters with age, we divide the sample into different bins according to their relative probabilities for the thick-disk-to-thin-disk $(TD/D)$. 
Then we calculate the frequencies of giant planets with the following formula:
\begin{equation}
  F = \frac{\sum\limits_{i=1}^{N_{\rm p}}{\frac{1}{\rm GE}}}{\sum\limits_{i=1}^{N_{\rm S}}  {\rm DE}} \equiv \frac{\sum\limits_{i=1}^{N_{\rm p}}{\frac{1}{\rm GE}}}{N_{\rm S} <\rm DE>},
\end{equation}
where $N_{\rm p}$ and $N_{\rm S}$ are the numbers of planets (hot Jupiters or warm/cold Jupiters) and stars in the given bin.
$\rm DE$ denotes the planetary detection efficiency detected by a given star, regardless of
whether the star has actually detected planets or not.
GE is the geometric effect of a given planet, taken as follows:
\begin{eqnarray}
{\rm GE}= \begin{cases} {1}&\mbox{For RV}\\ \frac{R_*}{a}&\mbox{For Transit}, \end{cases}
\end{eqnarray}
where $a$ is the semi-major axis of the planet and $R_*$ is the radius of the host star.
   
As shown in Table \ref{tab:planetnumberdisc}, our sample contains giant planets discovered by different methods and facilities, which can be classified into three categories, i.e., ground-based RV subsample, ground-based transit subsample, and space-based transit subsample. Specifically, most of warm/cold Jupiters are discovered by RV, while ground-based transit facilities discover the majority of hot Jupiters. 
Space-based facilities (i.e., {\it Kepler}, K2, CoRoT) also contribute a small part of giant planets. 
Since the parent stellar samples for giant planets discovered by different methods and facilities are different, in the subsequent subsections, we calculate their frequencies respectively.

\subsection*{3.1 Ground-based radial velocity sub-sample}
\label{sec.obs.RV}
In this subsection, we calculate the frequencies of hot Jupiters and warm/cold Jupiters \textbf{discovered by Ground-based facilities with RV method (RV)} and quantitatively explore their evolution with age.

\subsubsection*{3.1.1 Collecting the parent stellar sample and planetary sample}
\label{sec.obs.RV.sample}
A homogeneous stellar sample to search and characterize extrasolar planetary systems with RV is required to calculate the frequencies.
Here we collect public databases for stars observed by RV projects (i.e., Keck, Lick, HARPS), which provide homogenized radial velocity time series for 4,592 stars \citep{2014ApJS..210....5F,2017AJ....153..208B,2020A&A...636A..74T}.

We then crossmatch them with Gaia DR2 catalog by using the X-match service of
the the Centre de Donnees astronomiques de Strasbourg (CDS;
http://cdsxmatch.u-strasbg.fr) to obtain their astrometry and radial velocity data.
We make the
following quality cuts:
$\rm astrometric\_excess\_noise\_sig<2 \& parallax\_over\_error>10$ \citep{2018A&A...616A...2L}.
Besides, we only kept stars with Sun-like stars belonging to the Galactic disk with the same criteria as \S~\ref{sec.meth.sample}.
The final parent Ground-based RV stellar sample contains 1,662 stars.
We then crossmatch the parent stellar sample with planets discovered by RV, yielding a planetary sample of 17 hot Jupiters and 85 warm/cold Jupiters.

\subsubsection*{3.1.2 Qualifying the detection bias}
\label{sec.obs.RV.debias}
Stars with different kinematic ages could differ significantly in the precision
of their RV measurements, causing an observational bias that stars with poorer RVs bias against finding smaller planets with longer periods.
In order to quantify the detection bias with age, we perform the following investigation and correction.

For a planet with with a given minimum mass ($M_{\rm p} \sin{i}$) and
orbital period ($P$), ignoring the effect of orbital eccentricity, the amplitude of the generated RV signal is as follows:
\begin{equation}
  K = 28.4 {\rm m \ s^{-1}} \times \frac{M_{\rm p} \sin{i}}{M_{J}} \left(\frac{P}{1 \rm yr}\right)^{-1/3} \left(\frac{M_*}{M_\odot}\right)^{-2/3},
\end{equation}
where $M_*$ and $M_{\rm p}$ are the masses of star and planet, $i$ is the inclination between
the normal of the planet’s orbit and the line of sight.

We calculate the RV root mean square (RMS) and observation time baseline for each star in the parent RV stellar sample.
Then we divide the stellar sample into ten bins at the same interval as the giant planetary sample.
The typical detection limits are set as follows:
\begin{equation}
  K \ge RMS_{\rm median} \&
  P \le \rm baseline_{\rm median},
\end{equation}
where $RMS_{\rm median}$ and
$\rm baseline_{\rm median}$ are the median values of $RMS$ and time baseline for the ten bins.

\begin{figure}[!t]
\centering
\includegraphics[width=\linewidth]{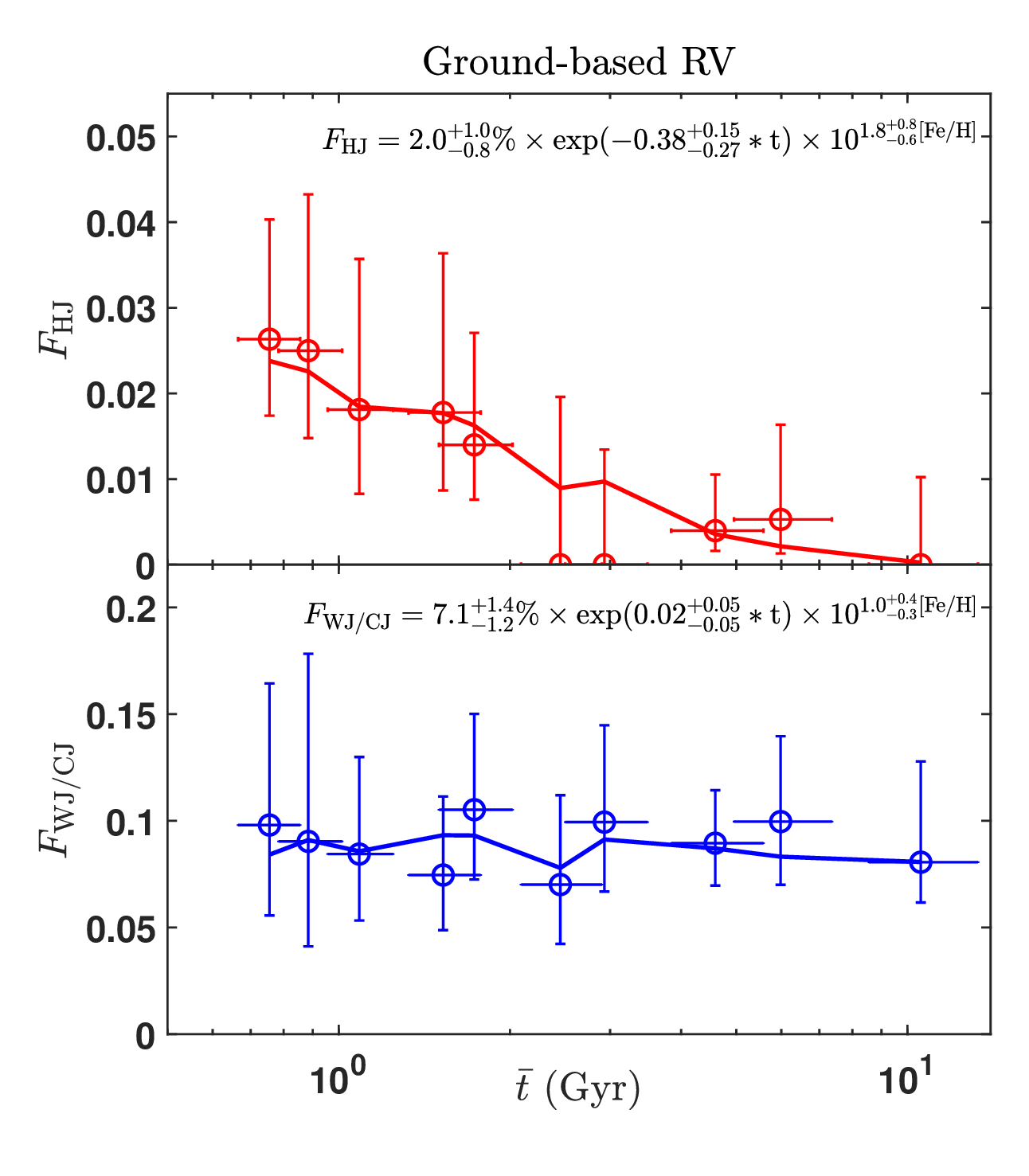}
\caption{The evolution of the frequencies of hot Jupiters $F_{\rm HJ}$ (Top panel) and warm/cold Jupiters $F_{\rm WJ/CJ}$ (Bottom panel) after qualifying the detection efficiency derived from Ground-based RV subsample.
The solids lines represent the best fits of Equation S8.
\label{figGBRVOccurrencerateDE}}
\end{figure}

To eliminate the detection bias, for a given planet, for each star with RMS and time baseline data at the same bin of kinematic age, we calculate the RV signal with Equation S6, and then take the proportion of stars with $K \ge \rm RMS \& time baseline \ge P$ as the detection efficiency of the planet, $\rm DE$.
Then we calculate the frequencies of hot Jupiters and warm/cold Jupiters for each bin with Equation S4.
Figure \ref{figGBRVOccurrencerateDE} displays the evolution of $F_{\rm HJ}$ and $F_{\rm WJ/CJ}$.

\subsubsection*{3.1.3 Model fitting}
\label{Sec.obs.RV.fitting}
\begin{figure}[!t]
\centering
\includegraphics[width=\linewidth]{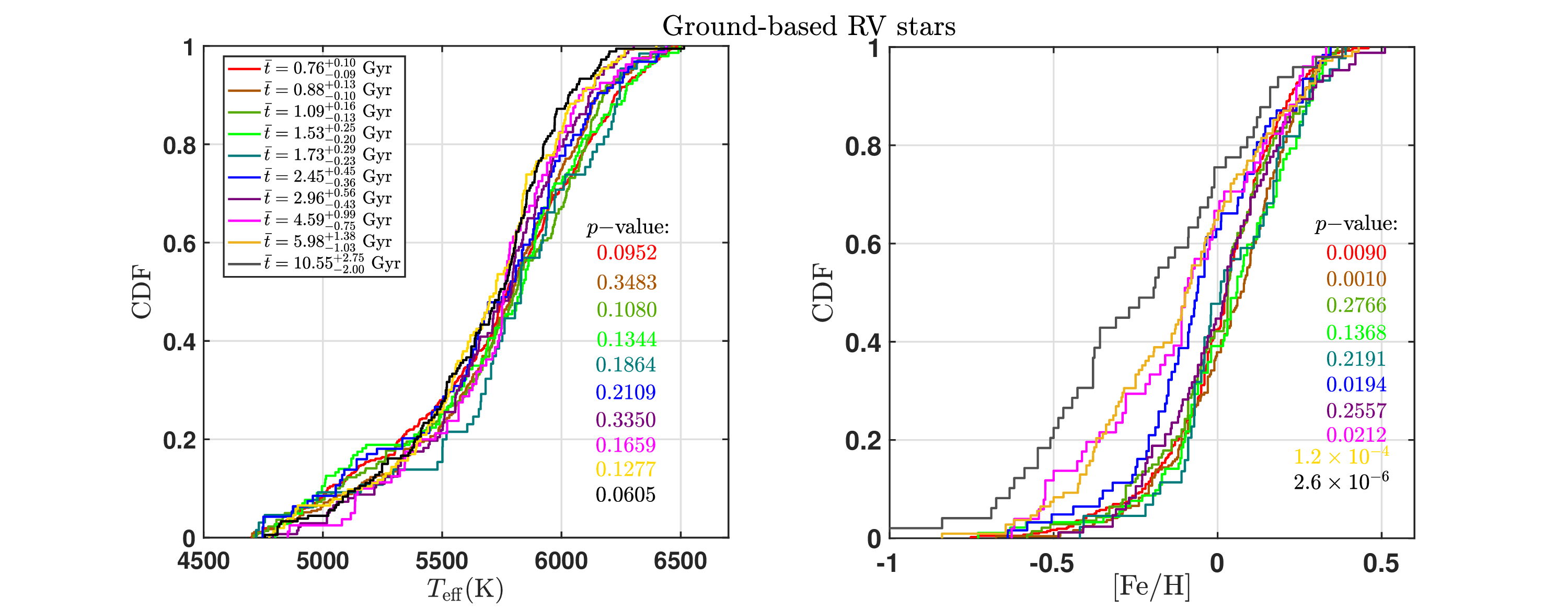}
\caption{The cumulative distributions of effective temperature ($T_{\rm eff}$, Left panel) and metallicities (Right panel) for parent RV stars of different kinematic ages. In each panel, we print the two sample K-S test $p-$values for the distributions of RV parent stars of each bin comparing to the total RV parent sample.
\label{figGBRVTeffFeH}}
\end{figure}

In Figure \ref{figGBRVOccurrencerateDE}, we show the evolution of the frequencies of hot Jupiters and warm/cold Jupiters with kinematic age.
Nevertheless, various stellar properties are not independent but found to be interrelated with kinematic age \citep[e.g..][]{2014A&A...562A..71B,2022AJ....163..249C}.
In Figure \ref{figGBRVTeffFeH}, we compare the cumulative distributions of stellar effective temperatures (proxy of masses) and metallicities, which are found to be correlated to the frequencies of giant planets (especially hot Jupiters) \citep{1997MNRAS.285..403G,2005ApJ...622.1102F,2010PASP..122..905J,2016A&A...587A..64S}.
As can be seen, different bins of different ages do not differ significantly (with K-S $p-$ values $>0.05$) in the distribution of effective temperatures since we only select Sun-like stars.
While the stellar metallicities generally decrease with increasing age.
Therefore, the trend we obtain in Figure \ref{figGBRVOccurrencerateDE} could be a combination of the effects of growing age and decreasing $\rm [Fe/H]$.

Referring to previous studies \citep{2010PASP..122..905J,2016A&A...587A..64S}, we fit the frequencies of hot Jupiters and warm/cold Jupiters as functions of age and metallicity (i.e., hybrid model) with the following formula:
\begin{equation}
  F({\rm [Fe/H]}, t) = C 10^{\beta \rm [Fe/H]} {\rm exp}(\gamma*t).
\end{equation}

To obtain the values and uncertainties of $C, \beta, \gamma$ (hereafter, we denote these coefficients as $X$ for conciseness), we fit them from the measured number of planets $H$ drawn from a larger sample of $T$ targets using Bayes’ theorem with the similar procedure described in \S~4.2 of Johnson et al. (2010) \citep{2010PASP..122..905J}.
For a given set of $X$, the probability of a detection around star $i$ (of $H$ total detections) is given by:
\begin{equation*}
  F ({\rm [Fe/H]}_i, t_i) {\rm GE}_i {\rm DE}_{i},
\end{equation*}
where GE and DE are the geometry effect and detection efficiency of each planet.
The probability of the $j-$th nondetection is given by:
\begin{equation*}
  1-(F ({\rm [Fe/H]}_j, t_j) {\rm GE}_i {\rm DE}_{j}),
\end{equation*} 
here we generate the GE and DE with similar methods as \S~4.2 of Dong \& Zhu 2013 \citep{2013ApJ...778...53D}.
For each detection or nondetection, the stellar metallicity of $\rm [Fe/H]$ is collected from public databases (e.g., RAVE, LAMOST and NASA exoplanet archive).
For stars belonging to the $m-$th bin, we assume their ages obey a Normal distribution $N (t, \sigma_t)$, which satisfies that the average age and uncertainty of stars in $m-$th bins are equal to the average kinematic age and uncertainty.
Thus, the possibility/likelihood of detecting $H$ planets from $T$ targets for a given $X$ is:
\begin{equation}
  P \propto \prod \limits_{i}^{H} (F ({\rm [Fe/H]}_i, t_i) {\rm GE}_i {\rm DE}_{i}) \times \prod \limits_{j}^{T-H} [1-(F ({\rm [Fe/H]}_j, t_j) {\rm GE}_j {\rm DE}_{j})].
\end{equation}
The marginalized log-likelihood is:
\begin{equation}
\begin{split}
  \log L & = \sum \limits_{i}^{H} \log(F ({\rm [Fe/H]}_i, t_i) {\rm GE}_i {\rm DE}_{i}) \\
  & + \sum \limits_{j}^{T-H} \log[1-(F ({\rm [Fe/H]}_j, t_j) {\rm GE}_j {\rm DE}_{j})].
\end{split}
\end{equation}
The best-fits of $X=[C, \beta, \gamma]$ have the maximum $L$ conditioned on the data and the confidence intervals
($1-\sigma$: 68.3\%, $2-\sigma$: 95.4\% and $3-\sigma$: 99.7\%) are estimated using a spline function.

\begin{figure}[!t]
\centering
\includegraphics[width=\linewidth]{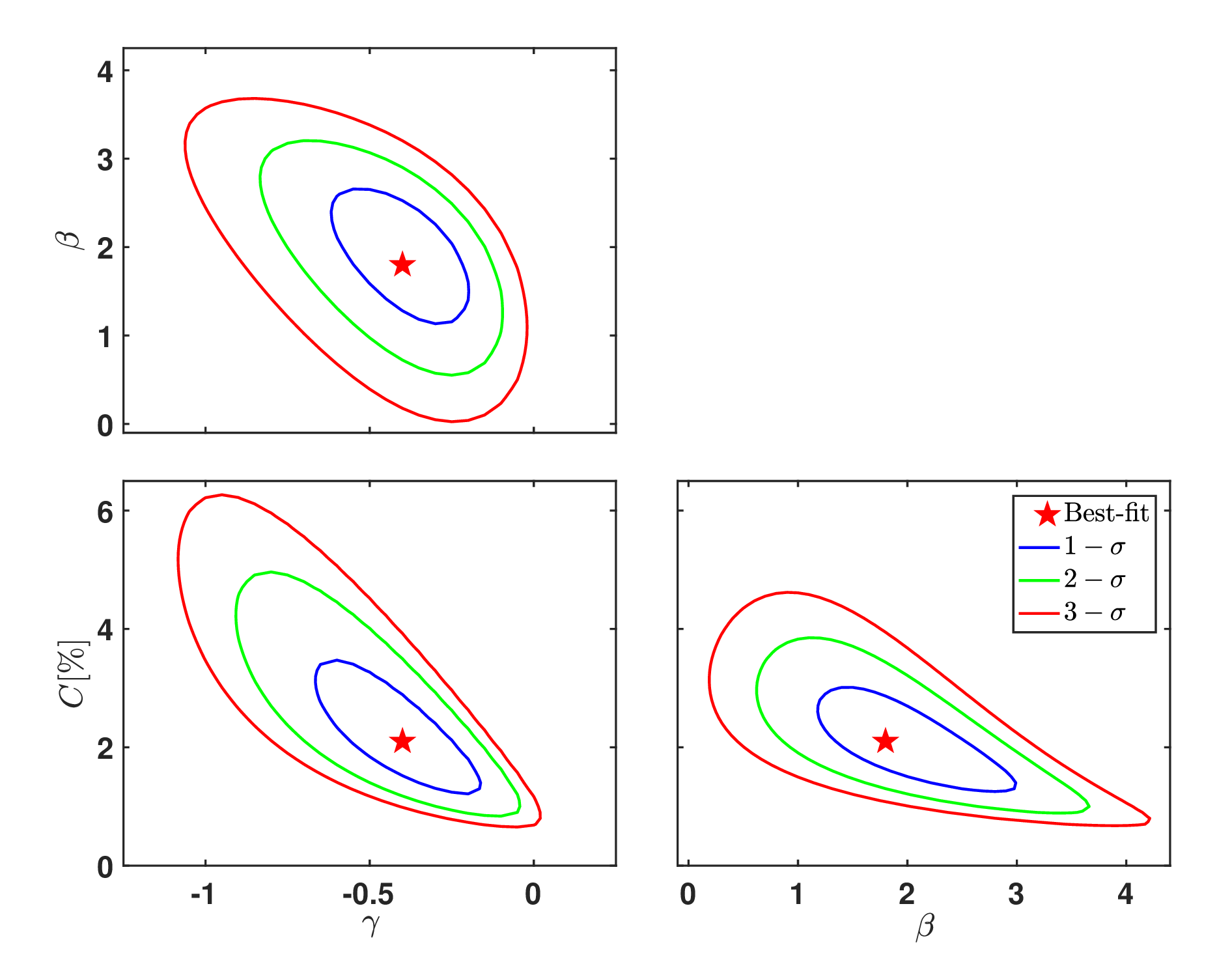}
\caption{Marginalized posterior probability density distributions of the parameters $(C, \beta, \gamma)$ for the frequency of hot Jupiters derived from the ground-based RV data.
\label{figFHJfittingFeHAgeGBRV}}
\end{figure}

\begin{figure}[!t]
\centering
\includegraphics[width=\linewidth]{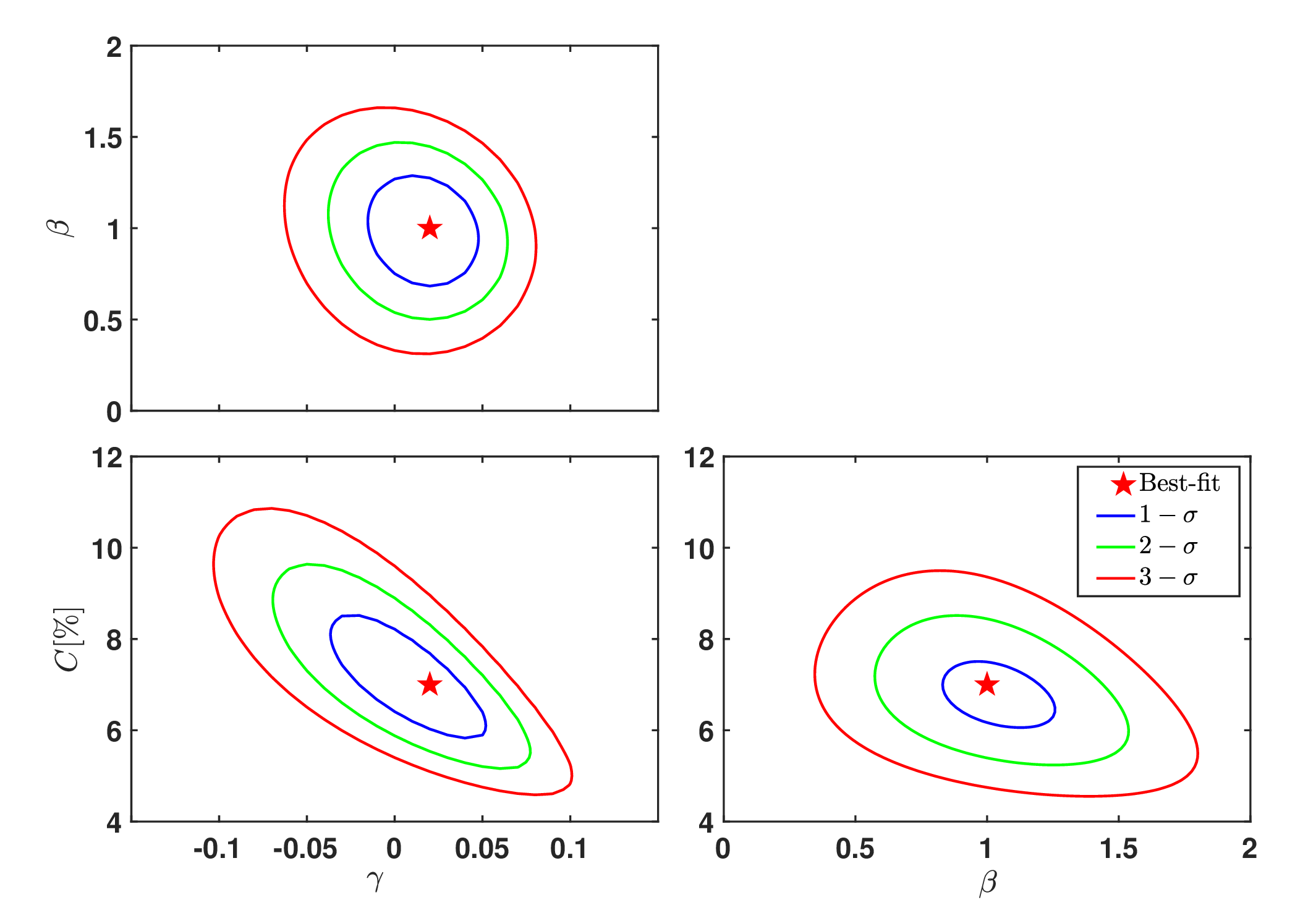}
\caption{Marginal posterior probability density functions for the model coefficients $(C, \beta, \gamma)$ for the frequency of warm/cold Jupiters conditioned on the ground-based RV data.
\label{figFCJfittingFeHAgeGBRV}}
\end{figure}

Figure \ref{figFHJfittingFeHAgeGBRV} and \ref{figFCJfittingFeHAgeGBRV} show the probability density functions of $X$ for the evolution trends of hot Jupiters and warm/cold Jupiters, respectively.
The comparisons between the best-fitting relationships and
the data are shown in Figures \ref{figGBRVOccurrencerateDE}.
The best-fitting parameters and their $1-\sigma$ intervals are listed in Table \ref{tab:fittingparamodel}.
As can be seen, both the frequencies of hot Jupiters and warm/cold Jupiters are positively correlated ($\beta$ of $1.8^{+0.8}_{-0.6}$ and $1.0^{+0.4}_{-0.2}$) to the stellar metallicity with confidence levels (i.e., the probabilities for $\beta>0$) $\gtrsim 3-\sigma$.
For the dependence on stellar age, the frequency of hot Jupiters decreases with increasing age ($\gamma = -0.38^{+0.17}_{-0.25}$), and the probability for $\gamma<0$ (corresponding to confidence level) is 98.45\%.
For comparison, the frequency of warm/cold Jupiters shows a weak dependence ($\gamma = 0.02^{+0.05}_{-0.05}$) with stellar age since the probabilities for $\gamma>0$ or $<0$ are both less than $1-\sigma$ confidence level.
In order to further test the dependence of $F_{\rm HJ}$ and $F_{\rm WJ/CJ}$ on stellar age/$\rm [Fe/H]$, we also adopt a single-age (i.e., $\beta = 0$) model and a single-metallicity (i.e., $\gamma = 0$) model to fit the observation data.
The differences in AIC ($\rm \Delta AIC = 2 \Delta \log L - 2$) between the hybrid model and single-age model/single-metallicty model are 11.3/6.1 and 22.2/3.4 for $F_{\rm HJ}$ and $F_{\rm WJ/CJ}$ respectively, suggesting that the hybrid model is more preferential. 

\begin{figure}[!t]
\centering
\includegraphics[width=\linewidth]{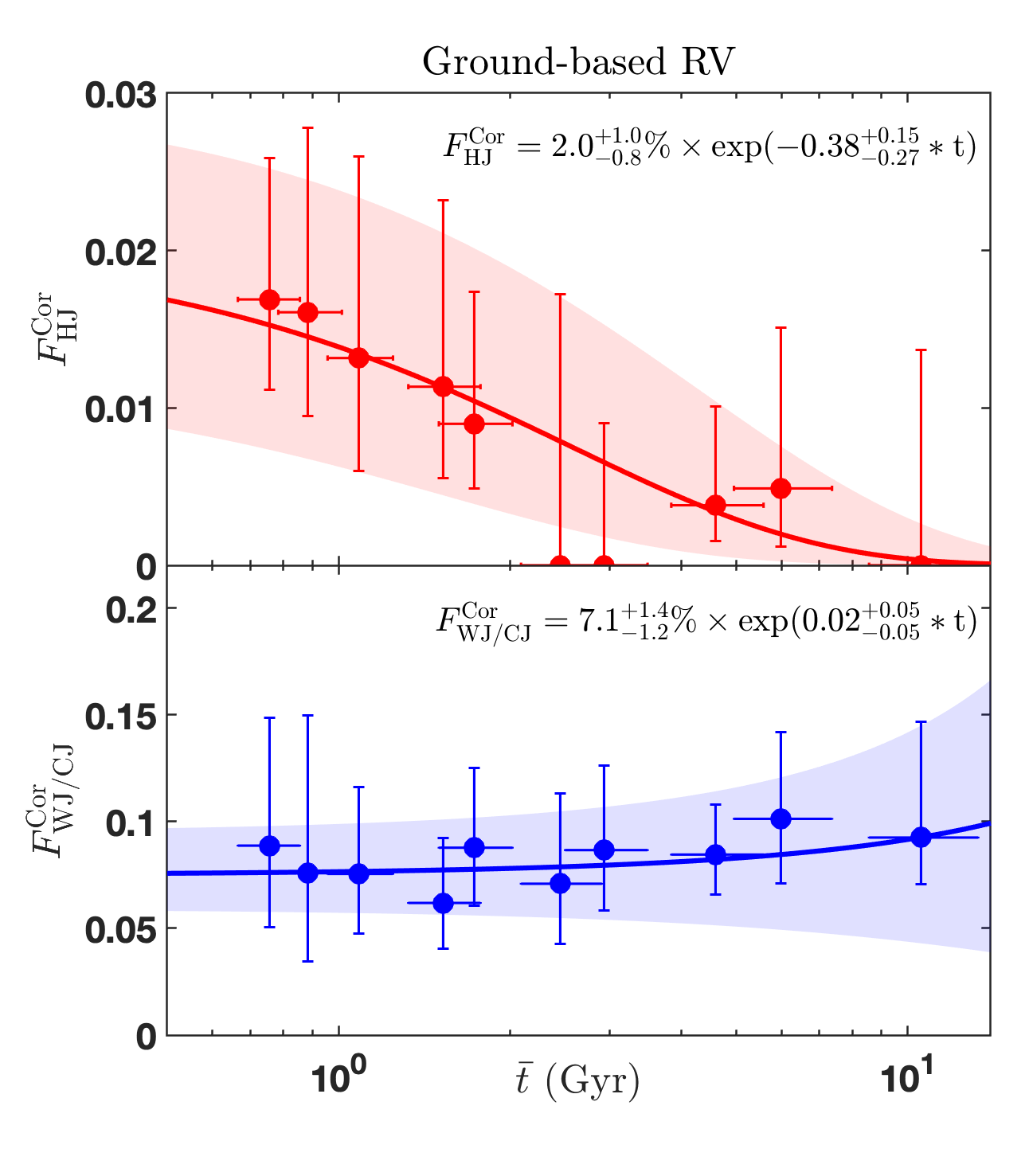}
\caption{The age evolution of the frequencies of hot Jupiters $F^{\rm Cor}_{\rm HJ}$ (Up panel) and warm/cold Jupiters $F^{\rm Cor}_{\rm WJ/CJ}$ (Bottom panel) after qualifying the detection efficiency and \textbf{eliminating the effect of metallicities} derived from Ground-based RV subsample.
The solids lines and regions represent the best fits and $1-\sigma$ intervals of Equation S8.
\label{figGBRVOccurrencerateDEFeH}}
\end{figure}

In order to intuitively show the separate evolution of planet frequencies with age, we normalize the results to solar metallicity to remove the metallicity dependence.
Specifically, due to the dependence on $\rm [Fe/H]$, the frequency of $i-$th bin differ from that with solar metallicity (i.e., $\rm [Fe/H] =0$) by a factor of:
\begin{equation}
  \rm Factor_{[\rm Fe/H]} = 
  \frac{\sum\limits_{i} 10^{\beta \times [\rm Fe/H]}}{\sum\limits_{i} 10^{\beta \times 0}}
  = \frac{\sum\limits_{i} 10^{\beta \times [\rm Fe/H]}}{N_i},
  \label{normfeh}
\end{equation}
where $i$ denotes stars in the i-th bin. 
Then we eliminate the effect of metallicity by dividing $\rm Factor_{[\rm Fe/H]}$ , i.e.,
\begin{equation}
  F^{\rm cor} = F \times \frac{1}{\rm Factor_{[\rm Fe/H]}}.
   \label{corrfeh}
\end{equation}
Figure \ref{figGBRVOccurrencerateDEFeH} shows the evolution of frequencies of hot Jupiters and warm/cold Jupiters with age.
As can be seen, there is an obvious declining trend for hot Jupiter with increasing kinematic age, while the frequency of warm/cold Jupiters remains broadly unchanged.

\subsection*{3.2 Ground-based transit sub-sample}
\label{sec.obs.GBTS}
In this subsection, we calculate the frequency of giant planets \textbf{discovered by Ground-based facilities with Transit method (GT)} and quantitatively explore the evolution with age.
As shown in Table \ref{tab:planetnumberdisc}, there are only 3 warm/cold Jupiters in our sample, which are too few to carry out detailed research. 
Therefore, in this subsection, we only focus on the age evolution of hot Jupiters.

In the following, we derive the relative frequency of hot Jupiters as a function of age from GT.  First of all,  inferring the absolute planet frequency from GT is not essential for our main science. Secondly, as discussed in Gould et al. (2006) \citep{2006AcA....56....1G}, factors such such as by-eye vetting of photometric candidates and evaluating the feasibility of RV follow-up are needed to derive the absolute planet frequency, but these require tight co-operations with transit survey groups, which are not available to us.

\subsubsection*{3.2.1 Constructing the parent stellar sample}
\label{sec.obs.GBTS.sample}
GT planetary subsample is discovered by various surveys, which have searched millions of stars for transits. Many surveys have not published their parent stellar samples or the selection criteria necessary to reconstruct the parent samples from stellar catalogs, making it difficult to directly infer the detection efficiency. One exception is (Super)WASP, which have published their first public release with light curves \citep{2010A&A...520L..10B} and stellar target selection criteria \citep{2007ASPC..366..187W, 2007MNRAS.381..851C}. Our analysis is primarily based on the information from the (Super)WASP survey.

Here we construct a parent stellar sample based on the Hipparcos/Tycho-2 catalog \citep{2000A&A...355L..27H}.
The Tycho-2 catalog provides parallax, celestial coordinates (RA and DEC), and proper motions and magnitude for 2,539,913 stars, which are used to select the searching parent star samples for multiple surveys (e.g., SuperWASP).
Here we select stars from Tycho-2 catalog by adopting the same criteria as SuperWASP \citep{2007ASPC..366..187W}: \\
(1) V-band magnitude in the range $8 \leq V \leq 13$; \\
(2) Dwarf stars by excluding giant stars with reduced proper motion (RPM) suggested by Gould \& Morgan (2003) \citep{2003ApJ...585.1056G} when relevant parallax measurements were not available to most ground-based transit targets.  Note that Tycho-2, USNOB1.0 and UCAC catalogs were used in SuperWASP RPM \citep{2007ASPC..366..187W}, and we only adopt Tycho-2 for homogeneity.   
We use the selection method adopted by SuperWASP based on $\rm RPM-(V-Ks)$ as described in Clarkson et al. (2007) \citep{2007MNRAS.381..851C} with proper motion and magnitude data from Hipparcos/Tycho-2 \citep{2000A&A...355L..27H} and Two-Micron All-Sky Survey (2MASS) catalogs.

We then crossmatch the selected stars with Gaia DR2 catalog to obtain astrometric data.
For the radial velocity, similar to PAST \uppercase\expandafter{\romannumeral1}, we collect RV data from APOGEE, RAVE, LAMOST and Gaia with the same criteria.
Then we calculate their kinematic properties (i.e., Galactic position, velocity, and components).
We make quality control and keep Sun-like stars in the Galactic disk with the same criteria as \S~\ref{sec.obs.RV.sample}.
The final parent GT stellar sample contains 536,661 stars.
To obtain the planetary sample, we crossmatch the parent stellar sample with planets discovered by transit method and ground-based facilities.
Referring to Smith et al . (2006) \citep{2006MNRAS.373.1151S}, transiting planets in SuperWASP  selected as high-priority transit
candidates are vetted for RV follow-ups.
For these with lower signal-to red noise $\rm SNR_{\rm red}$, only part of them \citep[e.g., WASP-22b, WASP-45 b with similar periods in data of different years;][]{2010AJ....140.2007M,2012MNRAS.422.1988A} had RV follow-ups and the selection conditions are not uniform, making their selection bias difficult to corrected.
Thus, we only keep hot Jupiters with high Signal-to-red noises.
Specifically, for planet hosts with SuperWASP DR1 light curves, we derive their red noises and numbers of transit from their light curves and calculate their $\rm SNR_{\rm red}$.
For those planet hosts with no SuperWASP DR1 light curves, we select their nearest neighbors and obtain their $\rm SNR_{\rm red}$ (see the detailed description of the calculation of $\rm SNR_{\rm red}$ at following subsection 3.2.2).
After the signal-to-red noise selection, there are 43 hot Jupiters left in the parent sample. 

\begin{figure}[!t]
\centering
\includegraphics[width=\linewidth]{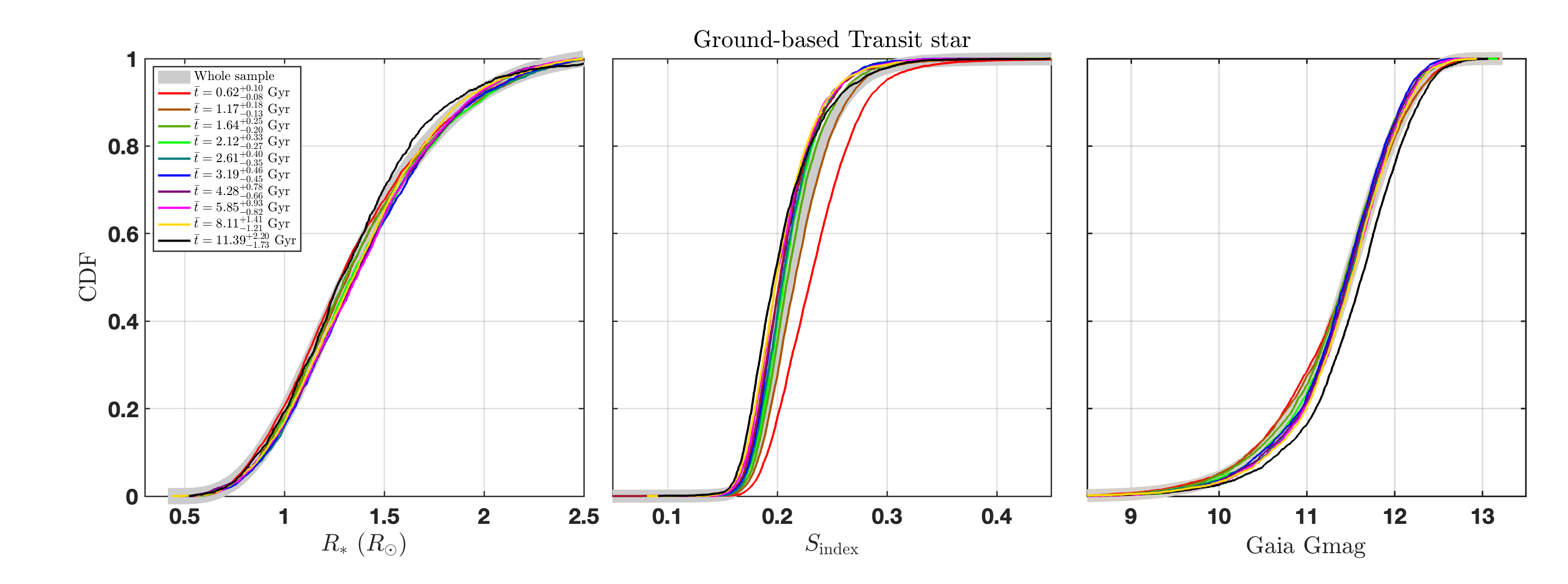}
\caption{The cumulative distributions of stellar radii, S-indexes, and Gaia G-band magnitudes for parent GT stellar sample of different kinematic ages.
\label{figGBTSstellarproperties}}
\end{figure}

\begin{figure}[!t]
\centering
\includegraphics[width=\linewidth]{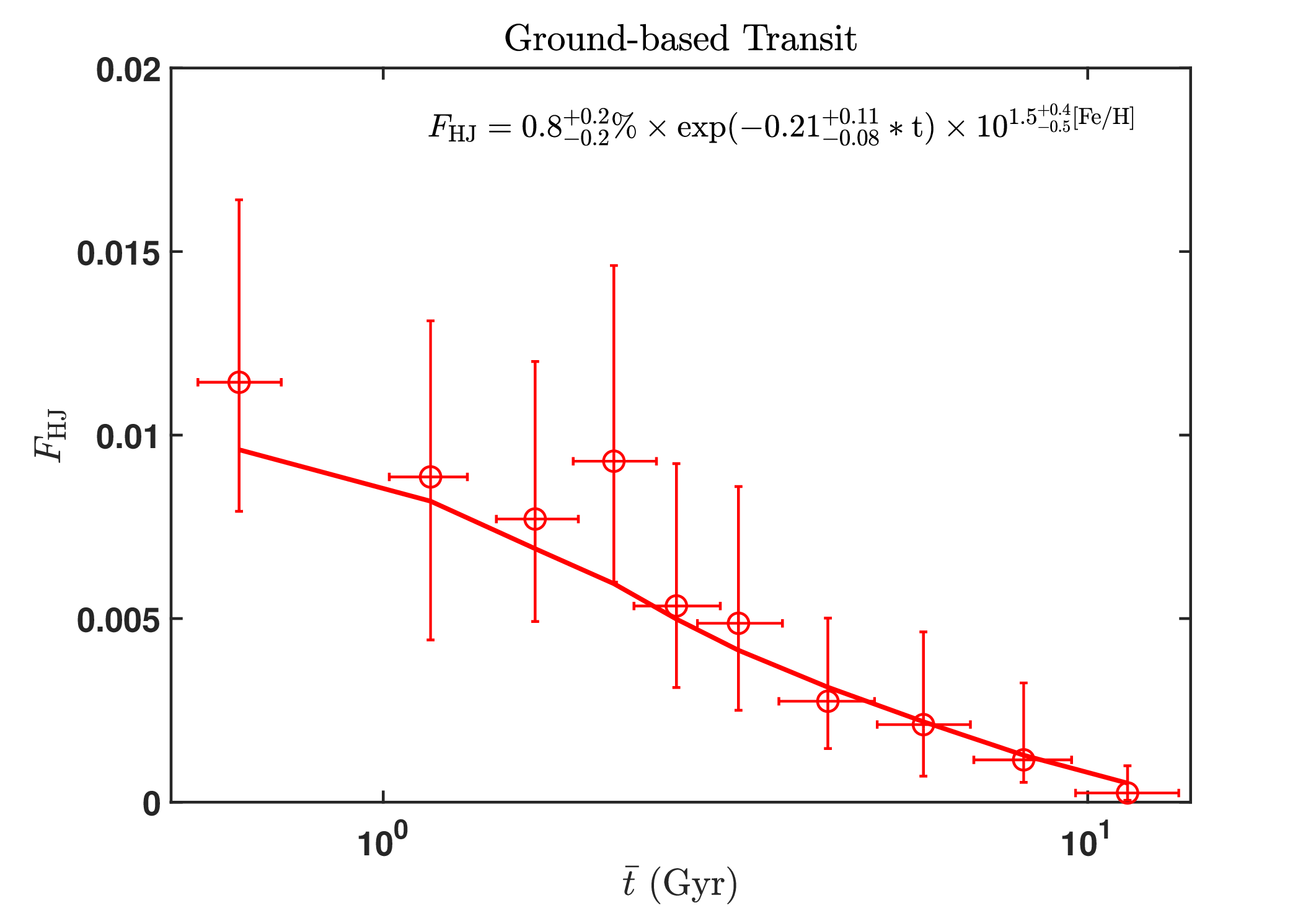}
\caption{The evolution of the frequency of hot Jupiters $F_{\rm HJ} = \sum\limits_{i=1}^{N_{\rm p}}{\frac{a}{R_*} \times \frac{1}{\rm DE}} \times \frac{1}{N_{\rm S}}$ (Low panel) derived from Ground-based Transit subsample.
The solids line represents the best fits of Equation S8.
\label{figFHJGBTSGCDE}}
\end{figure}

\begin{figure}[!t]
\centering
\includegraphics[width=\linewidth]{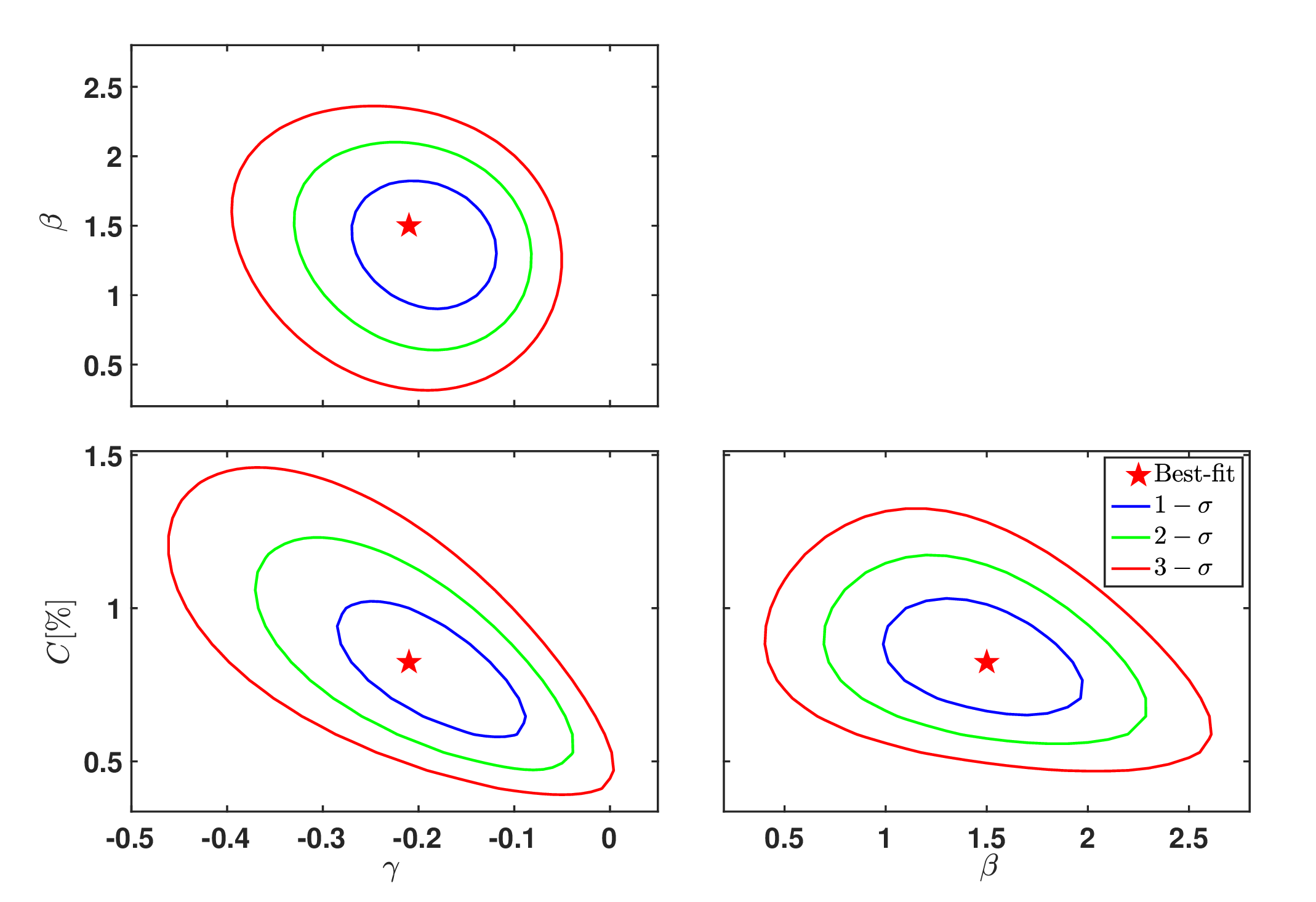}
\caption{Marginal posterior probability density functions for the model coefficients $(C, \beta, \gamma)$ for the frequency of hot Jupiters conditioned on the ground-based Transit data.
\label{figFHJfittingFeHAgeGBTS}}
\end{figure}

\subsubsection*{3.2.2 Correcting for the detection efficiency and geometric probability}
\label{sec.obs.GBTS.debias}
We divide the GT planets and parent stars into ten bins and calculate their frequencies of hot Jupiters with Equation S4:
\begin{enumerate}
  \item \textbf{Correcting the geometry probability} 
 For a given planet, we make the geometric correction by multiplying the ratio of the semi-major axis to the radius of the host star, $\frac{a}{R_*}$. The observing window function of GT surveys can significantly reduce detection efficiency with increasing planetary semi-major axes $a$. We examine the distributions in semi-major axes $a$ of GT planets of different kinematic ages (as shown in Figure \ref{figGBTSadis}) and find that they are statistically indistinguishable (with K-S $p-$values $>0.2$).
For our primary purpose of understanding the relative frequency as a function of age, the effect of observing window function does not differ significantly for bins of different ages, and thus slopes ($\beta$ and $\gamma$) should not be (significantly) affected. Therefore, we neglect the effect of the window function in our analysis.

  \item \textbf{Correcting the detection efficiency} 
  The NASA Exoplanet Archive provides access to the public SuperWASP light curve data release 1, which contains RA, DEC, apparent magnitudes, and light curves.
  We crossmatch the SuperWASP light curve data with our parent GT stellar sample using the X-match of CDS and obtain the light curves for 217,626 stars in our parent sample.
  Then we download the light curves of these stars.
  For each light curve, we record the number of data points ($\rm npts$) and calculate the $\sigma_{\rm red}$ for duration of 1 hour to 7 hour (the typical duration interval for hot Jupiters).
  Here $\sigma_{\rm red}$ represents the uncertainty of transit depth binned on the expected transit duration, which can be assessed by computing the sliding average of the out-of-transit data over the data points contained in a transit-length interval as proposed by Pont et al. (2006) \citep{2006MNRAS.373..231P}.
  For each star in the parent sample but not with SuperWASP DR1 light curves, we select the 100 closest neighbors of it in these stars with SuperWASP light curves in the space of the controlled parameters (i.e., stellar radius, $\rm [Fe/H]$, $TD/D$, distance and magnitude) by adopting the nearest neighborhood method from scikit-learn \citep{10.5555/1953048.2078195} and take the average $\sigma_{\rm red}$ and $\rm npts$ of the 100 neighbors as its red noise and number of data points.

  For a planet transiting a star , the Signal-Noise-Ratio ($\rm SNR_{red}$) is as follows:
  \begin{equation}
    {\rm SNR_{red}} = \sqrt{Nt} \frac{\delta}{\sigma_{\rm red}},
  \end{equation}
  where $\delta = 1.3 \left(\frac{R_{\rm p}}{R_*} \right)^2$ denotes the depth of transit.
  $R_{\rm p}$ and $R_*$ are the radius of planet and star, and the factor 1.3 is caused by the effect of stellar limb-darkening \citep{2005ApJ...627.1011T}.
  $\sigma_{\rm red}$ is the red noise of a given duration $D = \frac{R_*}{\pi a} P$.
  $Nt$ is the number of transit and approximately obeys a Poisson distribution. 
  The expectation of $Nt$ is $T_{\rm cov}/P$ and $T_{\rm cov}$ is the observational coverage time \citep{2004PASP..116..985D}.
  Here we take $T_{\rm cov}$ as $\rm npts \times cadence$ and the cadence is set as a typical value (8 minutes) of SuperWASP observations \citep[e.g.,][]{2006MNRAS.373.1151S,2006MNRAS.372.1117C}.

  For each hot Jupiter, for all stars at the same bin of kinematic age, we calculate their $\rm SNR_{red}$, and then take the proportion of stars with $\rm SNR_{red}>10$ \citep{2006MNRAS.373.1151S} as the detection efficiency of the given planet, $\rm DE$.
  Then we correct the detection efficiency and obtain the frequency of hot Jupiters for each bin, $F_{\rm HJ} = \frac{\sum\limits_{i=1}^{N_{\rm p}}{\frac{1}{\rm GE}}}{\sum\limits_{i=1}^{N_{\rm S}} \rm DE}$. 
   
\end{enumerate}

\begin{figure}[!t]
\centering
\includegraphics[width=\linewidth]{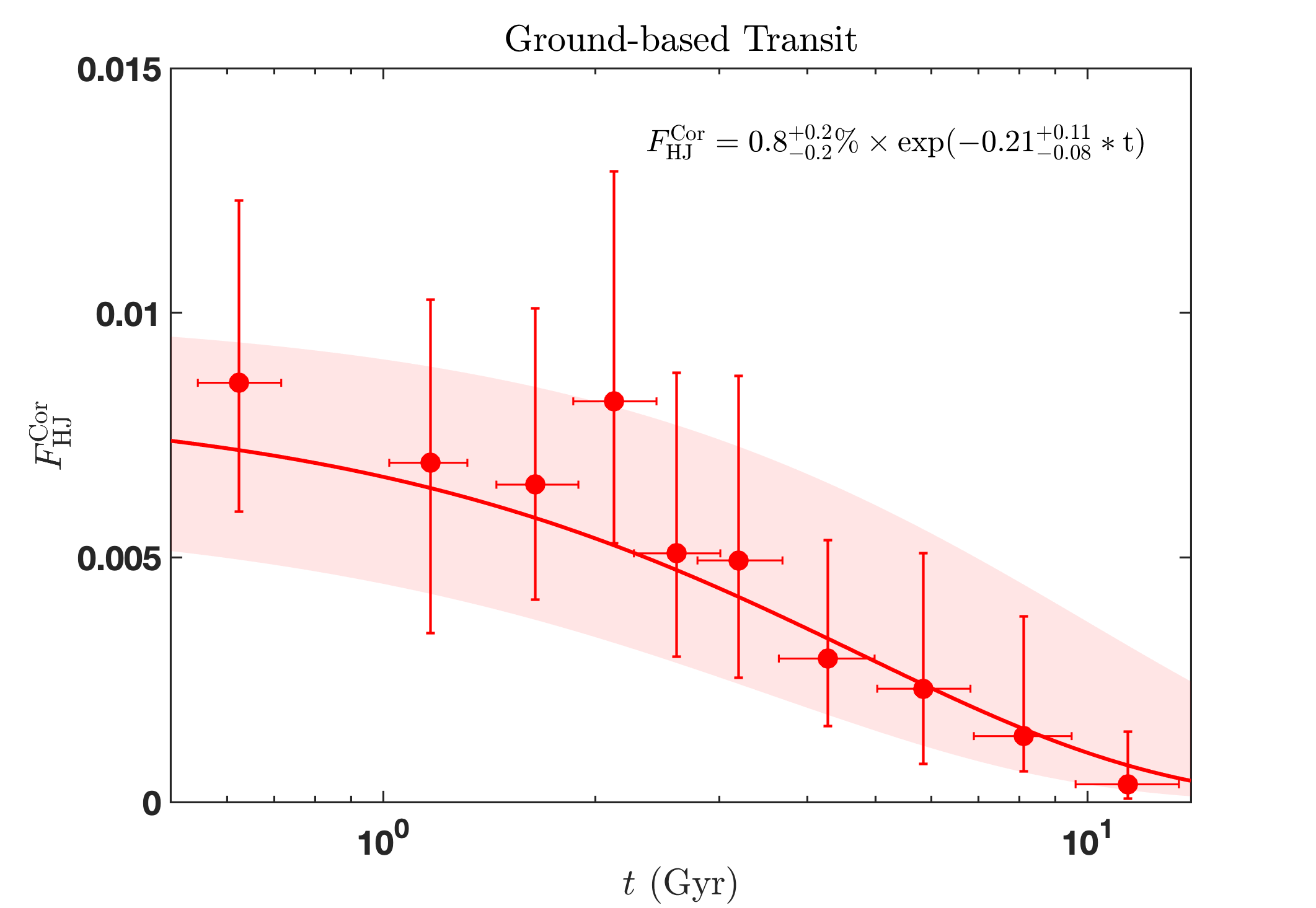}
\caption{The age evolution of the frequency of hot Jupiters $F^{\rm Cor}_{\rm HJ}$  derived from Ground-based Transit subsample after unifying $\rm [Fe/H]$ to solar metallicity.
The solids lines and regions represent the best fits and $1-\sigma$ intervals of Equation S8.
\label{figGBTSOccurrencerateDEFeH}}
\end{figure}

\begin{figure}[!t]
\centering
\includegraphics[width=\linewidth]{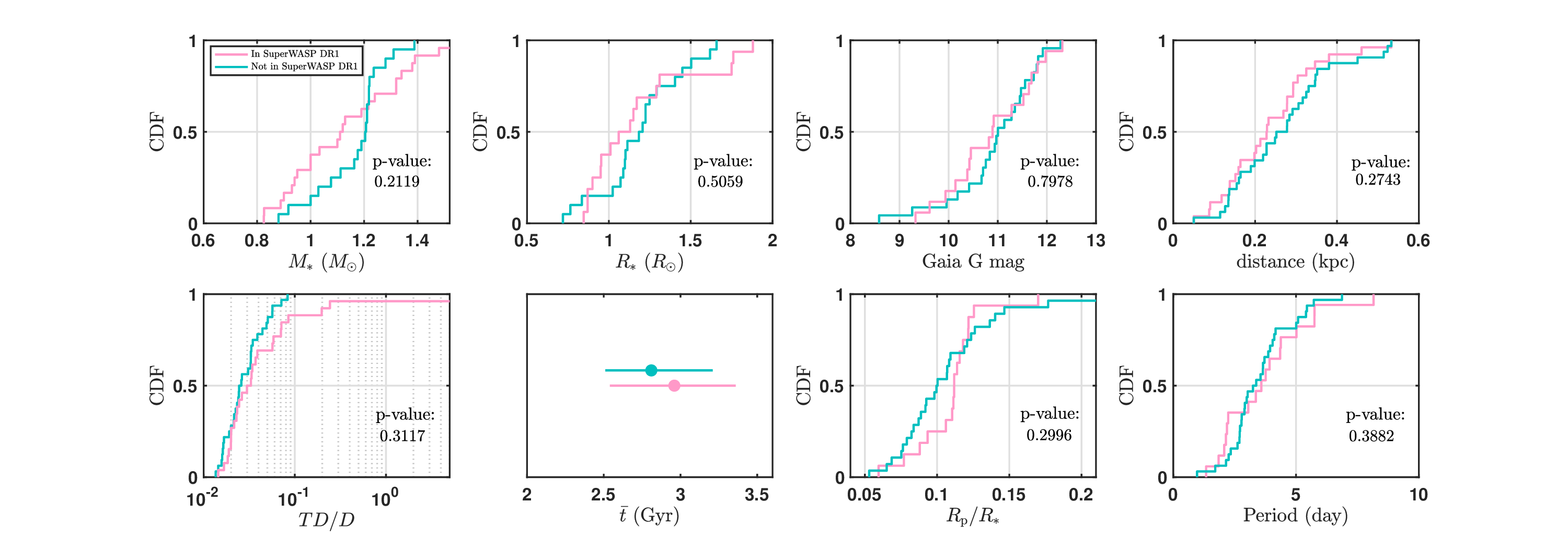}
\caption{The cumulative distributions of stellar mass $M_*$, $\rm [Fe/H]$, Gaia G magnitude, distance $TD/D$, average kinematic age $\bar{t}$ and planetary radius over stellar radius $R_{\rm p}/R_*$ and orbital period for the hot Jupiters in SuperWASP DR1 (pink) and not in SuperWASP DR1 (blue).
The two-sample K-S test $p-$ values are plotted at the lower right corner of each panel.
\label{figGTSuperWASPProperties}}
\end{figure}

\begin{figure}[!t]
\centering
\includegraphics[width=\linewidth]{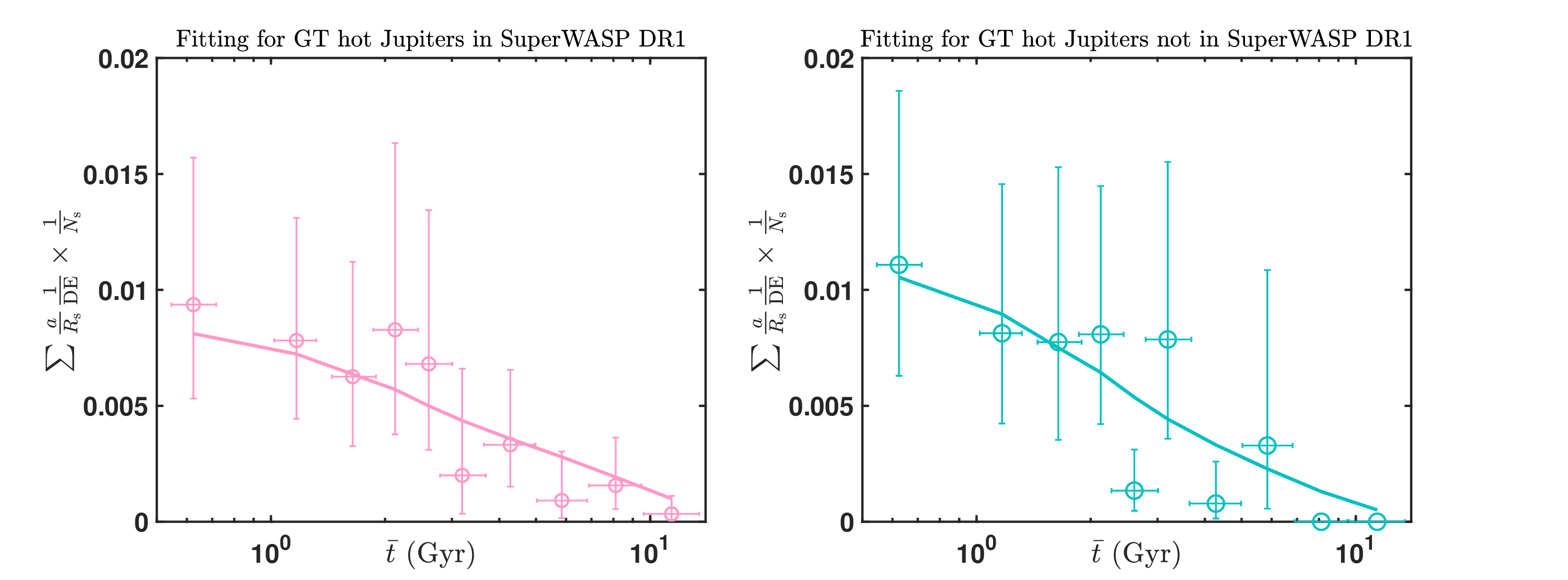}
\caption{The evolution of the number ratio of hot Jupiter over stars after correcting geometric effect and detection efficiency $\sum\limits_{i=1}^{N_{\rm p}}{\frac{a}{R_*} \times \frac{1}{\rm DE}} \times \frac{1}{N_{\rm S}}$ derived from Ground-based transit hot Jupiters in SuperWASP DR1 (Left panel) and not in SuperWASP DR1 (Right panel).
The solids line represents the best fits of Equation S8.
\label{figFHJGBTScategory}}
\end{figure}

\begin{figure}[!t]
\centering
\includegraphics[width=\linewidth]{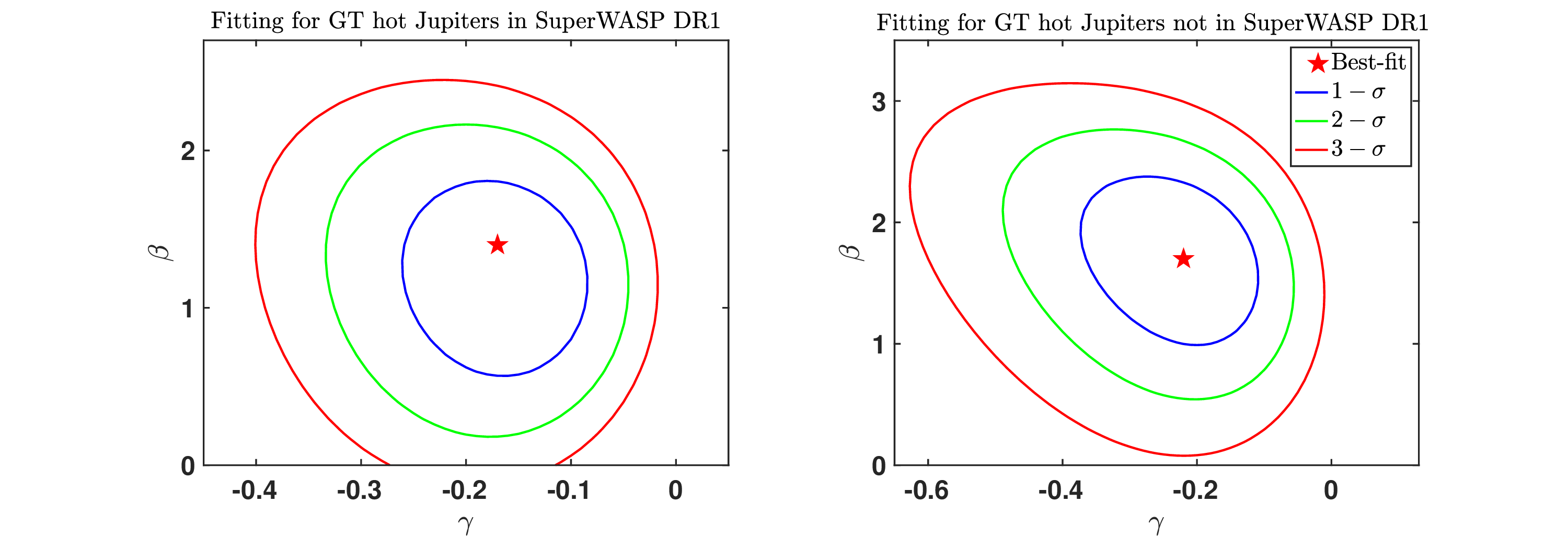}
\caption{Fitting for the Marginal posterior probability density functions for the model coefficients $(\beta, \gamma)$ for the frequency of hot Jupiters conditioned on the Ground-based transit hot Jupiters in SuperWASP DR1 (Left panel) and not in SuperWASP DR1 (Right panel).
\label{figFHJfittingGBTScategories}}
\end{figure}

\begin{figure}[!t]
\centering
\includegraphics[width=\linewidth]{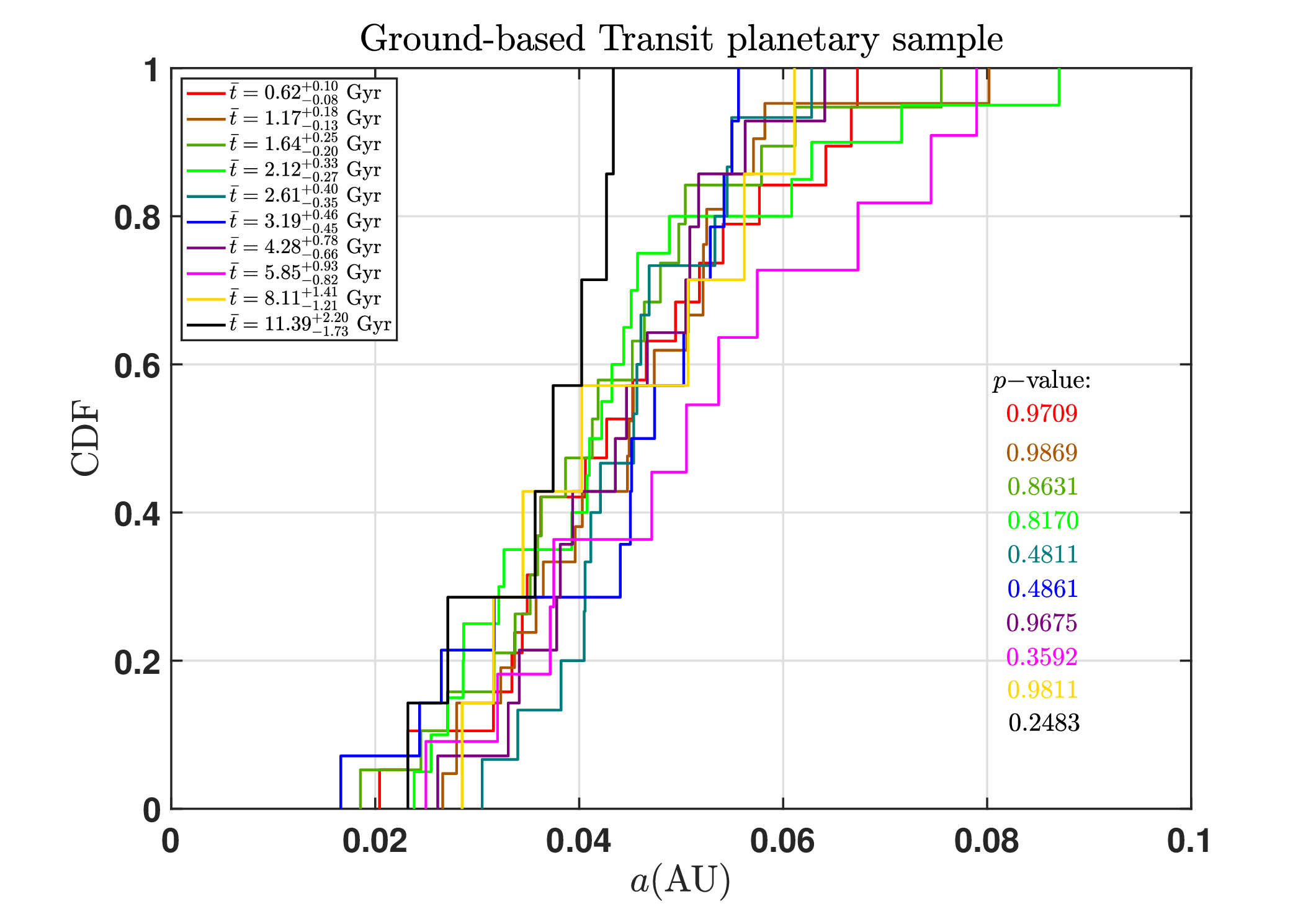}
\caption{The cumulative distributions of semi-major axis $a$ of GT planets of different kinematic ages. We print the two sample K-S test $p-$values for the distributions of GT planets of each bin comparing to the total GT planetary sample.
{The typical uncertainty in semi-major axis is $\sim 0.002$ AU.}
\label{figGBTSadis}}
\end{figure}

In Figure \ref{figGBTSstellarproperties}, we compare the cumulative distributions of stellar radii, S-indexes, and Gaia G-band magnitudes for parent stars of different kinematic ages.
As can be seen, different bins of different ages do not differ significantly in the distributions of stellar magnitudes and radii.
Whereas, the stellar activity S-index decreases as stars age, which is consistent with previous studies \citep{2021AJ....162..100C} and could lead to a decrease in the stellar noise \citep{2011ApJS..197....6G,2021AJ....162..100C} and an increase of detection efficiency with increasing age.
In Figure \ref{figFHJGBTSGCDE}, we show the temporal evolution of derived hot Jupiter frequency after correcting the geometric effect and detection bias.

\subsubsection*{3.2.3 Model fitting} 
\label{sec.obs.GBTS.fitting}
We then fit the frequency of hot Jupiters $F_{\rm HJ}$ with Equation S8 by adopting the same procedure as described in \S~3.1.3.
The differences in AIC ($\rm \Delta AIC = 2 \Delta \log L - 2$) between the hybrid model and single-age model/single-metallicty model are 15.7/13.9 for $F_{\rm HJ}$, demonstrating that the hybrid model is preferred. 
Figure \ref{figFHJfittingFeHAgeGBTS} shows the probability density functions of $X$ for the evolution trend of hot Jupiters.
The comparison between the best-fitting relationship and
the data is shown in Figures \ref{figFHJGBTSGCDE}.
The best-fitting parameters and their $1-\sigma$ intervals are listed in Table \ref{tab:fittingparamodel}.
As can be seen, the frequency of hot Jupiters is positively correlated ($\beta$ of $1.5^{+0.4}_{-0.5}$) to the stellar metallicity with a confidence level (i.e., the probability for $\beta>0$) $\gtrsim 3-\sigma$.
For the dependence on stellar age, the frequency of hot Jupiters decreases with increasing age ($\gamma = -0.21^{+0.11}_{-0.08}$) and the confidence level (i.e., the probability for $\gamma<0$) is 99.60\%.
Figure \ref{figGBTSOccurrencerateDEFeH} displays the frequency of hot Jupiter after unifying $\rm [Fe/H]$ to the solar metallicity, $F^{\rm Cor}_{\rm HJ}$, which decreases significantly as star ages.

Among the 43 hot Jupiters, 15 of them are discovered by Super-WASP DR1.
Thus it is necessary to verify whether parent GT stellar sample based on the selection criteria similar to SuperWASP DR1 could be applicable for the other 28 hot Jupiters.
We first compare the properties (i.e., stellar mass, $\rm [Fe/H]$, $TD/D$, magnitude, distance kinematic age, planetary orbital period and radius over stellar radius) between hot Jupiters in SuperWASP DR1 and not in SuperWASP DR1.
As shown in Figure \ref{figGTSuperWASPProperties}, they have statistically indistinguishable distributions (with K-S test $p-$values $>0.2$) in stellar and planetary properties.
We then obtain the temporal frequencies of hot Jupiters for the two categories (as shown in Figure \ref{figFHJGBTScategory}) and fit their frequencies with Equation S8.
In Figure \ref{figFHJfittingGBTScategories}, we show the probability density functions of $(\beta, \gamma)$
for the evolution trends of the two categories.
As can be seen, ($\beta$, $\gamma$) are ($-0.17^{+0.08}_{-0.09}$, $1.4^{+0.6}_{-0.6}$) and ($-0.22^{+0.11}_{-0.15}$, $1.7^{+0.6}_{-0.7}$) for GT hot Jupiters in SuperWASP DR1 and other GT hot Jupiters respectively, which are consistent with each other as well as the whole GT planetary sample.
Based on the above analyses, we conclude that the stellar sample based on SuperWASP criteria can be applicable for all the 43 GT hot Jupiters.

\subsection*{3.3 Space-based transit subsample}
\label{sec.obs.SBTS}
In this subsection, we calculate the frequencies of giant planets \textbf{discovered by Space-based facilities with Transit method (ST)} and explore their evolution with age quantitatively.

\subsubsection*{3.3.1 Collecting the parent stellar sample and planetary sample}
\label{sec.obs.SBTS.sample}
Here we adopt the {\it Kepler} targets to represent the ST subsample since {\it Kepler} mission provides a remarkable legacy, homogeneous parent stellar sample and contributes the majority (47/69) of the ST subsample.
We initialize the parent stellar sample with the LAMOST-Gaia-Kepler Kinematic catalog of PAST \uppercase\expandafter{\romannumeral2} \citep{2021AJ....162..100C}, which provides kinematic and physical properties for 35,835 {\it Kepler} stars with no bias toward planets.
Then we make quality control and kept Sun-like stars in the Galactic disk with the same criteria as \S~1 and yielding a final sample of 19,268 stars and 33 giant planets.
Due to the small size of the planetary sample, we only divided the {\it Kepler} sample into five bins with approximately equal numbers of stars.

\begin{figure}[!t]
\centering
\includegraphics[width=\linewidth]{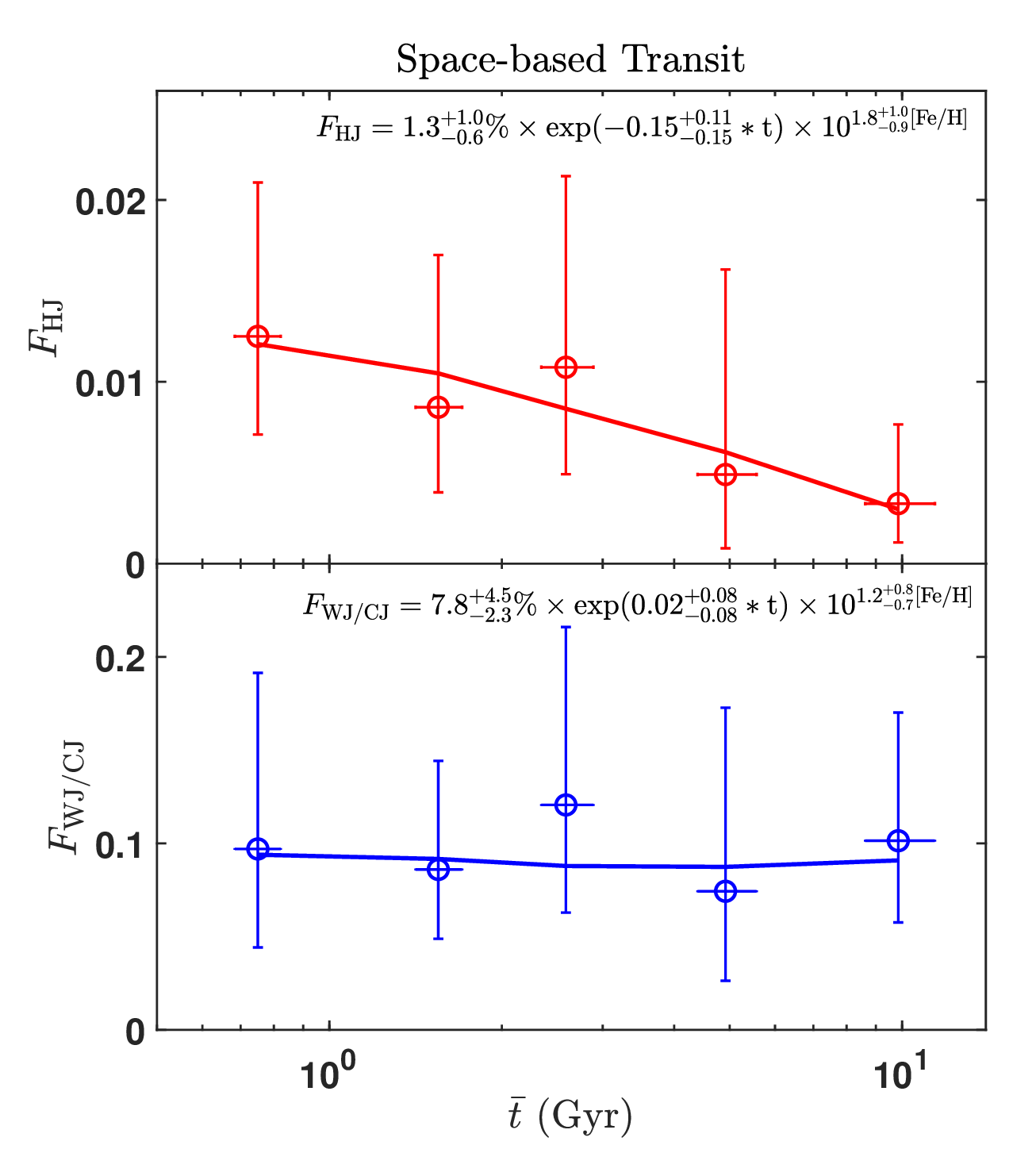}
\caption{The age evolution of the frequencies of hot Jupiters $F_{\rm HJ}$ (Up panel) and the warm/cold Jupiters $F_{\rm WJ/CH}$ (Bottom panel) after qualifying the detection efficiency derived from Space-based Transit subsample.
The solids lines represent the best fits of Equation S8.
\label{figKeplerOccurrencerateDE}}
\end{figure}

\subsubsection*{3.3.2 Correcting for the detection efficiency and geometric probability}
\label{sec.obs.SBTS.Debias}
In this subsection, we calculate the frequencies of hot Jupiters and warm/cold Jupiters with Equation S4:
\begin{enumerate}
  \item \textbf{Correcting the geometry probability} 
  For a given planet, we make the geometric probability correction by multiplying the ratio between the semi-major axis and the radius of the host star, $\frac{a}{R_*}$.
   
  \item \textbf{Qualifying the detection efficiency} 
  We calculate the transit detection efficiency using the KeplerPORTs \citep{2017ksci.rept...17B} and the detection
  metrics available from the NASA exoplanet archive (https://exoplanetarchive.ipac.caltech.edu/docs/). 
\end{enumerate}

Figure \ref{figKeplerOccurrencerateDE} displays the evolution of hot Jupiters and cold Jupiters with age derived from {\it Kepler} sample.

\begin{figure}[!t]
\centering
\includegraphics[width=\linewidth]{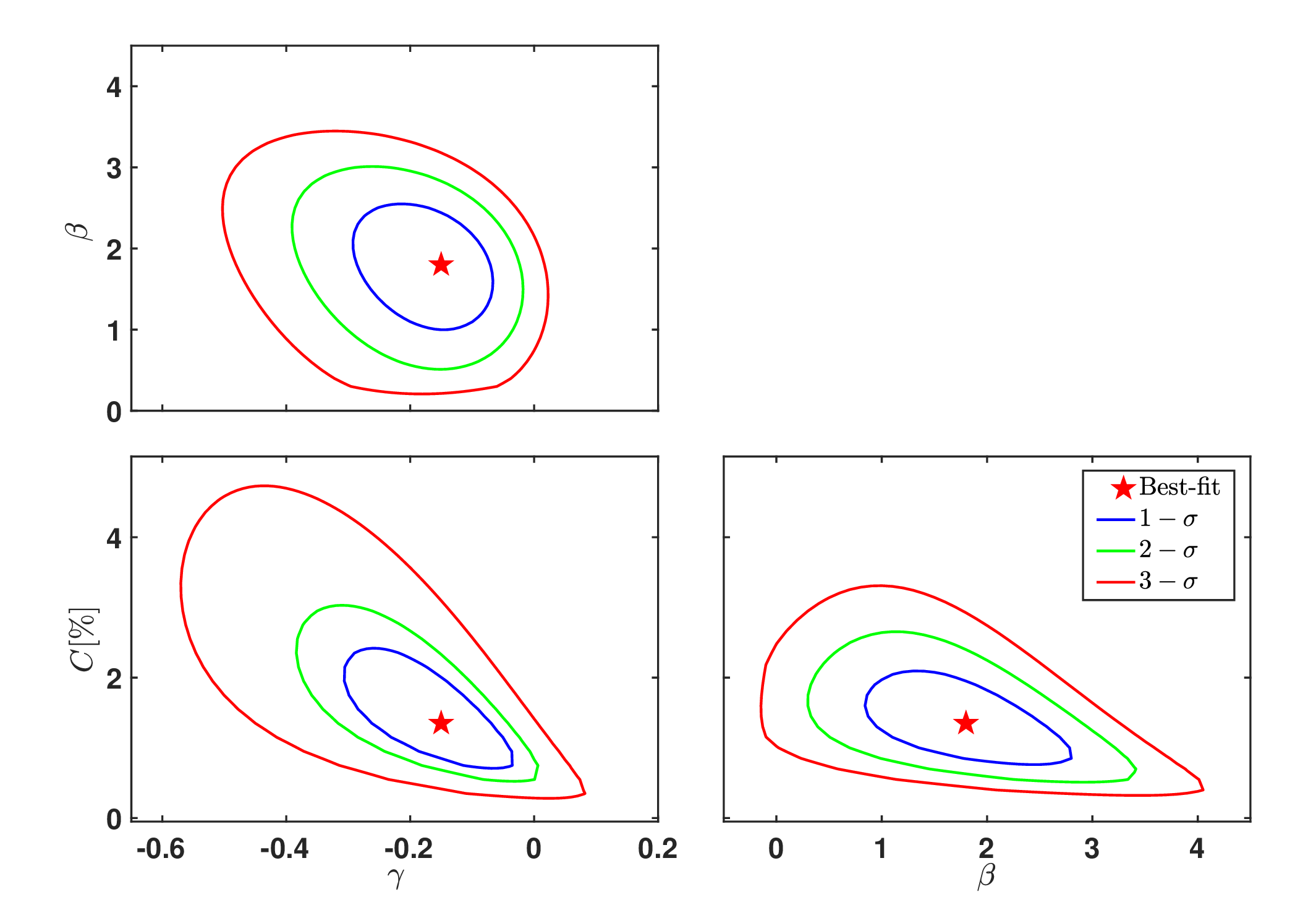}
\caption{Marginal posterior probability density functions for the model coefficients $(C, \beta, \gamma)$ for the frequency of hot Jupiters conditioned on the Space-based Transit data.
\label{figFHJfittingFeHAgeKepler}}
\end{figure}

\begin{figure}[!t]
\centering
\includegraphics[width=\linewidth]{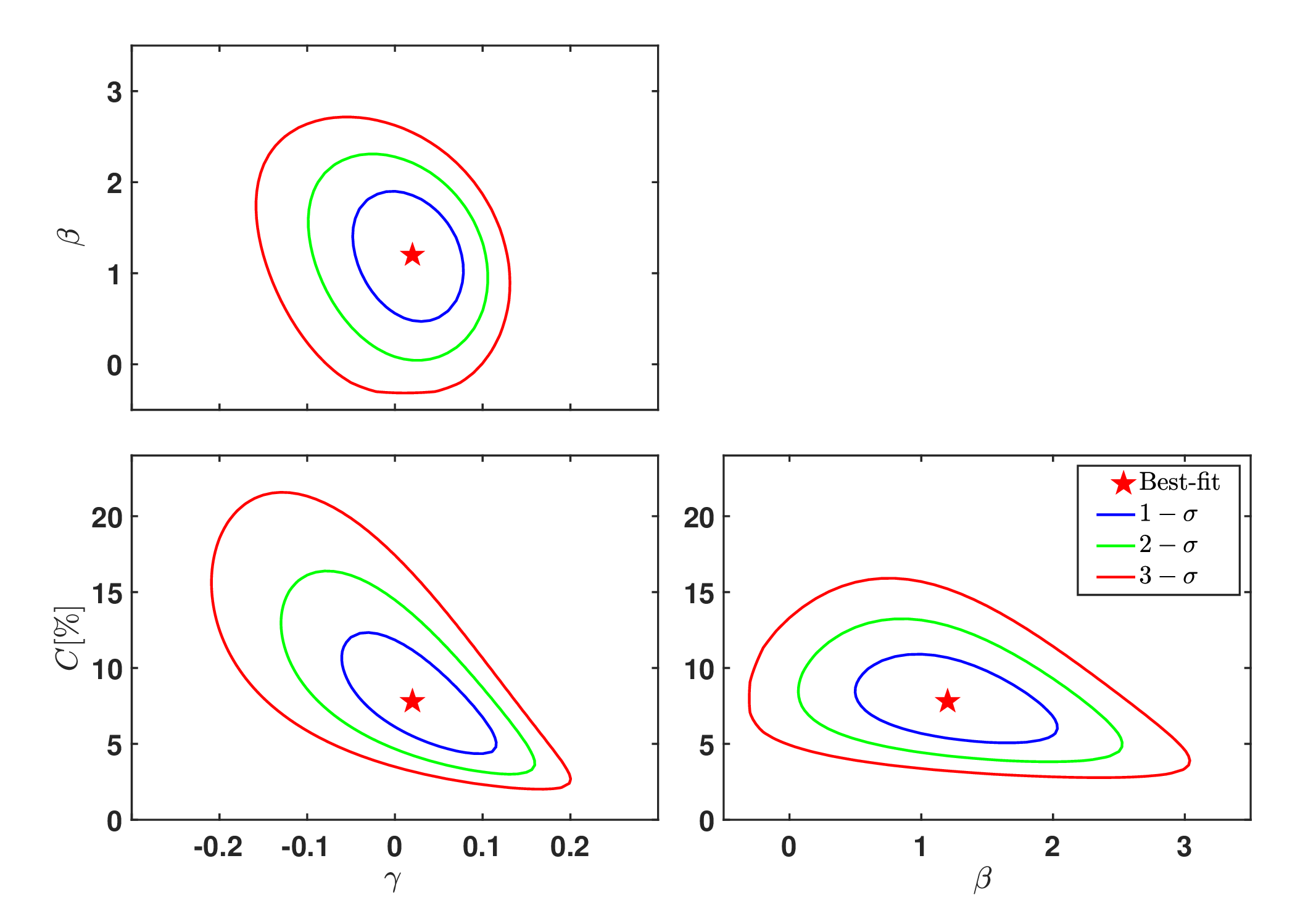}
\caption{Marginal posterior probability density functions for the model coefficients $(C, \beta, \gamma)$ for the frequency of warm/cold Jupiters conditioned on the Space-based Transit data.
\label{figFCJfittingFeHAgeKepler}}
\end{figure}

\subsubsection*{3.3.3 Model fitting}
\label{sec.obs.SBTS.fitting}
\begin{figure}[!t]
\centering
\includegraphics[width=\linewidth]{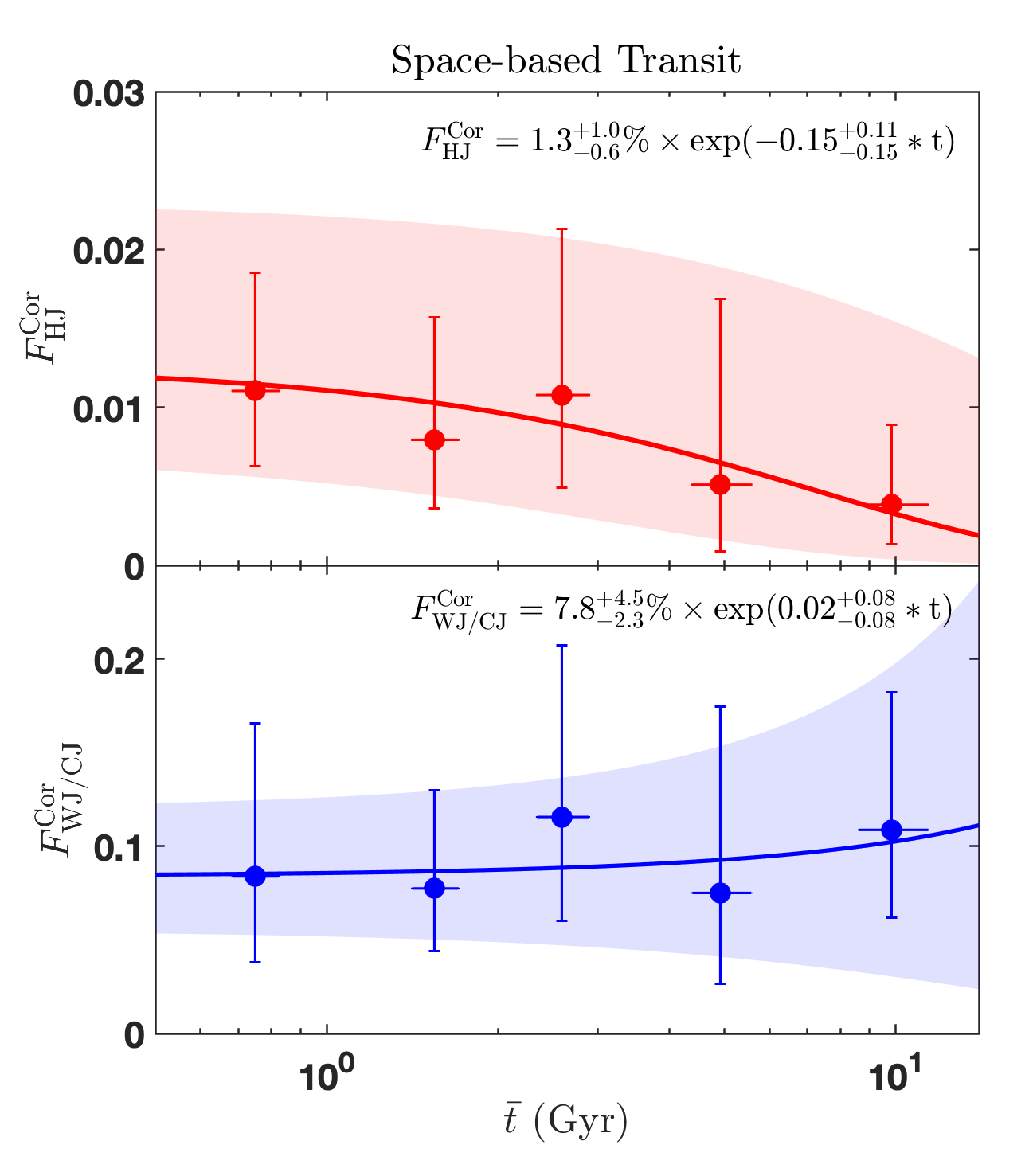}
\caption{The age evolution of the corrected frequencies of hot Jupiters $F^{\rm Cor}_{\rm HJ}$ (Up panel) and warm/cold Jupiters $F^{\rm Cor}_{\rm WJ/CJ}$ (Bottom panel) after eliminating the effect of metallicities derived from Space-based Transit subsample.
The solids lines and regions represent the best fits and $1-\sigma$ intervals of Equation S8.
\label{figKeplerOccurrencerateDEFeH}}
\end{figure}

We then fit the frequencies of hot Jupiters $F_{\rm HJ}$ and warm/cold Jupiters $F_{\rm WJ/CJ}$ with Equation S8 by adopting the same procedure as described in \S~3.1.3.
The hybrid models are preferred comparing to the single-age model/single-metallicty with $\Delta AIC$ of 6.8/4.2 and 7.2/2.1 for $F_{\rm HJ}$ and $F_{\rm WJ/CJ}$, respectively.  
Figure \ref{figFHJfittingFeHAgeKepler} and \ref{figFCJfittingFeHAgeKepler} show the probability density functions of $X$ for the evolution trends of hot Jupiters and warm/cold Jupiters.
The comparisons between the best-fit model and
the data are shown in Figures \ref{figKeplerOccurrencerateDE}.
The best-fitting parameters and their $1-\sigma$ intervals are listed in Table \ref{tab:fittingparamodel}.
As can be seen, the frequencies of hot Jupiters and warm/cold Jupiters are positively correlated ($\beta$ of $1.8^{+1.0}_{-0.9}$ and $1.2^{+0.8}_{-0.7}$) to the stellar metallicity with confidence levels (i.e., the probabilities for $\beta>0$) $\sim 2-3\sigma$.
For the dependence on stellar age, the frequency of hot Jupiters decreases with increasing age ($\gamma = -0.15^{+0.11}_{-0.15}$) and the confidence level (i.e., probability for $\gamma<0$) is 94.23\%,
while the frequency of warm/cold Jupiters is weakly-dependent ($\gamma = 0.02^{+0.08}_{-0.08}$) with age.
In Figure \ref{figKeplerOccurrencerateDEFeH}, we show the frequencies of hot Jupiter and warm/cold Jupiters after unifying $\rm [Fe/H]$ to the solar metallicity, $F^{\rm Cor}_{\rm HJ}$.

\subsection*{3.4 Joint analysis of the full sample}
\label{sec.fitting.Joint}
In this subsection, we make joint fitting for the frequencies of hot Jupiters and warm/cold Jupiters by combining the data of the above three subsamples.
Previous studies have found that $F_{\rm HJ}$ from RV surveys are higher than those from transit surveys \citep[see details in \S~4;][]{2021ARA&A..59..291Z}.
Furthermore, the effect of observing window function has not been corrected for the ground-based transit subsample.
Thus, the amplitudes $C$ of different subsamples exist discrepancy.
Therefore, in the joint fitting,
we combine the three subsamples to fit $\beta$ and $\gamma$.
However, $C$ of the three subsamples are not required to be the same.
The fitting parameters $X_{\rm Joint}$ consist of $\beta$, $\gamma$, as well as $C_{\rm RV}$, $C_{\rm GT}$, and $C_{\rm ST}$.

\begin{figure}[!t]
\centering
\includegraphics[width=\linewidth]{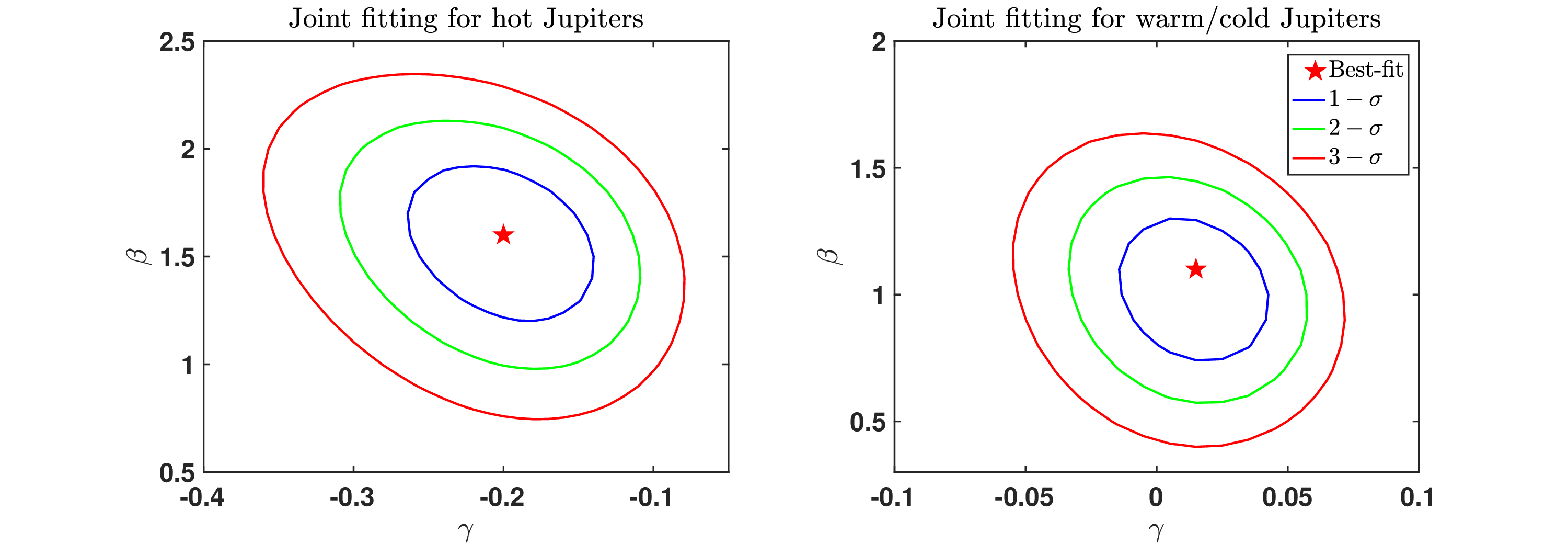}
\caption{Joint fitting for the Marginal posterior probability density functions for the model coefficients $(\beta, \gamma)$ for the frequency of hot Jupiters (Left panel) and warm/cold Jupiter (Right panel) conditioned on the whole sample.
\label{figFHJCJJointfitting}}
\end{figure}

With similar procedure described in \S~3.1.3, we calculated the joint likelihoods of detecting $H_{\rm RV}$ ($H_{\rm GT}$, $H_{\rm ST}$) from $T_{\rm RV}$ ($T_{\rm GT}$, $T_{\rm ST}$) stars.
Figure \ref{figFHJCJJointfitting} shows the joint probability density functions of $\beta$ and $\gamma$ for the evolution trends of $F_{\rm HJ}$ and $F_{\rm WJ/CJ}$.
The results from the joint fit are in general agreement with those of the three subsamples.

\section{On the discrepancy between the hot Jupiter frequencies inferred from RV and transit surveys}
\label{sec.discrepancy.RVTS}
There is an unresolved puzzle on the inconsistent frequencies of hot Jupiters inferred from RV and transit surveys \citep[e.g.,][]{2021ARA&A..59..291Z}.
Specifically, transit surveys (e.g., Kepler, OGLE-III) yield a hot Jupiter frequency of $\sim 0.3\%-0.8\%$ \citep[e.g.,][]{2006AcA....56....1G,2012ApJS..201...15H,2016A&A...587A..64S}, whereas previous studies of the RV survey report a frequency of $\sim 0.9\%-1.5\%$ that are typically a factor of $\sim 2$ higher \citep[e.g.,][]{2008PASP..120..531C,2011arXiv1109.2497M,2012ApJ...753..160W}.  
In this work, our samples yield hot Jupiter frequencies of $ 0.55^{+0.11}_{-0.10}\%$ (without correcting the weather window effect), $0.75^{+0.19}_{-0.19}\%$ and $1.26^{+0.31}_{-0.30}$ for the Ground-based transit, {\it Kepler} and RV surveys respectively, which are in agreement with previous studies.
It has been suggested that the difference in the close binary fraction between RV and transit samples can potentially explain the discrepancy in hot Jupiter frequencies if the formation of hot Jupiters is strongly suppressed in such close binary systems \citep{2021MNRAS.507.3593M,2022MNRAS.516...75B}.
If this is the case, the formation of warm Jupiters should also be strongly suppressed.
Previous studies have obtained the frequencies of warm Jupiters from RV surveys, which are $\sim 4\%-5\%$ \citep{2019ApJ...874...81F,2022AJ....164....5Z}.
To make a comparison, we select warm Jupiters from the LAMOST-Gaia-Kepler catalog with the same criteria as Zhu 2022 \citep[i.e., $0.1-1$ AU, $8-20 R_\oplus$;][]{2022AJ....164....5Z} and yield a frequency of warm Jupiters as $4.1 \pm 1.2\%$, which exists no significant difference with those from RV surveys. 

\begin{figure}[!t]
\centering
\includegraphics[width=\linewidth]{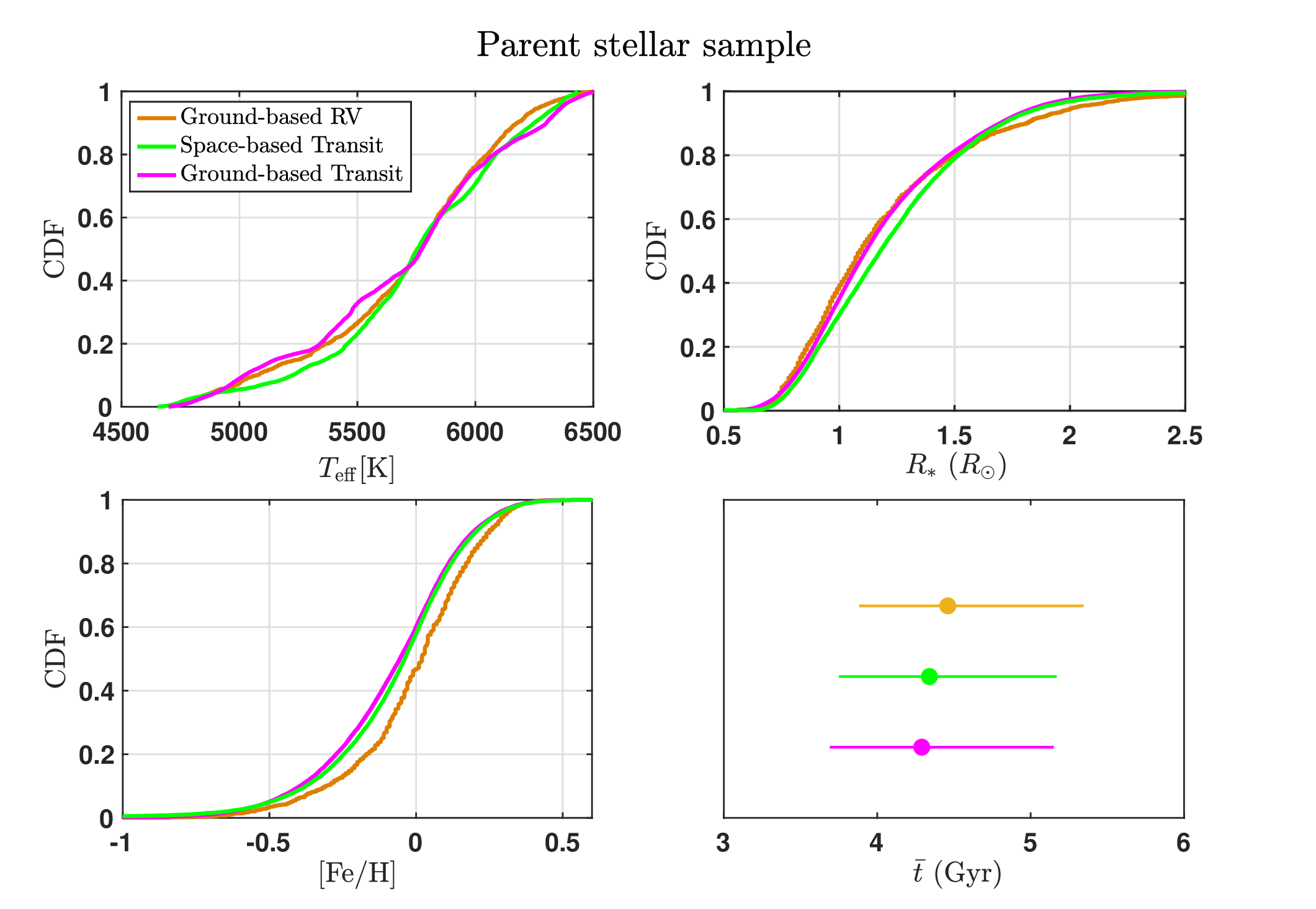}
\caption{The cumulative distributions of stellar effective temperatures ($T_{\rm eff}$), radii, $\rm [Fe/H]$ and average kinematic ages for the parent stellar sample of Ground-based RV (brown lines), Space-based transit (green lines) and Ground-based transit (purple lines) subsamples.
{The typical (median) uncertainties in $T_{\rm eff}$, $R_*$ and $\rm [Fe/H]$ are $\sim 135$ K, $\sim 7\%$ and 0.10 dex, respectively.}
\label{figParentstarspropertiesSPGBTS}}
\end{figure}

\begin{figure}[!t]
\centering
\includegraphics[width=\linewidth]{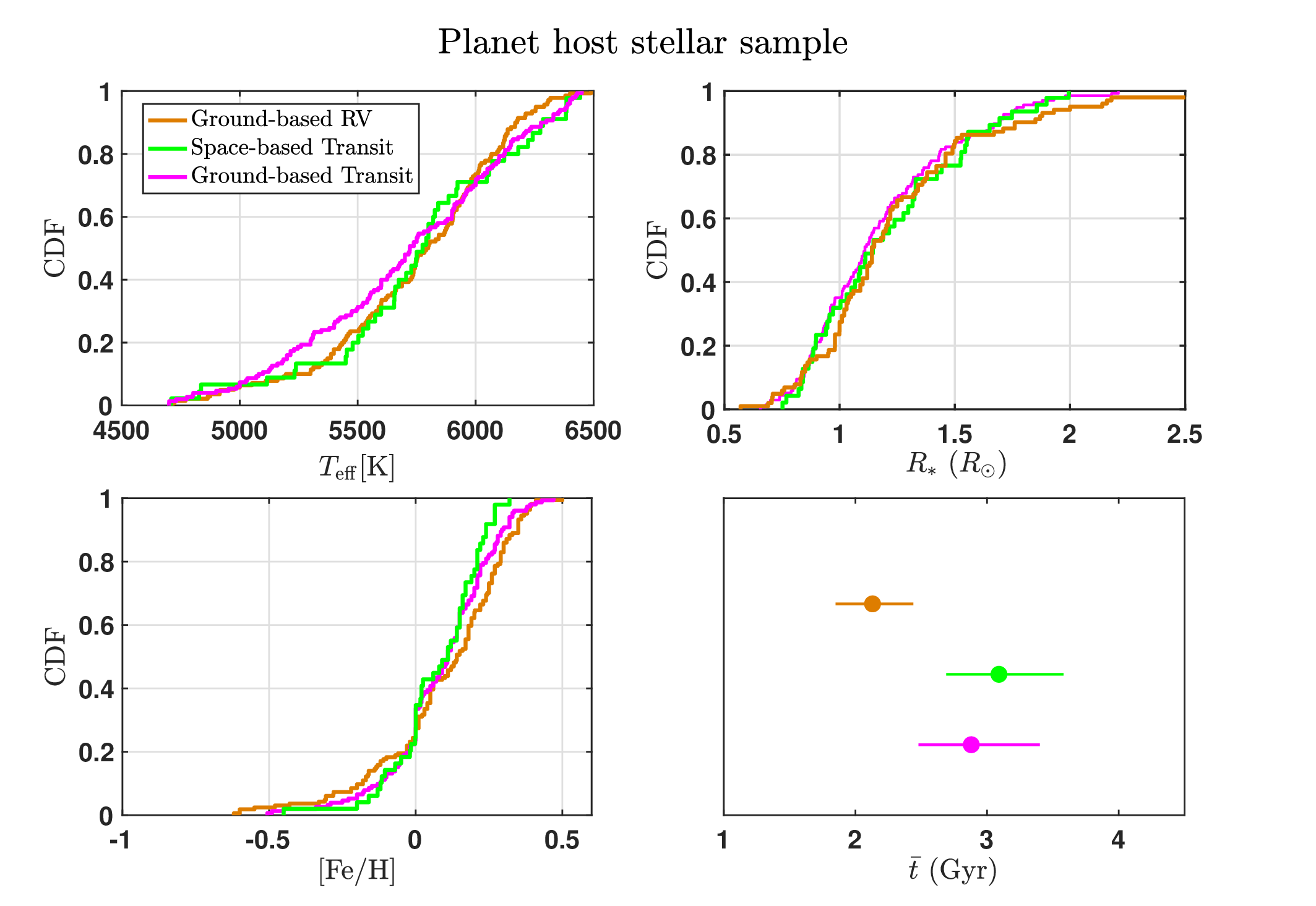}
\caption{The cumulative distributions of stellar effective temperatures ($T_{\rm eff}$), radii, $\rm [Fe/H]$ and average kinematic ages for the planet host stars of Ground-based RV (brown lines), Space-based transit (green lines) and Ground-based transit (purple lines) subsamples.
{The typical (median) uncertainties in $T_{\rm eff}$, $R_*$ and $\rm [Fe/H]$ are $\sim 120$ K, $\sim 6\%$ and 0.10 dex, respectively.}
\label{figHoststarspropertiesSPGBTS}}
\end{figure}

To explore the mechanisms leading to the above differences, we compare the distributions of stellar properties (i.e., effective temperature, radius, metallicity, and kinematic age) for RV and transit survey targets.
For the parent stellar sample, as shown in Figure \ref{figParentstarspropertiesSPGBTS}, RV targets do not differ significantly in the
distributions of $T_{\rm eff}$ and radii from those of Transit targets.
The average kinematic age of RV parent stars ($4.46^{+0.89}_{-0.58}$ Gyr) is slightly larger but statistically indistinguishable compared to {\it Kepler} stars ($4.34^{+0.83}_{-0.59}$ Gyr) and Ground-based transit target ($4.29^{+0.81}_{-0.60}$ Gyr).
Whereas, RV targets ($\rm [Fe/H]_{\rm RV} \sim 0.02$) are on average metal-richer than {\it Kepler} stars ($\rm [Fe/H]_{\rm Kepler} \sim -0.04$) and ground-based transit targets ($\rm [Fe/H]_{\rm Kepler} \sim -0.05$) as RV samples are biased toward more metal-rich star, which is in good agreement of previous studies \citep[e.g.,][]{2014ApJ...789L...3D,2017ApJ...838...25G}.
The surplus in the metallicity of RV stars will induce a higher frequency of hot Jupiters:
\begin{equation}
  K_{\rm [Fe/H]} = \frac{\left(\sum\limits_{\rm RV} 10^{\alpha \times [\rm Fe/H]}\right)/N_{\rm RV}}{\left(\sum\limits_{\rm Transit} 10^{\alpha \times [\rm Fe/H]}\right)/N_{\rm Transit}}.
\end{equation}
Using Equation 3 and using error propagation, we obtain the increasing factors caused by the surplus in metallicity:
\begin{equation}
  K^{\rm RV-ST}_{\rm [Fe/H]} = 1.16^{+0.11}_{-0.10}, \ 
  K^{\rm RV-GT}_{\rm [Fe/H]} = 1.20^{+0.13}_{-0.11}. 
\end{equation}

\begin{figure}[!t]
\centering
\includegraphics[width=\linewidth]{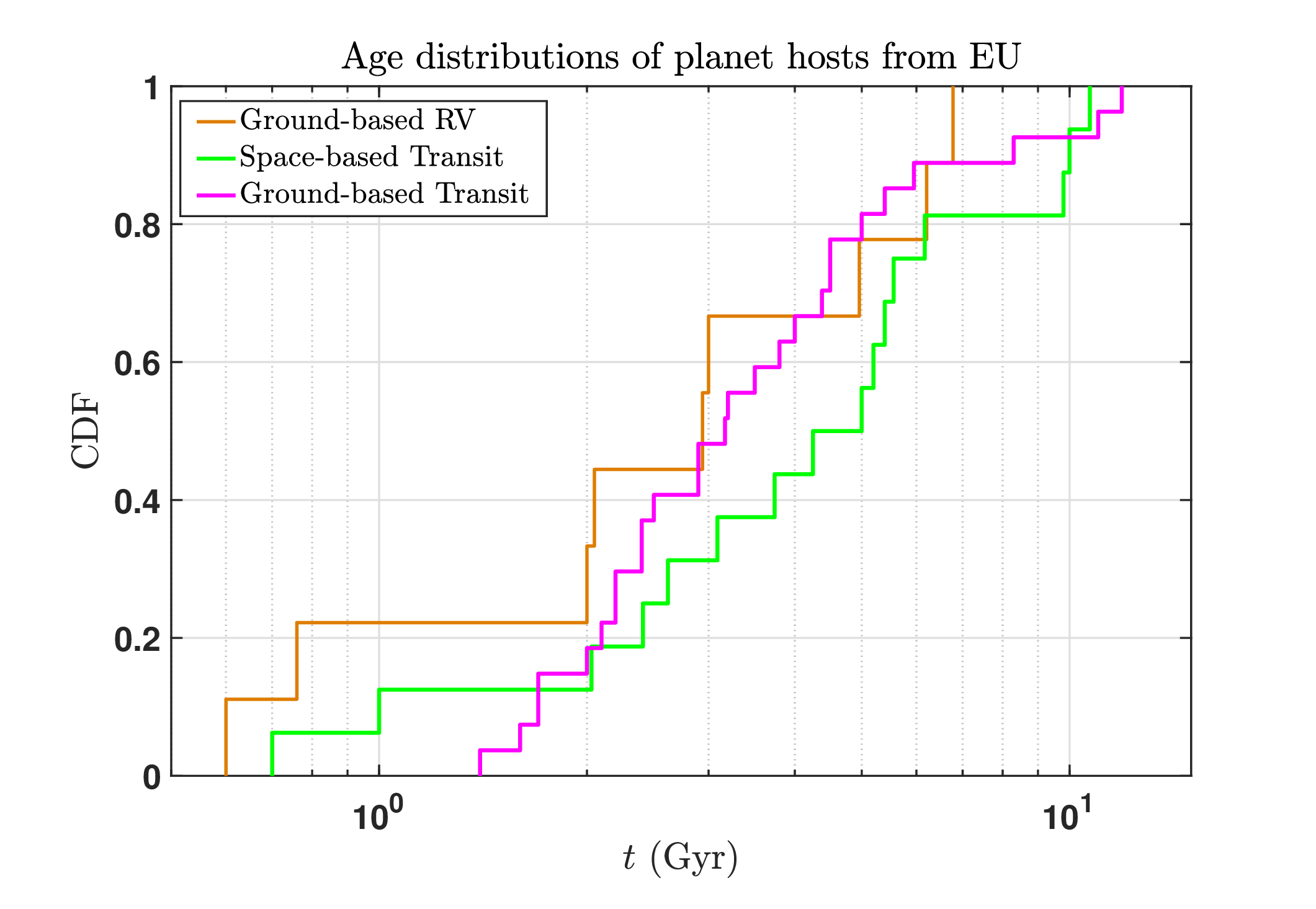}
\caption{The cumulative distributions of stellar ages for stars hosting hot Jupiters discovered by RV (brown), ST (green) and GT (purple) collected from The Extrasolar Planets Encyclopaedia.
{The typical uncertainty in Age from EU is $\sim 28\%$.}
\label{figAgeHJRVTSEU}}
\end{figure}

For the planet host stellar sample, Figure \ref{figHoststarspropertiesSPGBTS} compares the cumulative distributions of stellar properties for RV, Space-based Transit (ST) and Ground-based Transit (GT) subsamples.
As can be seen, the three subsamples have similar
distributions of $T_{\rm eff}$, radius and $\rm [Fe/H]$.
While the average kinematic age of RV planet host stars ($2.13^{+0.31}_{-0.28}$ Gyr) is significantly smaller than the Space-based Transit planet host stars ($3.09^{+0.49}_{-0.40}$ Gyr) and Ground-based Transit host stars ($2.88^{+0.45}_{-0.40}$ Gyr).
To further confirm the age difference between RV and Transit subsamples, we collect the ages of host stars from EU and only keep Sun-like stars hosting hot Jupiters with relative uncertainties in ages less than $50\%$.
As shown in Figure \ref{figAgeHJRVTSEU}, RV hot Jupiter hosts have a younger average age by $\sim 1$ Gyr than transit hot Jupiter hosts, which is in good agreement with the kinematic age difference.
{We note that stellar activity tends to decrease when the age increases, and since stellar activity can sometimes contribute substantially to the stellar photometric variability, so such an effect needs to be evaluated when assessing how planet frequency depends on age derived from transit surveys.}
{Chen et al. 2021b \citep{2021AJ....162..100C} showed that the photometric variability (Combined Differential Photometric Precision, CDPP) decreases as stars age using the {\it Kepler} data (see Fig.5 and 6 of that paper). 
For the ground-based transit subsample (SuperWASP),  we calculate the distributions of photometric variability as a function of stellar age and find that older stars tend to have smaller red noises}
We find that the frequency of hot Jupiters declines with age and thus, the age difference could cause a higher rate for RV surveys.
With the fitting coefficients and considering the error of fitting coefficients as well as the kinematic ages of different subsamples (Equation 1 and Figure \ref{figFHJCJ_JointGBSBRVTS_ABFittting}), we obtain the increasing factors and uncertainties caused by the age differences:
\begin{equation}
  K^{\rm RV-ST}_{\rm Age} = 1.25^{+0.41}_{-0.25}, 
  K^{\rm RV-GT}_{\rm Age} = 1.17^{+0.33}_{-0.21}. 
\end{equation}

The combination of the influence of stellar metallicity ($\beta$) and long-term evolution ($\gamma$) can account for the discrepancy by factors of:
\begin{equation}
  K^{\rm RV-ST}_{\rm [Fe/H]+Age} = 1.45^{+0.64}_{-0.38}, 
  K^{\rm RV-GT}_{\rm [Fe/H]+Age} = 1.40^{+0.55}_{-0.37}.
\end{equation}
which could explain (the bulk of) the discrepancy between hot Jupiter frequencies inferred from RV and transit surveys within $\sim 1-\sigma$ errorbars (see Figure \ref{figFHJRVTSdiscrepency}).
{We also note that the frequency of hot Jupiters derived from the CoRoT survey ($\sim 0.95 \pm 0.26\%$, see Santerne 2012, Ph.D. Thesis, Aix-Marseille University) is found to be higher than those of other transit surveys (0.3\%-0.8\%). 
We have estimated the average kinematic ages and found that the eight CoRoT hot Jupiters in our sample are younger ($1.96^{+1.33}_{-0.97}$ Gyr) than hot Jupiters in other transit surveys ($3.43^{+0.81}_{-0.66}$ Gyr). So we speculate that the higher frequency of hot Jupiters from  CoRoT might be due to its relatively young stellar population, as broadly expected from our age-frequency relation in Equation 2.}
Future searches and analyses for hot Jupiters in different Galactic environments and close stellar companions of transiting hot Jupiters will test our results and other potential mechanisms.

\section{On the frequencies of hot Jupiters from searches in stellar clusters}
\label{sec.HJ.globularcluster}
Stellar clusters offer powerful probes into the effects of age and metallicities on the formation and evolution of hot Jupiters since stars in clusters can be assumed to be born at the same time with identical chemical compositions. Thus their properties, such as age, mass, and metallicities, can be precisely determined.
Among them, globular clusters are composed of old stars with low metallicities compared to the field stars of the Milky Way galaxy, while stars in open clusters have metallicities similar to that of the Sun and are typically younger than the field stars, which can potentially be used to constrain the early formation and migration of planets.

Plenty of research groups have performed planet surveys in clusters.
To date, tens of hot Jupiters (candidates) have been found in open clusters.
Specifically, using RV method, two Hot Jupiters, Pr0201 b and Pr0211 b have been discovered in the Praesepe open cluster \citep{2012ApJ...756L..33Q}; one eccentric Hot Jupiter, HD 285507 b has been found in the Hyades open cluster \citep{2014ApJ...787...27Q}; three hot Jupiters have been reported in M67 open cluster \citep{2014A&A...561L...9B,2016A&A...592L...1B,2017A&A...603A..85B}.
Recently, with transiting method, several hot Jupiters (e.g., TOI 837b) have been reported in open clusters \citep[e.g., IC 2602, NGC752, NGC 7789; Fang et al. 2022, In prep.,][]{2021MNRAS.505.3767N}.
With these data, several studies have estimated the frequencies of hot Jupiters around young stars, which are summarized in Table \ref{tab:FHJClusters}.
As can be see, Nardiello et al. (2021) infer a frequency much smaller than other works \citep{2021MNRAS.505.3767N}, which may be caused by an overestimate of detection efficiency/false positive probability and/or missing detection of hot Jupiters around active stars in their sample (Fang et al. 2022, In prep).
Expect it, the derived frequencies from other works are generally consistent within $1-\sigma$ errorbar and are substantially higher than those of field stars (after unifying to solar metallicities), which is consistent with our results that younger stars have higher frequencies of hot Jupiters and may suggest that part of hot Jupiters have been destroyed during the early evolution stage driven by the tidal force of the host stars or dynamic interactions of other planets/companions.

However,  for the globular clusters, several groups have observed 47 Tucanae (NGC 104), $\omega$ Centauri (NGC 5139) and NGC 6397 to search for transiting hot Jupiters \citep{2000ApJ...545L..47G,2008ApJ...674.1117W,2012A&A...541A.144N,2017AJ....153..187M}, but none has been detected, presenting an important puzzle whether or why globular clusters have a lower frequency of hot Jupiters.
A potential reason is that globular cluster stars have a relatively lower mass compared to field stars, yielding a smaller frequency of hot Jupiters since the frequency of hot Jupiters is positively correlated to stellar mass \citep{2010PASP..122..905J}.
Some other studies proposed that the absence of hot Jupiters could be explained by the low intrinsic metallicities of globular clusters \citep[e.g., NGC GC and NGC6397 with metallicities of about -1 to -2;][]{2008ApJ...674.1117W,2012A&A...541A.144N} since the frequency of hot Jupiter is found to be highly dependent on stellar metallicity from {\it Kepler} data \citep[e.g.,][]{2017ApJ...838...25G}. 
However, the {\it Kepler} field only includes a small number of low-metallicity stars, and the dependence on metallicity remains challenging for globular clusters with $\rm [Fe/H] \lesssim -0.5$ \citep[e.g.,][]{2017AJ....153..187M}.
Besides, transit surveys for clusters of solar metallicity (NGC 6940, $\rm [Fe/H] = 0.01$) and of supersolar metallicity (NGC 6791, $\rm [Fe/H] = 0.3$) also report null results of hot Jupiters \citep{2003AJ....125.1397C,2009AJ....137.4949B}.

\begin{figure}[!t]
\centering
\includegraphics[width=\linewidth]{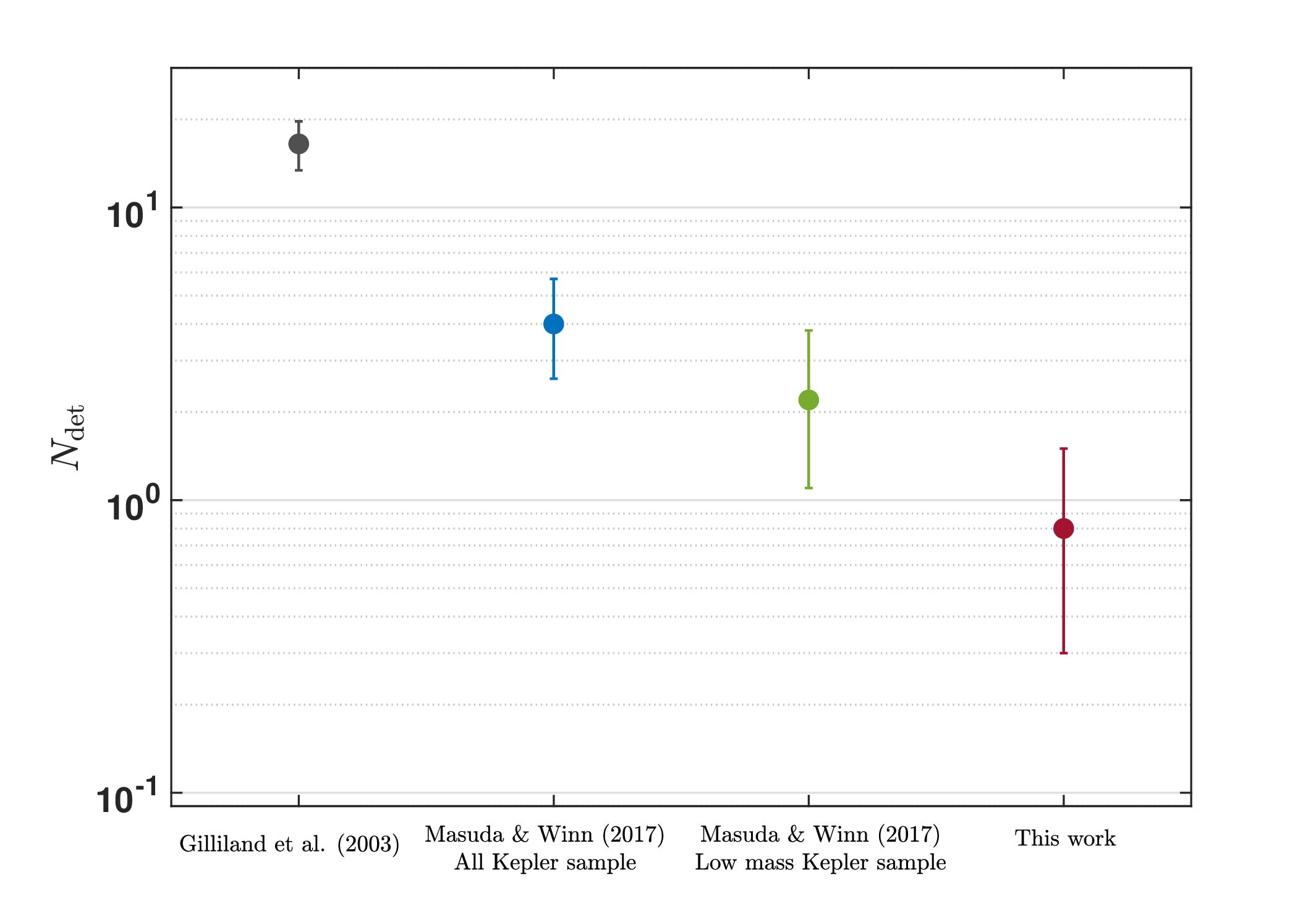}
\caption{The expected numbers of detected hot Jupiters $N_{\rm det}$ in the HST survey of 47 Tucanae.
From left to right, we show the results from Gilliland et al. (2000), Masuda \& Winn (2017) with all the {\it Kepler} stars, Masuda \& Winn (2017) with mass-controlled {\it Kepler} subsample, and our result which based on the former one and further consider the  decay of $F_{\rm HJ}$ with age.
\label{figNdet47Tuc}}
\end{figure}

Gilliland et al. (2003) presented a Hubble Space Telescope survey to search hot Jupiters by observing $\sim 34,000$ main-sequence stars in the 47 Tucanae (age of 11.6 Gyr,
$\rm [Fe/H] = -0.69$) nearly continuously for 8.3 days \citep{2003ApJ...585.1056G}.
The authors expected to find $\sim 17$ hot Jupiters if the 47 Tucanae stars have the same frequencies of hot Jupiters as those of the surveyed nearby stars.
Masuda \& Winn (2017) revisited the HST survey of hot Jupiters, and the expected detection number of hot Jupiter reduced to $4.0^{+1.7}_{-1.4}$ by assuming that 47 Tucanae and {\it Kepler} stars have identical planet populations \citep{2017AJ....153..187M}.
The expected number was further reduced to $2.2^{+1.6}_{-1.1}$ when considering the same range of stellar mass as 47 Tucanae ($0.568-0.876 M_\odot$), which can not fully explain the observed null result with $\gtrsim 2\sigma$ significance.

A factor not considered in previous works is that the globular clusters are old enough that some hot Jupiters have undergone tidal orbital decay \citep[e.g.,][]{2008ApJ...678.1396J,2009ApJ...698.1357J,2009ApJ...692L...9L}, leading to a decrease in the observed frequency.
To quantify the effect of stellar age on the expected number of hot Jupiters, we first select main-sequence stars ($\log g>4)$ belonging to galactic disk (same criteria in \S~1) with the same range of stellar mass ($0.568-0.876 M_\odot$) with 47 Tucanae from LAMOST-Gaia-Kepler catalog, yielding a sample of 3,556 stars.
Then we calculate their kinematic age from vertical velocity dispersion with Equation S1, $t=6.54^{+1.18}_{-0.93}$ Gyr.
Finally, using Equation 3 and the coefficients obtained from {\it Kepler} sample, we obtain the expected number of detected hot Jupiters by restricting {\it Kepler} stars with identical planet populations, stellar mass, and age as 47 Tucanae:
\begin{equation}
  N_{\rm det} = 0.8^{+0.7}_{-0.5},
\end{equation}
which is consistent with the observed null result within $\sim 1\sigma$ interval.
The expectations of the detection numbers of hot Jupiters in 47 Tucanae from different works are summarized in Figure \ref{figNdet47Tuc}.

\section{Further discussions}
\label{sec.dis}

\subsection*{6.1 The influence of stellar metallicity on the age difference between hot Jupiters and warm/cold Jupiters}
\label{sec.dis.FeH}

Thick disk stars are known to be more metal-poor than the thin disk \citep{1989ARA&A..27..555G,2003A&A...410..527B,2014A&A...562A..71B,2018MNRAS.478.4513B}. 
In other words, stars with larger $TD/D$ (kinematic age) are likely more metal-poor.
It has also been found that hot Jupiters' frequency has a stronger dependence on the stellar metallicity compared to warm/cold Jupiters \citep{2010PASP..122..905J,2015AJ....149...14W,2016A&A...587A..64S}.
The combination of the above two correlations could cause the preference of hot Jupiter for younger stars.
Thus it is necessary to clarify whether the age difference between hot Jupiter hosts and warm/cold hot Jupiter hosts is caused by the decrease of metallicity with age or not.

\begin{figure}[!t]
\centering
\includegraphics[width=\linewidth]{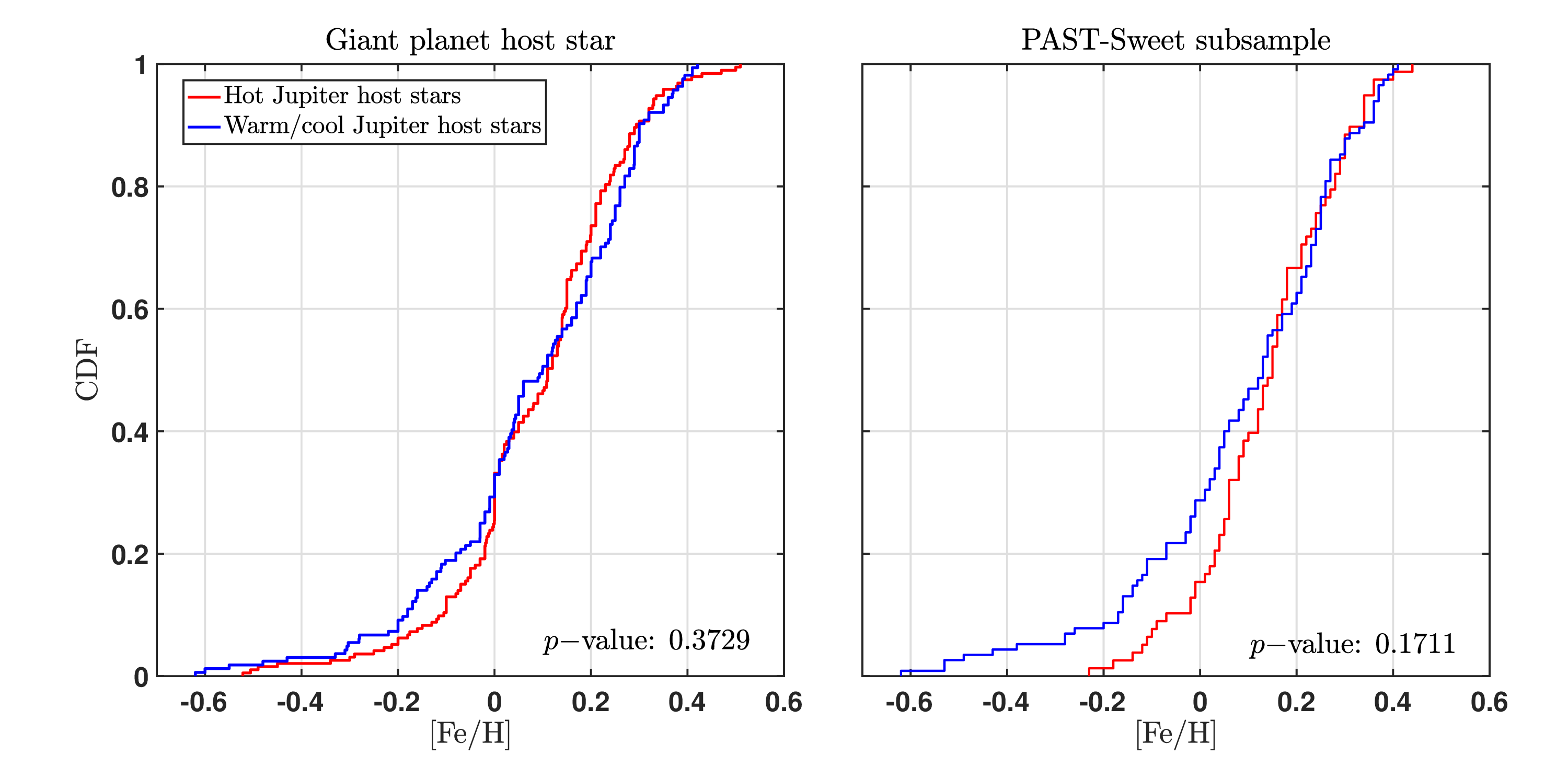}
\caption{The cumulative distributions of metallicity ($\rm [Fe/H]$) for hot Jupiter hosts (red) and warm/cold Jupiter hosts (blue) in the giant planet host sample (Left panel) and a homogeneous subsample by crossmatching with Sweet catalog (Right panel).
{The typical uncertainties in $\rm [Fe/H]$ are $\sim 0.1$ and $\sim 0.05$ dex for giant planet host stellar sample and PAST-sweet subsample, respectively.}
\label{figFeH_HJvsCJPAST}}
\end{figure}

To answer this question, we  compare the distributions of metallicities for the hot Jupiter hosts and warm/cold Jupiter hosts (shown in the left panel of Figure \ref{figFeH_HJvsCJPAST}) and the resulted $p-$value of Kolmogorov–Smirnov (K-S) test is 0.3729, demonstrating that their distributions in metallicities are statistically indistinguishable.

Furthermore, the metallicity data from PAST \uppercase\expandafter{\romannumeral1} is collected from various literature and thus inhomogeneous, which may lead to system differences from different sources.
To further clarify this, we rely on the Sweet catalog, which compiles sets of atmospheric parameters ($T_{\rm eff}$, $\log g$ and $\rm [Fe/H]$) for 2981 planet hosts \citep{2013A&A...556A.150S}.
Among these stars, Sweet team derives atmospheric parameters using the same uniform methodology for 643 stars and provides a homogeneous sample.
We crossmatch our sample in \S~\ref{sec.meth.sample} with the Sweet homogeneous sample and obtains $\rm [Fe/H]$ data from Sweet catalog for 214 giant planets (including 78 hot Jupiters).
In the right panel of Figure \ref{figFeH_HJvsCJPAST}, we compare the distributions of Sweet metallicity for hot Jupiter hosts and warm/cold Jupiter hosts.
As can be seen, the difference in metallicities is larger but still statistically insignificant (with a K-S $p-$ value of 0.1711).

Based on the above analyses, we conclude that the difference in metallicity is not the (dominant) reason that causes the difference in age distributions of hot Jupiters and warm/cold Jupiters.

\subsection*{6.2 The effect of phase space density on the age difference between hot Jupiters and warm/cold Jupiters}
\label{sec.dis.density}
Recently, Winter et al. (2020) \cite{2020Natur.586..528W} has estimated the probability that a planet host star belongs to over-density or field star in the position-velocity phase space using data from Gaia DR2 and explored the correlation between hot Jupiters and phase space density $P_{\rm high}$.
Their results show that hot Jupiters favor over-density stars.
However, Adibekyan et al. (2021) \cite{2021A&A...649A.111A} constructed a homogeneous sample of FGK dwarf stars with only RV-detected Jupiter using the SWEET-Cat catalog to test the former results, and they found no significant correlation between phase space density and hot Jupiter frequency.
This discrepancy indicates that the observation bias caused by the detection methods (e.g., RV and transits) may significantly affect the results of Winter et al. (2020) \cite{2020Natur.586..528W}. 
Besides, Adibekyan et al. (2021) \cite{2021A&A...649A.111A} and Mustill et al. (2021) \cite{2021arXiv210315823M} found that there exists a strong anti-correlation between the phase space density and stellar velocities/ages and suggested that the result of Winter et al. (2020) \cite{2020Natur.586..528W} may be caused by the bias towards detecting hot Jupiters around young stars due to tidal destruction. 

\begin{figure}[!t]
\centering
\includegraphics[width=\linewidth]{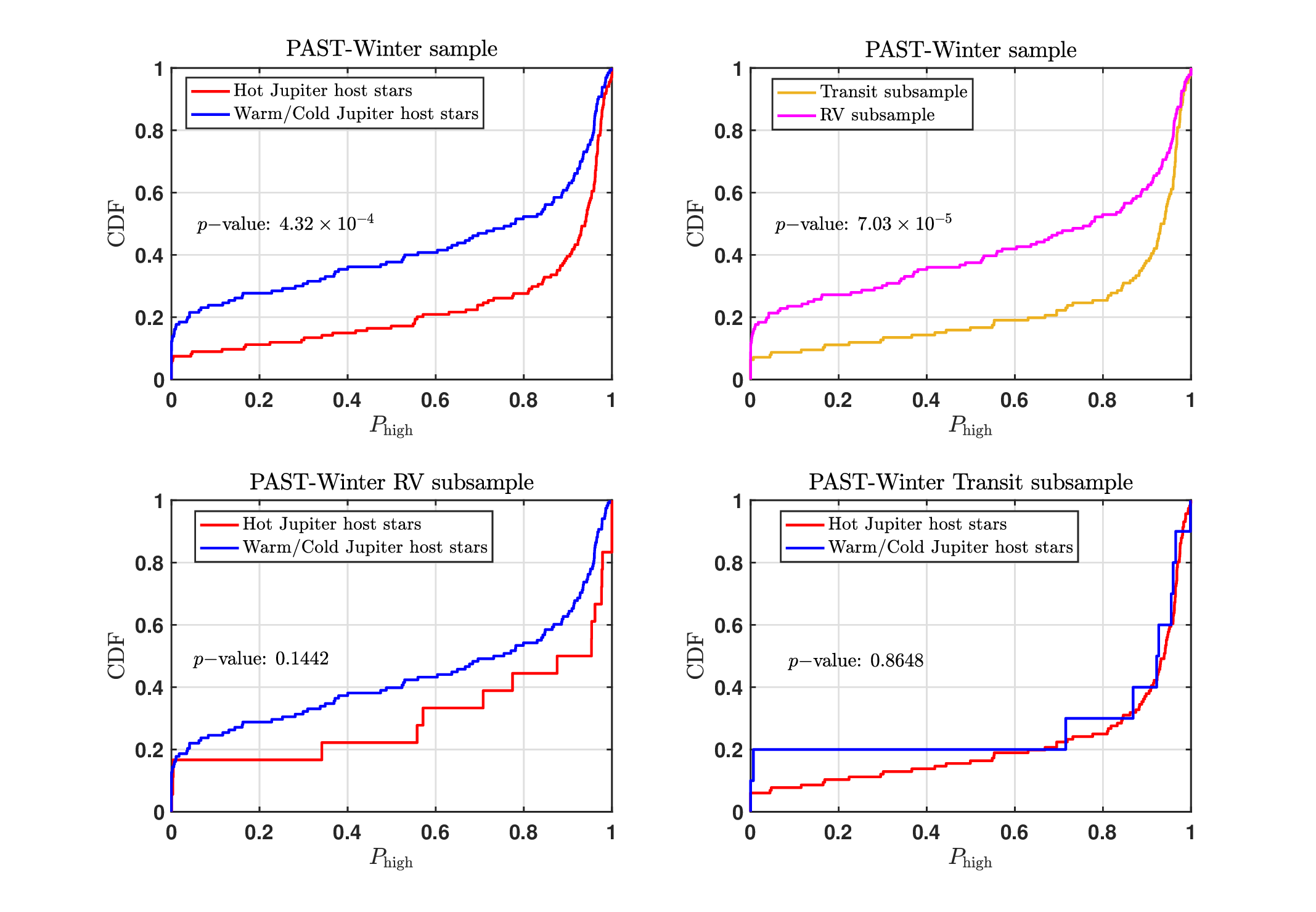}
\caption{The distributions of the phase space density $P_{\rm high}$ for giant planets with different discovery methods and orbital periods.
The K-S test $p-$values are printed at the left of each panel.
\label{figPhighPASTSweet}}
\end{figure}

\begin{figure}[!t]
\centering
\includegraphics[width=\linewidth]{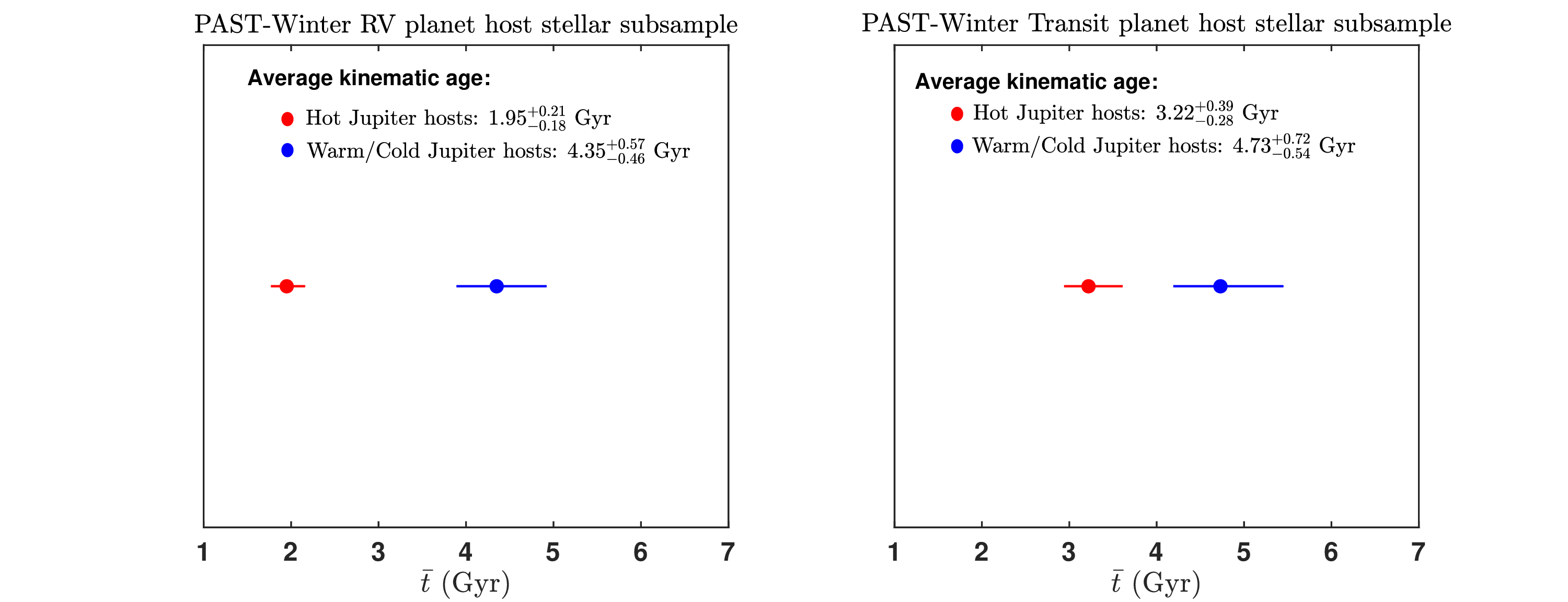}
\caption{The kinematic ages and uncertainties of hot Jupiter host stars (red) compared to warm/cold Jupiter host stars(blue) for RV detected subsample (top panel) and transit detected subsample (bottom panel) by crossmatching our giant planet host sample with that of \cite{2020Natur.586..528W}.
\label{figAgeHJCJRVTS_PASTWinter}}
\end{figure}

In order to clarify whether the frequency of hot Jupiters is (strongly) related to phase density or age, we crossmatch our giant planet host sample with that of Winter et al. (2020) \cite{2020Natur.586..528W} and obtain a PAST-Winter sample of 264 stars hosting 287 giant planets (including 134 Hot Jupiters) and then performed the following two exercises.

\begin{enumerate}

\item  
Firstly, we compare the distributions of the phase space densities for giant planets with different discovery methods and orbital periods.
As shown in Figure \ref{figAgeHJCJRVTS_PASTWinter}, hot Jupiters seem to be preferentially around over-density stars compared to warm/cold Jupiters.
However, this can be adequately be explained by the two combined factors:
transit detected planets favor over-density stars (even more significant with smaller $p-$value) and most of hot Jupiters are detected by transit while warm/cold Jupiters are mainly discovered by RV.
To clarify this, we compare the distribution of $P_{\rm high}$ of hot and warm/cold Jupiters detected by the same methods. 
As can be seen in the bottom two panels of Figure \ref{figAgeHJCJRVTS_PASTWinter}, there exists no significant difference (with $p-$values all larger than 0.15) in $P_{\rm high}$ between hot and warm/cold Jupiters both from RV and transit subsample.
This is well consistent with the result of \cite{2021A&A...649A.111A} and further supports that hot Jupiters have no (significant) correlation with $P_{\rm high}$ after eliminating the influence of detection methods.

\item Secondly, as discussed before, hot Jupiters and warm/cold Jupiters detected by the same methods (RV/Transit) have similar distributions in phase space density, eliminating the influence of both detection bias and phase space density on the stellar age.
Thus, whether the difference in their kinematic ages is still significant or not can verify whether hot Jupiters are related to age.
We calculate the kinematic ages of hot Jupiters and warm/cold Jupiter host subsample detected by RV and transit.
As shown in Figure \ref{figAgeHJCJRVTS_PASTWinter}, both RV and transit subsamples show that hot Jupiters are preferentially around younger stars with a confidence level of 99.99\% and 97.39\%, proving that phase space density is not the reason caused the tendency of hot Jupiter towards younger stars.

\end{enumerate}

The above explorations demonstrate that phase space density has little influence on the hot Jupiters and thus can not cause the age difference between hot Jupiter and warm/cold Jupiters or the decline of the frequency of hot Jupiters with age.

\section*{References}
\bibliography{main.bib}

\clearpage

\begin{table*}[!t]
\renewcommand\arraystretch{1.0}
\centering
\caption{Sample selection: Jovian planets.}
{\footnotesize
\label{tab:sampleprocedure}
\linespread{1.8}
\begin{tabular}{l|ccc} \hline
Selection criteria & $N_{\rm s}$ & $N_{\rm p}$ \\\hline
Planet host sample (PAST \uppercase\expandafter{\romannumeral1}) & 2174 & 2872 \\
$T_{\rm eff}:4700-6500$ K & 1805 & 2363 \\
$\log g>4$ &  1405 & 1886 \\
Belonging to Galactic disks & 1105 & 1489 \\ 
With giant planets & 357 & 385 \\ 
{Planet discovered by RV/Transit} & 355 & 383 \\
With hot Jupiters & 193  & 190 \\
\hline
\end{tabular}}
\flushleft
{\centering $N_{\rm s}$ and $N_{\rm p}$ are the numbers of host stars and planets during the process of sample selection in \S~1.}
\end{table*}

\begin{table*}[!t]
\renewcommand\arraystretch{1.2}
\centering
\caption{The number of planets discovered by different methods and facilities}
{\footnotesize
\label{tab:planetnumberdisc}
\begin{tabular}{lcc} \hline
     $N$    &  giant planets   & hot Jupiters  \\ \hline
   Space-based \& Transit &  69 & 29   \\ 
   Ground-based \& RV  & 164 & 17\\ 
   Ground-based \& Transit  &  150 & 147  \\ \hline

\end{tabular}}
\flushleft
{Discovery facilities: Space-based (Kepler, K2, CoRoT), Ground-based facilities; \\
Discovery methods: Radial velocity (RV), Transit.}
\end{table*}

\begin{table*}[!t]
\centering
\renewcommand\arraystretch{1.2}
\caption{fitting parameters of the Age-Velocity dispersion relation (AVR) from PAST \uppercase\expandafter{\romannumeral1} \citep{2021ApJ...909..115C}.}
{\footnotesize
\label{tab:AVRkb}
\begin{tabular}{l|cccccc} \hline
                & \multicolumn{2}{c}{---------~~$k \ \rm (km \ s^{-1})$~~---------} & \multicolumn{2}{c}{---------~~$b$~~---------}    \\ 
               &  value &  $1-\sigma$ interval &  value &  $1-\sigma$ interval  \\ \hline
    $U$  & $23.66$ & (23.07, 24.32) & $0.34$ & $(0.33,0.36)$  \\
    $V$  & $12.49$ & (12.05, 12.98) & $0.43$ & $(0.41,0.45)$  \\
    $W$  & $8.50$ & (8.09, 8.97) & $0.54$ & $(0.52,0.56)$ \\
    $V_{\rm tot}$ & $27.55$ & (26.84, 28.37) & $0.40$ & $(0.38, 0.42)$ \\ \hline
 
\end{tabular}}
\end{table*}  

\begin{table*}[!t]
\renewcommand\arraystretch{1.2}
\centering
\caption{fitting for the coefficients the evolutionary trends with Age and metallicities}
{\footnotesize
\label{tab:fittingparamodel}
\begin{tabular}{cc|cc|cc|cc} \hline
    &   &  \multicolumn{2}{c}{$C$}   &  \multicolumn{2}{c}{$\beta$} & \multicolumn{2}{c}{$\gamma$}\\
 &  & value  & $1-\sigma$ interval  & value  & $1-\sigma$ interval & value  & $1-\sigma$ interval \\ \hline
 \multirow{2}{*}{RV} & $F_{\rm HJ}$ & 2.0\% & (1.2\%,3.0\%) & 1.8 & (1.2, 2.6) & -0.38 & (-0.65, -0.23) \\
 & $F_{\rm WJ/CJ}$ 
   & 7.1\% & (5.9\%, 8.5\%) & 1.0 & (0.7, 1.4) & 0.02 & (-0.05,0.05)  \\
     \hline
  GT & $F_{\rm HJ}$ & 0.8\% & (0.6\%, 1.0\%) & 1.5 & (1.0, 1.9) & -0.21 & (-0.29, -0.10)   \\ \hline
  
  \multirow{2}{*}{ST} & $F_{\rm HJ}$ & 1.3\% & (0.7\%, 2.3\%) & 1.8 & (0.9, 2.8) & -0.15 & (-0.30, -0.04)\\
 & $F_{\rm WJ/CJ}$  & 7.8\% & (4.5\%, 12.3\%) & 1.2 & (0.5, 2.0) & 0.02 & (-0.06, 0.10)\\ \hline

 \multirow{2}{*}{Joint} & $F_{\rm HJ}$ &  & & 1.6 & (1.3, 1.9) & -0.20 & (-0.26, -0.14)\\
 & $F_{\rm WJ/CJ}$  &  &  & 1.1 & (0.8, 1.3) & 0.02 & (-0.01, 0.04)\\ \hline
\end{tabular}}
\end{table*}

\begin{table*}[!t]
\renewcommand\arraystretch{1.2}
\centering
\caption{The frequency of hot Jupiters in open culsters from previous literature}
{\footnotesize
\label{tab:FHJClusters}
\begin{tabular}{c|c|c|c|c|c} \hline
    Name & Age & $\rm [Fe/H]$ & $F_{\rm HJ}$ & $F^{\rm Cor}_{\rm HJ}$ & references \\
    Praesepe & $590-830$ Myr & $0.19 \pm 0.04$ & $3.8^{+5.0}_{-2.4}\%$ & $1.9^{+4.7}_{-1.0} \%$  & Quinn et al. (2012) \\
    Hyades \& Praesepe & $\sim 625$ Myr & $0.13 \pm 0.01$ & $1.97^{+1.92}_{-1.07}\%$ & $1.22^{+1.35}_{-0.70}\%$ & Quinn et al. (2014) \\
    Pleiades & $115 \pm 5$ Myr  & $0.03 \rm 0.05$ &  $\sim 3.7\%$  & $\sim 3.3\%$  & Takarada et al. (2020) \\
    \multirow{3}{*}{M67} & \multirow{3}{*}{3.5-4.8 Gyr} & \multirow{3}{*}{$0.03 \pm 0.01$} & $2.0^{+3.0}_{-1.5}\%$ & $1.8^{+2.6}_{-1.4}\%$ & Brucalassi et al., (2014) \\
    &  &  & $4.5^{+4.5}_{-2.5}\%$ & $4.0^{+4.1}_{-2.4}\%$ & Brucalassi et al., (2016) \\
    &  &  &  $5.7^{+5.5}_{-3.0}\%$ & $5.0^{+5.1}_{-2.8}\%$ & Brucalassi et al., (2017) \\
    Group &  $\sim 10-1000$ Myr & $0.0 \pm 0.3$ & $<0.1\%$ & $<0.1\%$ & Nardiello et al. (2021) \\
    Group & $\sim 10-200$ Myr & $\sim 0$ & $1.8^{+1.8}_{-1.8}\%$ & $1.8^{+1.8}_{-1.8}\%$ & Fang et al. (2022), In prep \\ \hline  
\end{tabular}}
\flushleft
{$F^{\rm Cor}_{\rm HJ}$ denotes the frequency of hot Jupiters after unifying $\rm [Fe/H]$ to solar metallicities. \\
Names labelled as 'Group' represent surveys detecting multiple clusters.}  
\end{table*}


\end{document}